\documentclass[onecolumn,aps,floatfix,nopacs,superscriptaddress]{revtex4}
\usepackage{amsmath}
\usepackage{amssymb}
\usepackage{graphicx} 
\usepackage{color}
\usepackage{ulem}

\definecolor{ca}{rgb}{0.9,0,0}    
\newcommand{\ca}[1]{\textcolor{ca}{#1}} 

\definecolor{cb}{rgb}{0,0.5,0}    
\newcommand{\cb}[1]{\textcolor{cb}{#1}} 

\definecolor{cc}{rgb}{0,0,1}    
\newcommand{\cc}[1]{\textcolor{cc}{#1}} 

\definecolor{mauve}{rgb}{0.75,0,0.6}

\definecolor{autre}{rgb}{0.1,0.6,0.8}

\begin{document} 

\title{The Exclusion Process: A paradigm for non-equilibrium behaviour}
\author{Kirone Mallick}
\address{Institut de Physique Th\'eorique CEA, IPhT, F-91191 Gif-sur-Yvette,
 France}

\begin{abstract} 
In these lectures, we shall present some remarkable results that have
been obtained for systems far from equilibrium during the last two
decades. We shall put a special emphasis on the concept of large
deviation functions that provide us with a unified description of many
physical situations. These functions are expected to play, for systems
far from equilibrium, a role akin to that of the thermodynamic
potentials. These concepts will be illustrated by exact solutions of
the Asymmetric Exclusion Process, a paradigm for non-equilibrium
statistical physics.
\end{abstract}

\maketitle

\begin{figure}[ht]
 \begin{center}
  \includegraphics[height=5.5cm,angle = 0]{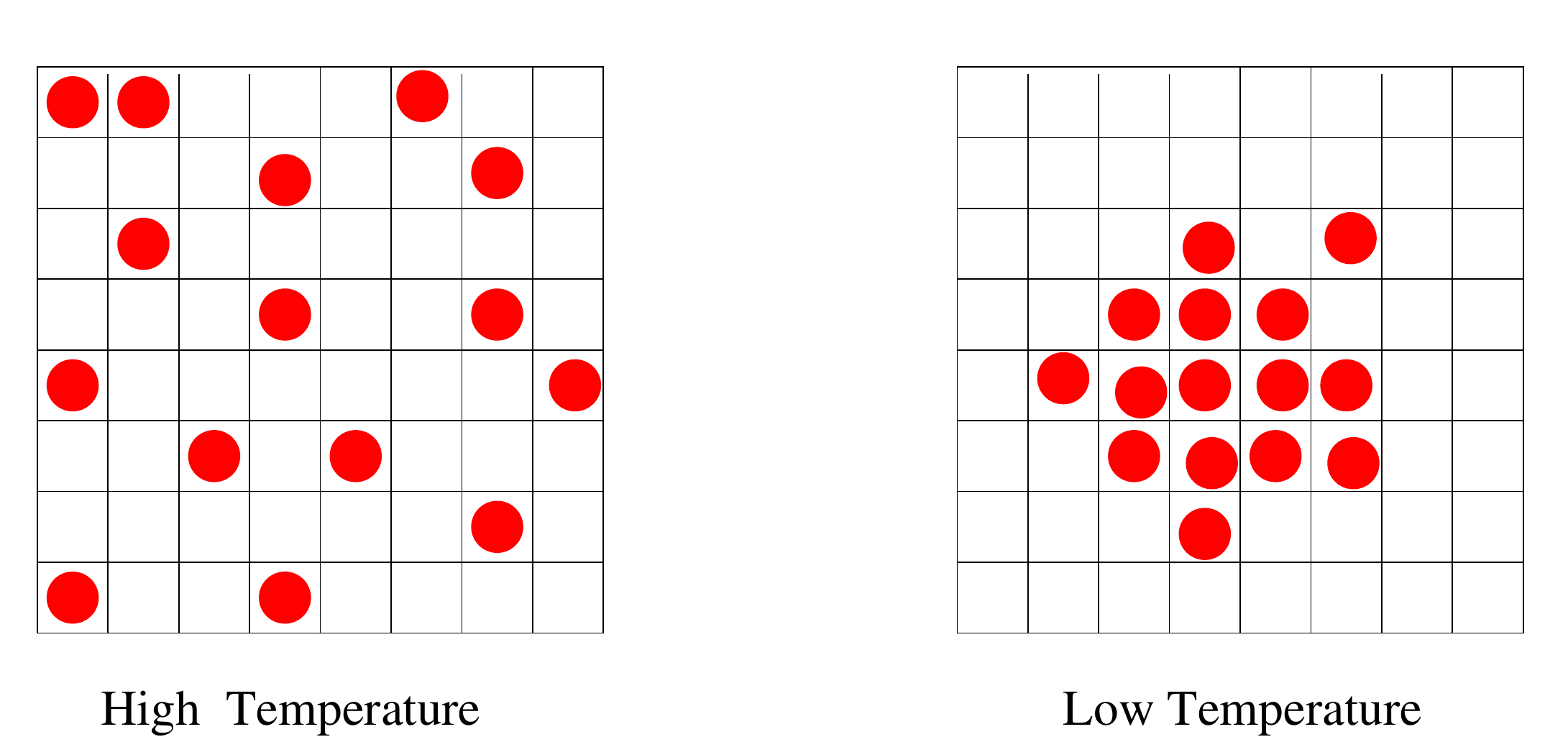} 
  \caption{The 2d  Ising Model: the   `SCORE' of a given
 configuration is defined 
as the  number of  particles without  a neighbour. Then,
  the probability to observe a configuration defined to be proportional to
 $e^{- \beta \, SCORE}$, where $\beta$ is proportional to the inverse temperature.
 This model displays a phase transition: At  High Temperature, $ \beta \to  0$,
 the system does not display any order,  we have  {gas;} 
  at low Temperature, $\beta \to \infty$
a  clustering occurs and the system is in a {condensed phase.}
A phase transition occurs at a Critical  Temperature  ${\beta_c}$. }
\label{fig:Ising}
 \end{center}
\end{figure}

  A  system at mechanical and at  thermal equilibrium
  obeys  the  principles of 
 thermodynamics that are  embodied in the laws of equilibrium
 statistical mechanics. The fundamental property  is that a system,
 consisting of a huge number of microscopic degrees of freedom, 
 can be described at equilibrium by only  a few  macroscopic
  parameters, called state variables.
  The values of these parameters can be determined by optimizing
 a  potential function (such as  the entropy,  the free energy,
 the Gibbs free  energy...) chosen according to the external constraints
 imposed upon the system. The connection between the macroscopic description
 and the microscopic scale is obtained through Boltzmann's formula
 (or one of its variants). Consider, for example,  a  system 
system at thermal equilibrium with a reservoir at temperature $T$.
Its thermodynamical properties are encoded 
by  Boltzmann-Gibbs canonical law:
\begin{eqnarray}
{ P_{{\rm eq}}({\mathcal C})  = \frac{{\rm e}^{ - E({\mathcal C}) /kT} }{Z}}
  \nonumber
\end{eqnarray}
where the  {Partition Function   Z}  is related 
 to the thermodynamic
  { Free Energy  F} via 
 $$  {  F = -kT Log\,Z  }  \, . $$  
 This expression (which is a consequence of Boltzmann's formula
 $s = k \log \Omega$) shows that 
 the  determination of the thermodynamic  potentials 
 can be expressed as  a combinatorial (or counting) problem, which of course,
 can be extremely  complex.  Nevertheless, equilibrium statistical physics
 provides us with a well-defined prescription to analyse thermodynamic systems:
  an explicit formula for  the canonical law is given;
 this defines a  probability measure on the
 configuration space of the system;  statistical properties of observables
(mean-values, fluctuations)
 can be calculated by performing averages with respect
 to this probability measure. The paradigm of equilibrium
 statistical physics is the Ising Model (see Figure~\ref{fig:Ising}).
 It was  solved  in two dimensions by L. Onsager (1944).
 We emphasize that  equilibrium  statistical mechanics
  predicts  macroscopic  fluctuations  (typically Gaussian) that are
  out of reach of classical  thermodynamics:   the  paradigm of  such
  fluctuations is the Brownian Motion.

  For systems far from equilibrium, a  theory that would
   generalize  the formalism of
   equilibrium  statistical mechanics  to  time-dependent processes 
     is  not yet available. However,  although  the theory is far from being
    complete,  substantial progress has been made, particularly during
     the last twenty years.  One line of research consists in
    exploring  structural properties of non-equilibrium systems: this
    endeavour   has led to celebrated results such  as  Fluctuation
     Theorems that generalize Einstein's fluctuation relation and linear response theory.
    Another strategy, inspired from the research devoted to the Ising model,  is 
   to gain insight into non-equilibrium physics   from analytical
   studies and from  exact solutions of some special  models. In the field of
 non-equilibrium statistical mechanics, the Asymmetric Simple
    Exclusion Process (ASEP) is reaching  the  status of  a
 paradigm.

  In these lecture  notes, we shall first review equilibrium properties
  in Section~\ref{Sec:Equil}. Using Markov processes, we shall give a dynamical
 picture of equilibrium in  Section~\ref{SubSec:Markov}. Then we shall introduce
  the  detailed balance condition and explain how it is related to time
 reversal (Section~\ref{SubSec:DetailedB}). This will allow us to give a precise definition
 of the concept of `equilibrium' from a dynamical point of view.

 The study of non-equilibrium processes will begin in Section~\ref{Sec:NonEq}. 
 We shall use as a  leitmotiv for  non-equilibrium, the picture of   rod (or pipe)
  in contact  with  two reservoirs at different temperatures, or  at
   different electrical (chemical) potentials (see Figure~\ref{fig:Courant}). This 
 simple  picture will allow us to formulate some of the basic  questions
  that have to be  answered in order to understand  non-equilibrium   physics. 
 The current  theory of non-equilibrium processes requires the use of some  mathematical
 tools, such as large-deviation functions, that are introduced, through
 various examples (Independent Bernoulli variables, random walk...),  in Section~\ref{SubSec:LDF};
 in particular, we explain how the thermodynamic Free Energy is connected to the large deviations
 of the density profile of a gas enclosed  in a vessel. In  Section~\ref{SubSec:Cumulants},
 we show the relations between the  large-deviation function and  cumulants of a random variable.
 Section~\ref{SubSec:GDB} is devoted to the very important concept of generalized detailed balance, 
  a fundamental  remnant of the  time-reversal 
 invariance of  physics, that prevails  even in situations  far from equilibrium. 
 Then,  in Section~\ref{SubSec:FT},   the Fluctuation Theorem is derived for Markov system
 that obey  generalized detailed balance. 

  From  Section~\ref{Sec:ASEP} on,  these lectures
 focus on the Asymmetric Exclusion Process  (ASEP) and on some of the techniques developed in 
 the last twenty 
 years to derive exact solutions for this model  and its variants. After a brief presentation
 of the model and of some of its simple  properties (Sections~\ref{SubSec:DefASEP} to
\ref{SubSec:Ptes}), we apply  the Mean-Field approximation to derive
 the hydrodynamic behaviour in Section~\ref{SubSec:MF}; in particular,
 this technique is illustrated on the  Lebowitz-Janowsky blockage model, a fascinating
 problem that has so far  eluded an exact solution. Finally, in Section~\ref{SubSec:Steady},
 the celebrated exact calculation of the steady state  of the ASEP with open boundaries, using
 the Matrix Representation Method, is described. 

 Section~\ref{Sec:Bethe} contains a crash-course on the Bethe Ansatz. We believe that the ASEP
 on a periodic ring, is the simplest model to learn how to apply this very important method.
 We try to explain the various steps that lead to the Bethe Equations
 in   Section~\ref{SubSec:crashcourse}.  These equations are analysed in the special TASEP
 case in  Section~\ref{SubSec:BATASEP}.

  We are now ready   to calculate large deviation  functions
 for  non-equilibrium  problems: this is the goal of Section~\ref{Sec:LDFNonEq}.
 Our aim is to derive large deviations of the  stationary current for  the pipe picture,
 modelled  by the ASEP.  This is done first for the periodic case (ASEP on a ring) in 
  Section~\ref{SubSec:periodic}, then for the open system in contact with two reservoirs
 (Section~\ref{SubSec:openASEP}). The similarities between the two solutions are emphasized.
 Exact formulae for cumulants and for the  large deviation functions are given. This Section
 is the most advanced part of the course and represents the synthesis of the concepts and techniques
 that were developed in earlier sections. Detailed calculations are not given and can be found
 in recent research papers.

The last section is devoted to concluding remarks and is followed by the Bibliography.
 We emphasize that  these lecture notes are not  intended  to be a review paper.
 Therefore, the   bibliography is   rather   succinct and   is restricted to some
 of the  books, review papers or articles that were used while  preparing this course.
 More precise references can be found easily from  these sources. Our major influences
 in preparing these lectures come from the review of B. Derrida \cite{DerrReview}
 and from  the book of P. L. Krapivsky, S. Redner and  E. Ben-Naim \cite{PaulK}.

\section{Dynamical  Properties of the  Equilibrium State} 
\label{Sec:Equil}

 The average macroscopic properties of
  systems  at  thermodynamic equilibrium are independent of time.
 However, one should not think that  thermodynamic equilibrium
 means absence of dynamical behaviour: at the microscopic scale,
 the system keeps on evolving from one micro-state to another.
 This never-ending  motion manifests itself as {\it fluctuations} at the
 macroscopic scale, the most celebrated example being the Brownian
 Motion. However, one  crucial  feature of a system at thermodynamic
 equilibrium is the absence of  currents in the system: there is no
 macroscopic transport of matter, charge, energy, momentum, spin or
 whatsoever within the system or between the system and its environment.
 This is a very fundamental   property, first stated by Onsager, that stems
 from the time-reversal symmetry of the microscopic equations of motion.
 This property is true both for classical and quantum dynamics.

\subsection{Markovian dynamical models}
\label{SubSec:Markov}

   We want to describe the evolution of a complex system consisting
 of a very large number $N$ of interacting degrees of freedom.
 In full rigour, one should write the $N$-body (quantum)  Hamiltonian that 
 incorporates  the full evolution of the system under consideration.
 However, it is often useful to consider effective dynamical descriptions
  that are obtained, for example,  by coarse-graining the phase-space
  of the system or by integrating-out fast modes. In the following, the 
   models we shall study will follow classical Markovian dynamics and we shall
 give a short presentation of  Markov systems. The interested reader can
 find more details, in particular about the underlying assumptions that lead to Markov dynamics,
  in e.g. the book by  N. G. Van Kampen \cite{VanKampen}.
 We also  emphasize that  many properties that we shall discuss here
  can be generalized to other
 dynamical systems.

  The classical   Markov processes  that we shall study here 
 will be fully specified by a   (usually finite or numerable) set of microstates
 $\{ {\mathcal C_1}, {\mathcal C_2} \ldots \}$. At a given time $t$,
 the system can be found  in one its  microstates. The evolution
 of the system is specified by the following rule:
  Between  $t$ and $t +dt$, the system can jump
 from  a configuration  ${\mathcal C}$ to a configuration  ${\mathcal C'}$.
 It is assumed that the transition rate from  ${\mathcal C}$ 
 to ${\mathcal C'}$ does not depend on the previous history of the system:
 this is the crucial  {\it Markov  hypothesis} in which 
 short time correlations are neglected.  The rate of transition per unit time
 will be denoted by  $M({\mathcal C'}, {\mathcal C})$ (or equivalently,
 by $M({\mathcal C}  \rightarrow  {\mathcal C'})\, ).$   Note that this rate
 may vary with  time and depend explicitly on $t$. This case will
  not be considered  in the present lectures. To summarize, the Markov dynamics
 is specified by the following rules:

$$ 
{\fbox  {$  \quad\quad  {\mathcal C}  \rightarrow  {\mathcal C'} 
  \,\,\,\,  \hbox{ {\it with probability}}  \,\,\,\, 
 M({\mathcal C'}, {\mathcal C})dt  \quad \quad$}  }
 $$

\begin{figure}[ht]
 \begin{center}
  \includegraphics[height=6.0cm]{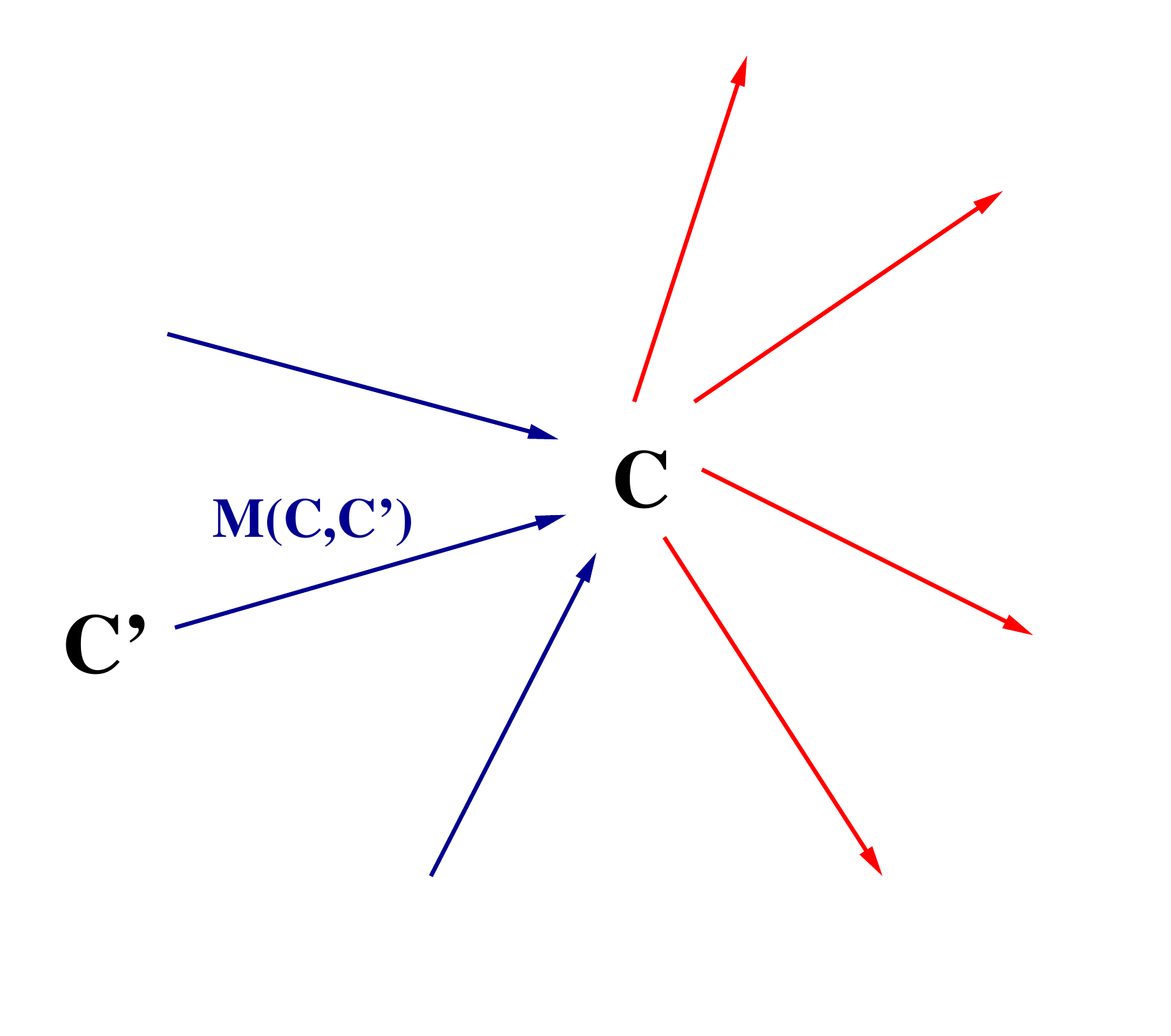}
  \caption{Representation of  a Markov process  as a network.}
\label{fig:Network}
 \end{center}
\end{figure}

 These dynamical rules can be illustrated by a network in the configuration
 space (see Figure~\ref{fig:Network}): the nodes of the graph are the
  microstates and oriented-edges, weighted by the Markov rates 
 $M({\mathcal C'}, {\mathcal C})$ represent possible transitions between
 configurations.

 For  a system with  Markov dynamics, one can define 
 $P_t({\mathcal C})$, the  probability
 of being in the micro-state ${\mathcal C}$ at time $t$. This probability
 measure varies with time: its  evolution is governed by the Master equation,
  given by 
 \begin{equation}
\textcolor{red}
{\fbox{$  \ca{ \frac{ d }{dt} P_t({\mathcal C})} \ca{  =  }
 \ca{ \sum_{ {\mathcal C'} \neq  {\mathcal C} } 
 M({\mathcal C}, {\mathcal C'}) P_t({\mathcal C'})
- \left\{\sum_{ {\mathcal C'} \neq  {\mathcal C} } 
 M({\mathcal C'}, {\mathcal C}) \right\} \,  P_t({\mathcal C}) }
 $} }
\label{eq:Master}
\end{equation}     

 This equation is fundamental. To derive it, one must 
 take into account   all possible 
  transitions  between time $t$ and $t + dt$ that involve a given
 configuration ${\mathcal C}$.  There are two types
 of moves:  (i) transitions   {\it into} 
 ${\mathcal C}$ coming from a different configuration ${\mathcal C'}$; 
  (ii) transitions {\it out of } ${\mathcal C}$ towards 
 a  different configuration ${\mathcal C'}$. The moves (i) and (ii) contribute
 with a different sign to the change of the   probability of occupying
  ${\mathcal C}$   between time $t$ and $t + dt$.
 Note that the Master equation  can well be interpreted  as a flux-balance
 equation on the network of Figure~\ref{fig:Network}.

 The way we have encoded the transition rates naturally suggests
 that the  Master Equation can be rewritten   in a  Matrix  form.
 Indeed,  reinterpreting the rate $ M({\mathcal C'}, {\mathcal C})$
 as matrix-elements and defining  the diagonal term  
 \begin{equation}
 M({\mathcal C}, {\mathcal C})  = 
 -  \sum_{ {\mathcal C'} \neq  {\mathcal C} } 
 M({\mathcal C}, {\mathcal C'}) \, ,
\label{Diagonal}
 \end{equation}
 allows us to rewrite Equation~(\ref{eq:Master}) as
\begin{equation}
 \frac{ d  P_t}{dt}   =  M .  P_t 
\label{eq:MarkovMatrix}
 \end{equation}

 We emphasize that the diagonal term  $M({\mathcal C}, {\mathcal C})$
 is a negative number: it represents {\it minus} the rate of leaving
 the configuration ${\mathcal C}$. This leads to  an important property:
 the sum of each column of $M$ identically vanishes. This property
 guaranties, by simple algebra, that the total probability is conserved:
$ \sum_{\mathcal C} P_t({\mathcal C}) = \sum_{\mathcal C}  P_0({\mathcal C}) =1$
 (the initial probability distribution being normalized).

  Note that there is a 
 formal analogy between Markov systems and quantum dynamics:
 the Markov operator $M$ plays the role of a quantum Hamiltonian.
 However, $M$ does not have to be a  symmetric or Hermitian matrix
 (and is not, in general).

\begin{figure}[ht]
 \begin{center}
  \includegraphics[height=3.5cm]{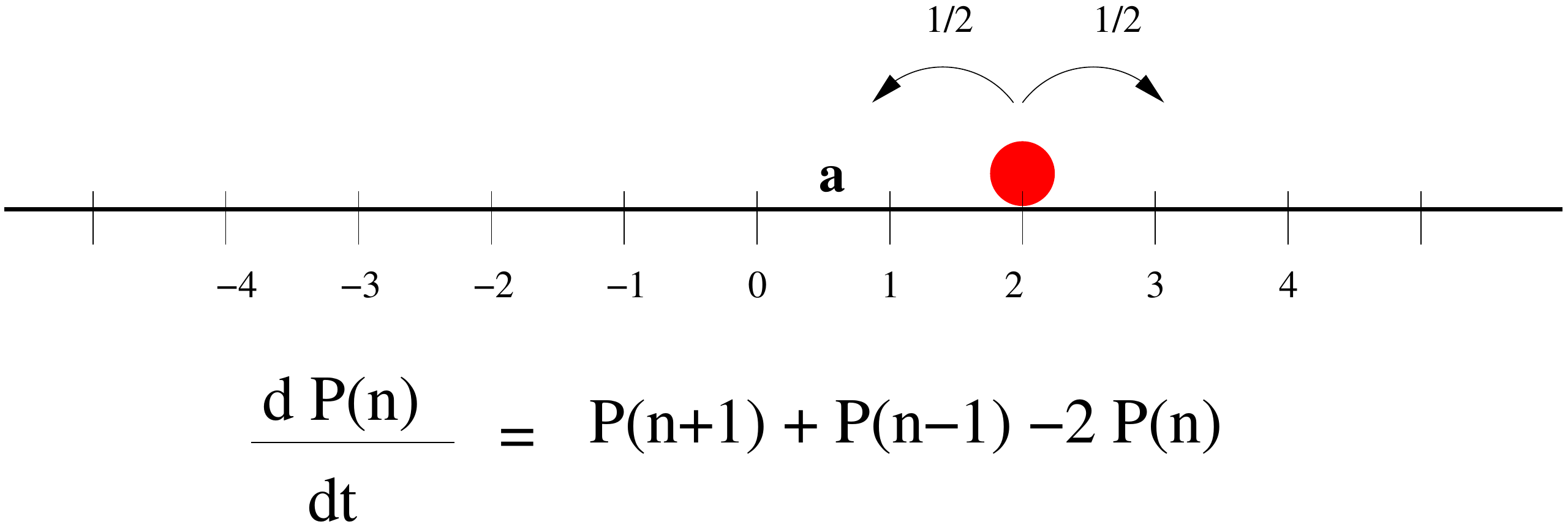}
  \caption{The simple random walk is a Markov process.}
\label{fig:Random Walk}
 \end{center}
\end{figure}

 There are numerous examples of Markov processes in statistical physics.
 The paradigm is certainly the simple symmetric  random walk
  on a discrete lattice  (see Figure~\ref{fig:Random Walk}). Here,
 the configurations are the lattice sites and the transition rates are
 constant and uniform (i.e. translation-invariant). The corresponding Markov
 equation is the discrete Laplace equation on the lattice.

\begin{figure}[ht]
 \begin{center}
  \includegraphics[height=4.0cm]{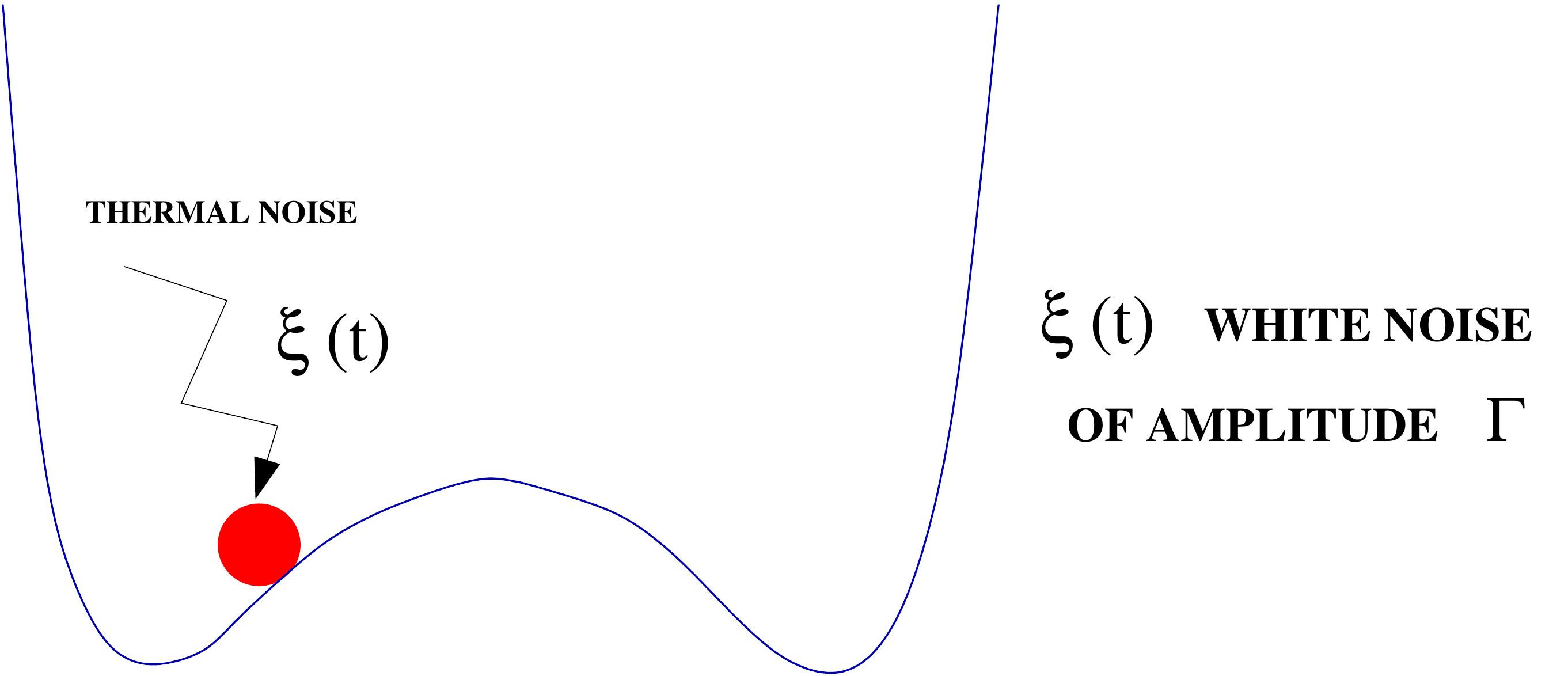}
\caption{Langevin dynamics in a double-well potential: this model
 leads to analytical calculations of reaction-rates and transition-times.}
\label{fig:2Well}
 \end{center}
\end{figure}

 Another important example is given by Langevin dynamics 
 (see Figure~\ref{fig:2Well}). 
 It was originally invented by Paul Langevin as  mechanical model
  for the Brownian Motion but stochastic dynamics has become
 a  widely studied  subject, that allows for instance to model
 the effects of noise  in mechanical and electric devices.
 The basic idea is to incorporate a random force that represents
 thermal noise into classical Newtonian dynamics:
$$  \,\,\,\, 
m \frac{d^2 x}{dt^2} = -\gamma \frac{d x}{dt} - \nabla{\mathcal U}(x) + 
 { \xi(t)}  \,\,\,\,   $$
Here $\xi(t)$ is a Gaussian  white noise of amplitude $\Gamma$.

 The state of a particle is a point in phase-space, i.e 
  a configuration is specified by the position and the velocity (or momentum)
 of the particle (the set of possible configurations is  continuous
 and non-enumerable).  The corresponding Markov
 equation for the probability distribution function $P_t(x,v)$,
 of being at $x$ with velocity $v$, is known as
 the Fokker-Planck equation:
$$ \frac{ d  P_t}{dt}   =
-\frac{\partial}{\partial x} \left\{ v P_t \right\}
 + \frac{\partial}{\partial v}
  \left\{ \frac{ \gamma v + \nabla U}{m}   P_t   \right\}
 + \frac{\Gamma}{m^2} \frac{\partial^2 P_t}{\partial v^2}
 =  {\mathcal L}_{FP} . P_t \, . $$
The role of the Markov matrix is played by the Fokker-Planck operator
 ${\mathcal L}_{FP}$. There are many formal  similarities between the 
 Fokker-Planck equation and the discrete Markov dynamics
 given by~(\ref{eq:MarkovMatrix}); however, subtle  mathematical issues can
 arise in the case of a continuous configuration space
 that  require the use of functional analysis.

 Both Markov and Fokker-Planck dynamics are mathematical models,
  that can be defined and studied without any specific reference
 to physical principles. However,  to be physically relevant, these dynamics
 should be connected to  the laws of thermodynamics and statistical physics.
  In particular, one can impose that the steady  state of these equations
  is an equilibrium-state: in other words, the stationary probability
  distribution must be identical to the  Boltzmann-Gibbs canonical law.

 For a discrete Markov dynamics~(\ref{eq:MarkovMatrix}), this condition
 reads 
 $$   \sum_{ {\mathcal C'} \neq  {\mathcal C} } 
 M({\mathcal C}, {\mathcal C'})  {\rm e}^{ - E({\mathcal C'}) /kT} 
=  {\rm e}^{ - E({\mathcal C}) /kT}
 \left\{\sum_{ {\mathcal C'} \neq  {\mathcal C} } 
 M({\mathcal C'}, {\mathcal C}) \right\} \, $$
 This is a set of global constraints on the rates, which is sometimes
 called the  `global balance' condition.

 Similarly, in the Langevin case, one imposes that the invariant measure
 of phase-space is given by 
$$ 
  P_{\rm{eq}}(x,v)= 
 \frac{1}{Z}{\rm e}^{ -\frac{1/2 mv^2 + {\mathcal U}(x)}{kT}}
$$
 Writing that this formula is  the stationary solution of the Fokker-Planck
 equation, one observes that this fixes the the noise-amplitude
  $\Gamma$ as a function of temperature
$$  { \Gamma =  \gamma k T}  \, .$$
 This is in essence the reasoning followed by Langevin in his study
 of the Free Brownian Motion (for which the external potential vanishes
${\mathcal U} =0$). Substituting the value of $\Gamma$ in the 
 corresponding Fokker-Planck equation leads to
 $$ \frac{ d P_t}{dt}   =
\frac{ \gamma }{m} \frac{\partial}{\partial v} ( v P_t)
 + \frac{\gamma k T }{m^2} \frac{\partial^2 P_t}{\partial v^2} $$

 Multiplying both sides of this equation by $x^2$ and integrating
 over phase-space allows us to show that 
 $$ \langle x^2 \rangle = 2 D t \quad \quad   {\rm with }  \quad \quad 
 { D = \frac{kT}{\gamma}} \, .$$
 Using Stokes' formula  $\gamma = 6 \pi \eta a$ for the friction-coefficient
 (coefficient of the linearized force felt by a sphere  of
  radius $a$,  dragged at velocity $v$,  in a liquid of viscosity $\eta$),
 leads to the celebrated formula of Einstein (1905) for the
 diffusion constant of the Brownian Motion, in terms of the Avogadro
 Number. 

\subsection{Time-reversal and Detailed Balance}
\label{SubSec:DetailedB}

\begin{figure}[ht]
\begin{center}
  \includegraphics[height=3.35cm]{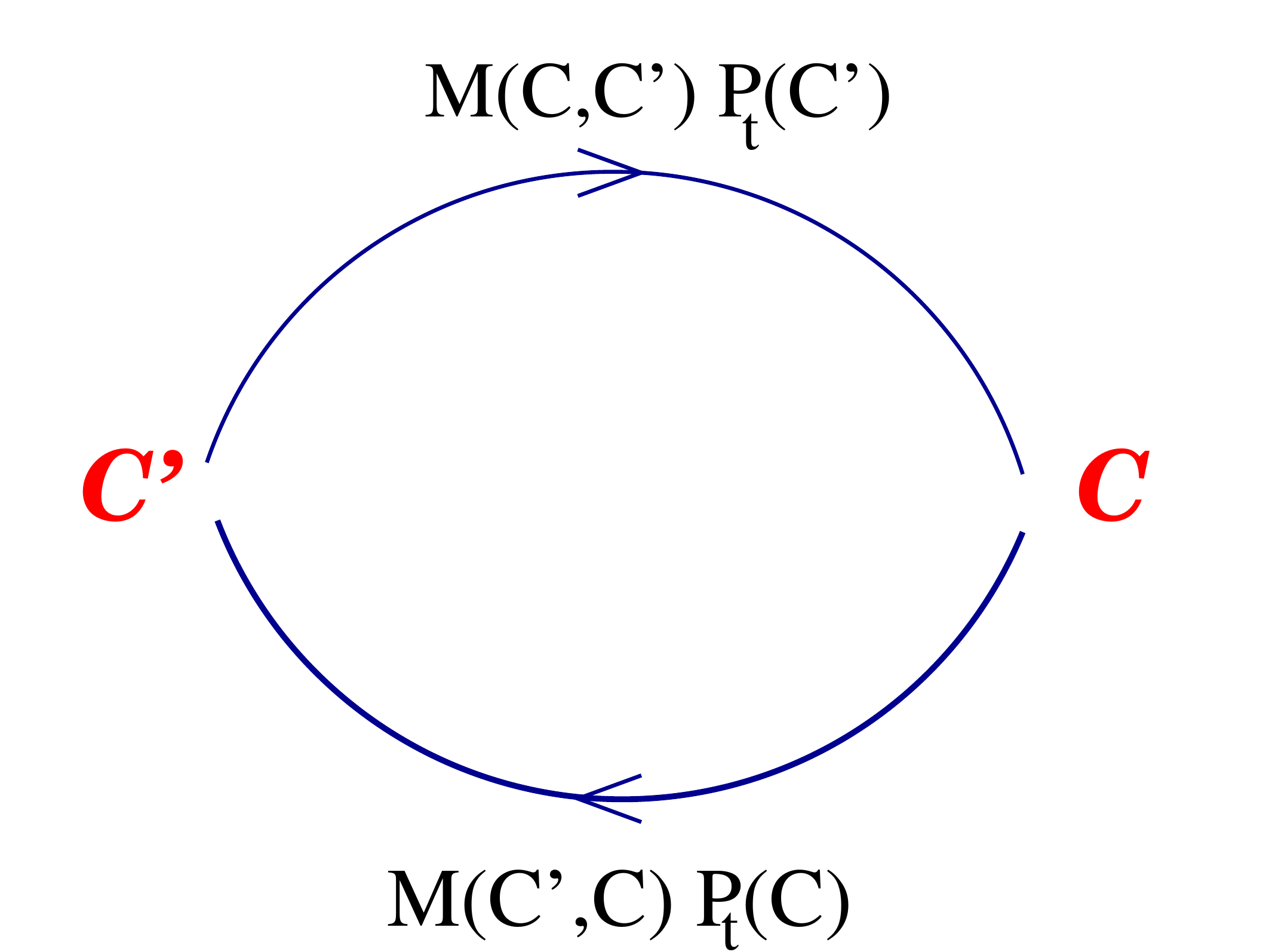}
 \caption{A graphical representation of the local 
probability current  $J_t({\mathcal C}, {\mathcal C'})$.}
\label{fig:ProbCourant}
\end{center}
\end{figure}

   We now discuss a fundamental characteristic   of equilibrium
 dynamics that was first investigated by L. Onsager. Again, the property
 discovered by Onsager is a very general one. We shall present  it here
 on Markov dynamics.
The master equation~(\ref{eq:Master})  can be written in the following manner
\begin{eqnarray}
 \nonumber 
\frac{ d }{dt} P_t({\mathcal C})  =  
  \sum_{ {\mathcal C'}} \left\{
 M({\mathcal C}, {\mathcal C'})  \, P_t({\mathcal C'}) - 
 M({\mathcal C'}, {\mathcal C})  \,  P_t({\mathcal C}) \right\}
 =  \sum_{ {\mathcal C'}} J_t({\mathcal C}, {\mathcal C'}) \, , 
\end{eqnarray}
   where we have introduced the local probability current 
$J_t({\mathcal C}, {\mathcal C'})$ between  ${\mathcal C}$  
 and  ${\mathcal C'}$ (See Figure~\ref{fig:ProbCourant}).

\hfill\break

 When the stationary
  state is reached,  we know that the right-hand
 side of this equation must vanish. However, 
 equilibrium  is a very particular stationary state: at 
  equilibrium  the microscopic dynamics of  the system is
  {\it time-reversible}. This symmetry property
  implies that all the  {\it local} currents
 $J_t({\mathcal C}, {\mathcal C'})$  vanish separately
 (Onsager):
\begin{equation}
\textcolor{red}
{\fbox{$ \,\,\,\,
\ca{ M({\mathcal C}, {\mathcal C'})  P_{{\rm eq}}({\mathcal C'}) = 
 M({\mathcal C'}, {\mathcal C})  P_{{\rm eq}}({\mathcal C}) }
\,\,\,\,   $} }
\label{eq:DetBal}
\end{equation}
 This   is  the 
 {\bf detailed balance equation.}  We emphasize that detailed balance
 is  a very {\it  strong property} goes beyond the laws
   of classical thermodynamics.

 We shall now explain the mathematical 
 relation between detailed balance
 and time-reversal.  The main-idea is to use the transition rates
 to construct  a probability measure on time-trajectories
 of the system.

 The two important mathematical properties we shall use are
the following:

1.  Probability of remaining in  ${\mathcal C}$
 during a time interval $\tau$:
  $$ \lim_{dt \to 0} 
\left( 1 + M({\mathcal C}, {\mathcal C}) dt\right)^{\frac{\tau}{dt}} =
 {\rm e}^{  M({\mathcal C}, {\mathcal C}) \tau}  $$

  2.   Probability of going from  ${\mathcal C}$
 to  ${\mathcal C}$  during $dt$:  
  $ { M({\mathcal C'}, {\mathcal C})dt }$

 Relation 1 can be derived by calculating the probability of staying
 in the same configuration between $t$ and $t + dt$ and integrating
 over $ 0 \le t \le \tau$. Relation 2 is simply the definition
 of the transition rates in a Markov process. 

\begin{figure}[ht]
 \begin{center}
  \includegraphics[height=5.0cm]{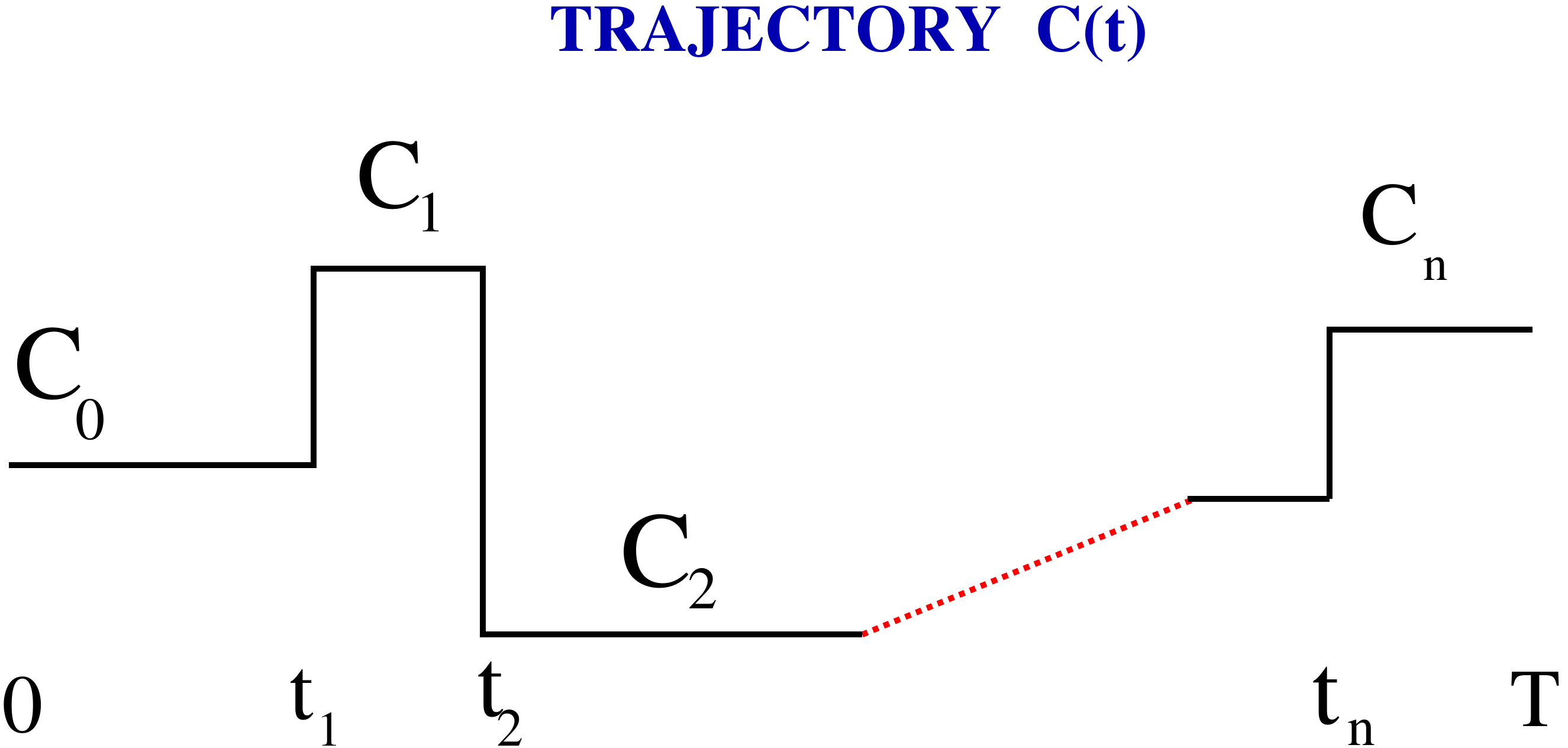}
  \caption{A trajectory of a Markov process during the time interval
 $[0,T]$.}
\label{fig:Trajectory}
 \end{center}
\end{figure}

Let us  now consider a `history' of the system between the initial time
0 and the final $T$. During this interval of time, the system follows
 a trajectory: it begins with a configuration ${\mathcal C}_0$, then
 at a date $t_1$ it jumps into  configuration ${\mathcal C}_1$, stays
 there till $t_2$ and jumps to ${\mathcal C}_2$ and so on.
 This special trajectory, denoted by $C(t)$ is depicted in
  Figure~\ref{fig:Trajectory}.

Using the relations 1 and 2 above, we can calculate the weight of this specific
 trajectory ${ C}(t)$, i.e 
  the probability ${\rm Pr} \{ { C}(t)\}$
  of observing  ${ C}(t)$, in the equilibrium state. 
 The only extra ingredient we need to  include is the fact that the initial
 condition ${ C}_0$ at $t =0$ is chosen according to the 
 equilibrium measure. We thus have

\begin{eqnarray}
 {\rm Pr} \{ { C}(t)\} =   
\cc{   {\rm e}^{ M({\mathcal C}_n, {\mathcal C}_n)(T-t_n)} }
    \, \ca{ M({\mathcal C}_n, {\mathcal C}_{n-1})dt_n }   \,  
 \,
\cc{    {\rm e}^{ M({\mathcal C}_{n-1}, {\mathcal C}_{n-1})(t_n-t_{n-1})}}
   \ldots
\cc{   {\rm e}^{ M({\mathcal C}_1, {\mathcal C}_1)(t_2-t_1)} }  \, 
 \ca{  M({\mathcal C}_1, {\mathcal C}_0)  dt_1 }   \, 
\cc{  {\rm e}^{ M({\mathcal C}_0, {\mathcal C}_0)t_1} }
 P_{{\rm eq}}({\mathcal C}_0)   
\nonumber
\end{eqnarray}

\begin{figure}[ht]
 \begin{center}
  \includegraphics[height=6.0cm]{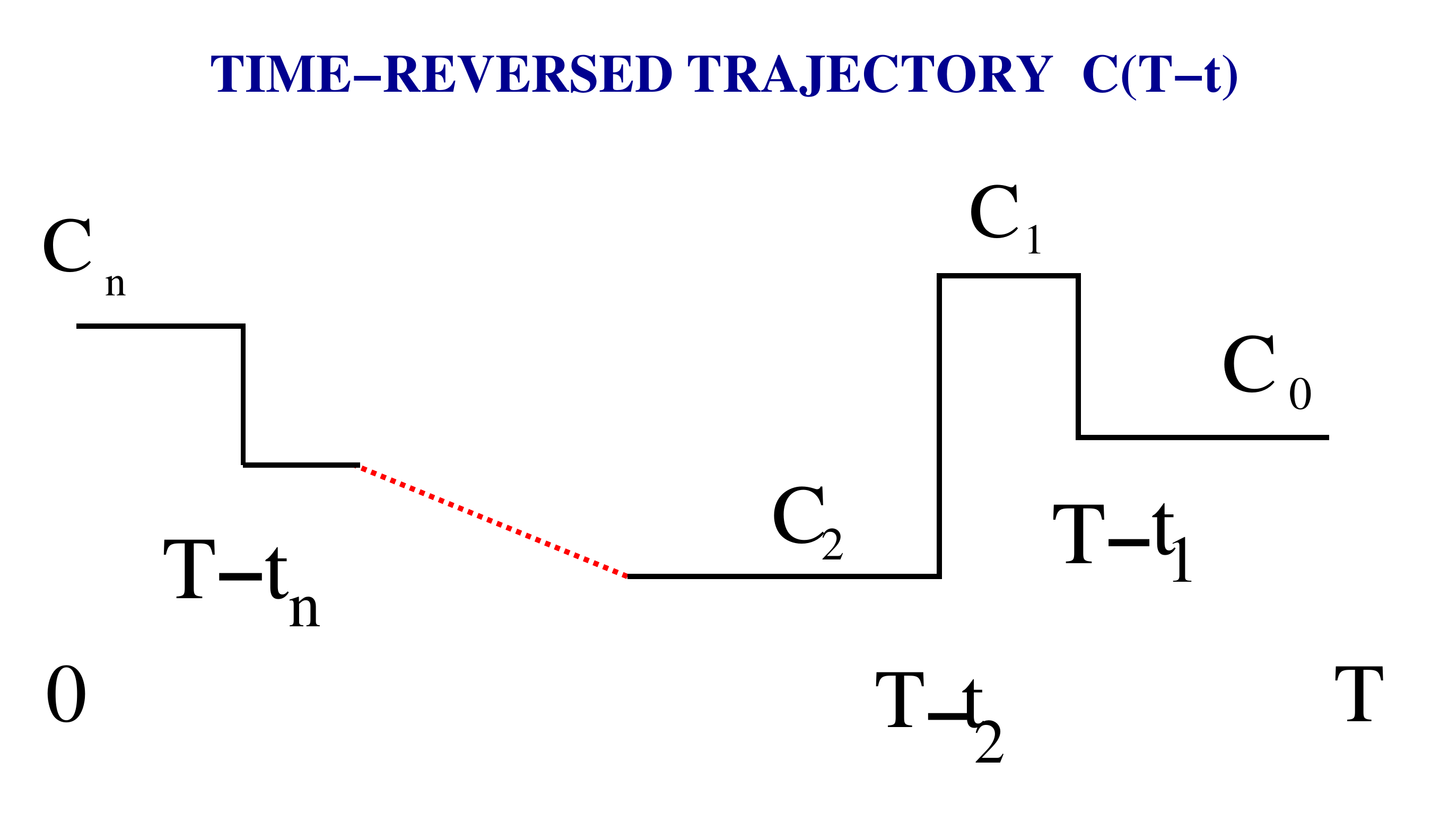}
  \caption{Calculating the  weight of the time-reversed trajectory.}
\label{fig:TrajectoryReversed}
 \end{center}
\end{figure}

For any given  trajectory ${ C}(t)$, a   time-reversed trajectory
 can be defined as  $\hat{C}(t) = C(T-t)$. This
 is a bona-fide history of the system (see Figure~\ref{fig:TrajectoryReversed})
  and one can  calculate
 the probability of observing it:

\begin{eqnarray}
 {\rm Pr}\{  \hat{C}(t) \} =
 \cc{  {\rm e}^{ M({\mathcal C}_0, {\mathcal C}_0)t_1} }  \, 
 \,   \ca{  M({\mathcal C}_0, {\mathcal C}_{1}) dt_1 }  \,\,   
   \cc{  {\rm e}^{ M({\mathcal C}_1, {\mathcal C}_1)(t_2-t_1)} } \, 
 \ldots 
\cc{ {\rm e}^{ M({\mathcal C}_{n-1}, {\mathcal C}_{n-1})(t_n-t_{n-1})}}  \,  
 \,  \ca{  M({\mathcal C}_{n-1}, {\mathcal C}_n) dt_n } \,\,  
\cc{ {\rm e}^{ M({\mathcal C}_n, {\mathcal C}_n)(T-t_n)}}  \, 
 P_{{\rm eq}}({\mathcal C}_n) \nonumber 
\end{eqnarray}

 If we now calculate the  ratio  of these two probabilities
 i.e. the ratio of the probability observing a  given  history  ${ C}(t)$
 by that of observing  the reversed history   $\hat{C}(t)$, we obtain 
{\it the  ratio between  the probabilities of forward and backward trajectories:}
\begin{eqnarray}
\frac{ {\rm Pr} \{ {\mathcal C}(t)\} }
{ {\rm Pr} \{\hat {\mathcal C}(t)\}} = \frac
 {  \ca{ M({\mathcal C}_n, {\mathcal C}_{n-1})}
  \ca{ M({\mathcal C}_{n-1}, {\mathcal C}_{n-2})} \ldots
 \ca{  M({\mathcal C}_1, {\mathcal C}_0)}   \, 
 P_{{\rm eq}}({\mathcal C}_0)}
 { \ca{ M({\mathcal C}_0, {\mathcal C}_{1})} \,
  \ca{ M({\mathcal C}_1, {\mathcal C}_{2})}\,   \ldots\, \, \quad
 \, \, \ca{  M({\mathcal C}_{n-1}, {\mathcal C}_n)} \, 
  P_{{\rm eq}}({\mathcal C}_n) }
 \label{eq:RatioFB}
\end{eqnarray}
Using  recursively the {\it detailed balance condition:} 
 $$ M({\mathcal C}_1,{\mathcal C}_0) P_{{\rm eq}}({\mathcal C}_0) 
= P_{{\rm eq}}({\mathcal C}_1)
 M({\mathcal C}_0, {\mathcal C}_1)  $$
 and zipping it through the previous result leads us to
 the following remarkable identity
$$ \large{ \ca{ \frac{ {\rm Pr} \{ {\mathcal C}(t)\} }
{ {\rm Pr} \{\hat {\mathcal C}(t)\}} =  1  }} $$ 
 {\it  Hence, detailed balance implies that the dynamics is time reversible.}
 The converse property is true: if we want that a dynamics to be
  time-reversal invariant, then the detailed balance relation must be satisfied
 (consider simply a history in which there occurs
  a single transition between two
 configurations ${\mathcal C}$ and  ${\mathcal C}'$).

 To conclude, the  detailed balance relation is a profound property
 of the equilibrium state that reflects time-reversal invariance
 of the dynamics. This relation is now taken as a definition for
 the concept of equilibrium:  {\bf  a stationary state is an  equilibrium state
 if and only if  detailed balance is satisfied.}

\section{Nonequilibrium  Processes}
\label{Sec:NonEq}

In Nature, many systems are far from thermodynamic equilibrium
 and keep on exchanging matter, energy, information with their surroundings.
 There is no general conceptual  framework to study such systems.

 A basic example of  a nonequilibrium process is a conductor, or a pipe,
 in contact with two  reservoirs at different temperatures, or 
 electrical or chemical  potential. In the stationary state, a 
 non-vanishing steady-state current will flow from the reservoir
 at higher  potential towards the one at lower potential. This
 current clearly breaks time reversal invariance. In the vicinity
 of equilibrium, linear response theories allow us to predict
 the statistical behaviour of this current and to derive analytically
 the response coefficients (conductance, susceptibilities) from the knowledge
 of equilibrium fluctuations. However, one may wonder if a general
 microscopic theory, not obtained by a perturbative expansion in the
 vicinity of equilibrium, may be constructed. At present no such framework
 exists. However, in the last two decades,  important progress and
 convincing proposals for a general description of non-equilibrium
 statistical mechanics have  been made. We shall describe some of these
 theories in these lectures. Our main inspiration in this section  is
  the two  review papers by B. Derrida \cite{DerrReview,DCairns}.

\begin{figure}[ht]
 \begin{center}
  \includegraphics[height=2.5cm]{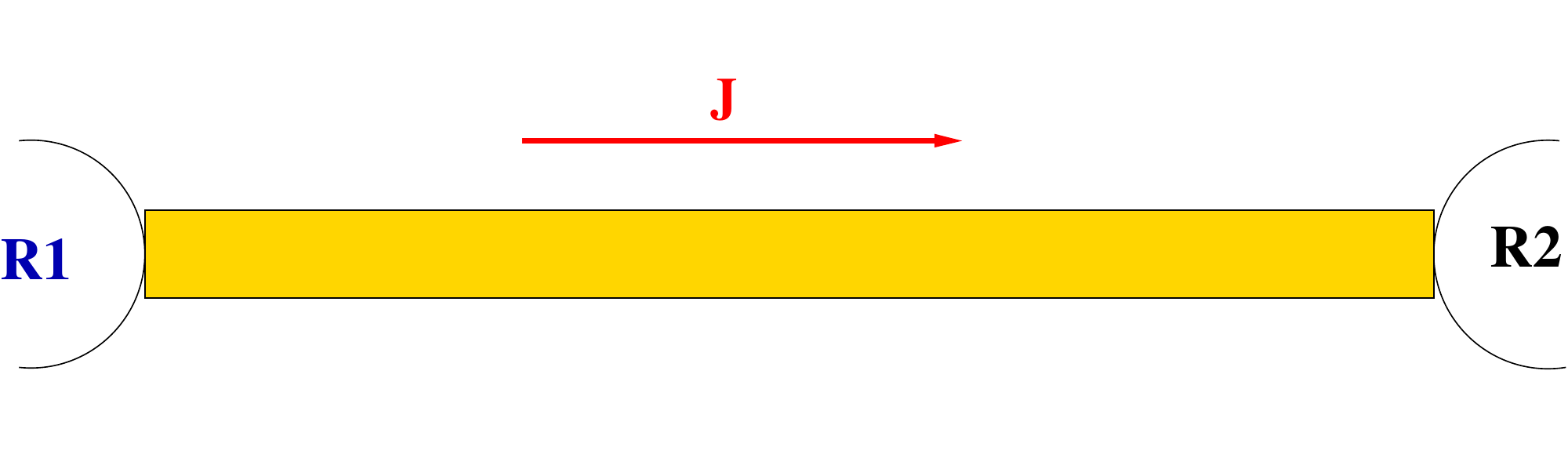}
  \caption{A  stationary driven system  in contact with 
 two reservoirs at different temperature and/or potential:
  In the steady state, a non-vanishing macroscopic current ${J}$
 flows. This pipe paradigm will be used as a leitmotiv throughout the text.}
\label{fig:Courant}
 \end{center}
\end{figure}

 For the moment being, we   use  the `pipe model' 
 (see Figure~\ref{fig:Courant}) as a  paradigmatic illustration   of stationary 
  non-equilibrium behaviour and let us 
  formulate some very basic questions:

\begin{itemize}
\item
 What are the {\it relevant macroscopic parameters}? How many macroscopic
 observables should we include to have a fair description of the system?
\item
 Which  {\it  functions} describe the state of a system? Can the
 stationary state be derived by optimizing a potential?
\item
 Do  {\it  Universal Laws}  exist? Can one define Universality Classes
 for systems out of equilibrium? Are there some general equations of state?
\item
 Can  one postulate a  general form for  the  {\it microscopic  measure}
 that would generalize the Gibbs-Boltzmann canonical Law?
\item
 What do the statistical properties of the current in the stationary state
 look like (In particular, are the current fluctuations Brownian-like)?
\end{itemize}

\subsection{Large Deviations and Rare Events}
\label{SubSec:LDF}

 Large Deviation Functions  (LDFs) are  important mathematical objects,
 used in  probability theory, that are becoming widely used
 in statistical physics. Large deviation functions are used to quantify
 rare events that, typically,    have  exponentially  vanishing
 probabilities.  We shall introduce this concept through
 an elementary example. A  very useful  review on large
 deviations has recently been written by H. Touchette \cite{Touchette}.
 
 Let   $\epsilon_1, \ldots ,\epsilon_N$ be  $N$ independent
 binary   variables, $\epsilon_k = \pm 1$,   with  probability  $p$  
 (resp. $q = 1 -p).$
 Their sum is denoted by  $S_N = \sum_1^N \epsilon_k$.
 We know, from the  {Law of Large Numbers} that
 $S_N/N \to p -q$ almost surely. Besides, 
 the {Central Limit Theorem} tells us that the fluctuations
 of the  sum  $S_N$ are of the order $\sqrt{N}$. More precisely,
 $[S_N -N(p -q)]/\sqrt{4pqN}$ converges towards a Normalized 
 Gaussian Law.

  One may ask a more refined question: how fast is the convergence implied by 
  the  Law of Large Numbers? In other  words, what does the probability
 that  $S_N/N$ assumes a non-typical value look like when $N \to \infty$?
  For the example, we consider, elementary  combinatorics
 shows  that for  $-1 < r < 1$, in the large $N$ limit, we have
 $$  {\rm Pr}\left( \frac{S_N}{N} = r \right) \sim
 {\rm e}^{ - N \, \Phi(r)}   $$ 
 where  the positive function   $\Phi(r)$ vanishes for  $r= (p -q)$.
 This is an elementary example of a large deviation behaviour.
 The  function  $\Phi(r)$ is a  called a rate function  or 
 a {\it large deviation function.} 
 It  encodes  the probability of rare events. A simple application
 of Stirling's formula yields
 $$ \Phi(r) = \frac{1+r}{2} \ln \left(\frac{1+r}{2p}  \right)
 + \frac{1-r}{2} \ln \left(\frac{1-r}{2q}  \right)  $$

\begin{figure}[ht]
 \begin{center}
  \includegraphics[height=3.5cm]{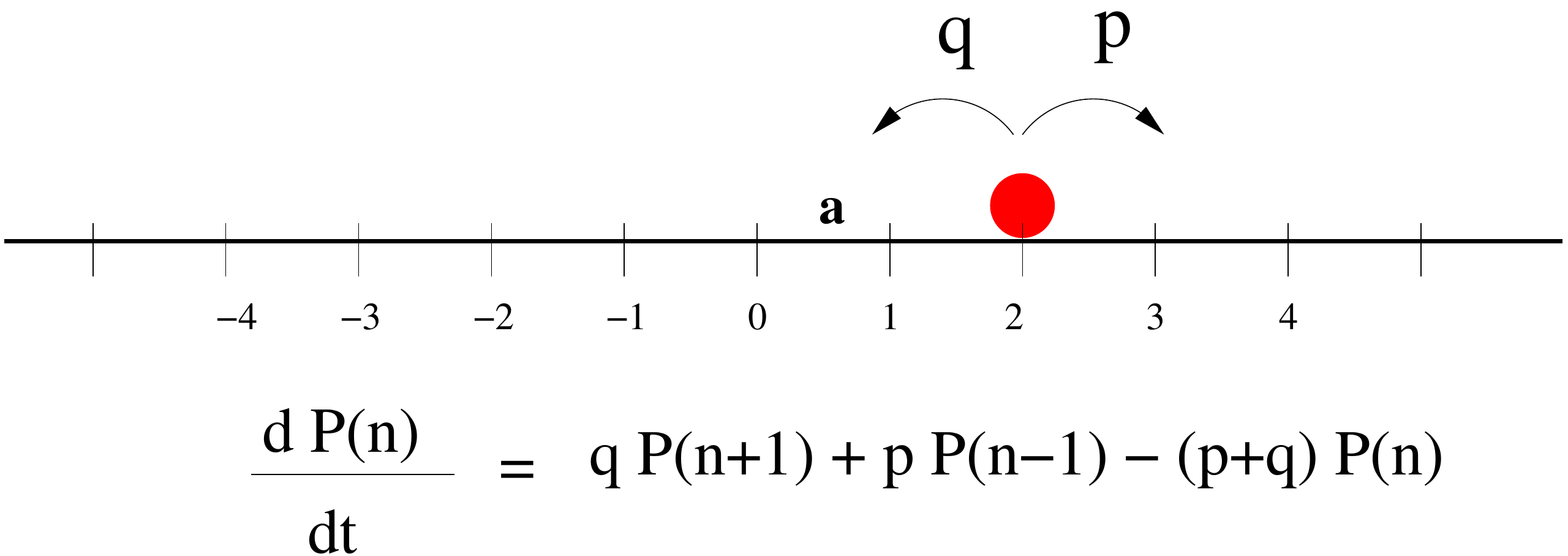}
  \caption{An asymmetric  random walker on a discrete line. The rates
 for right and left hopping are given by $p$ and $q$ respectively. The average
 speed  of the walker is given by $p-q$.}
\label{fig:rdmwlkASYM}
 \end{center}
\end{figure}

  We have discussed a very specific example but  large deviations 
  appear in many different contexts. We now  consider
   an asymmetric random walker on a one-dimensional lattice  with
 anisotropic  hopping rates to neighbouring sites, given by $p$ and $q$
 (see Figure~\ref{fig:rdmwlkASYM}). 
 The  average speed of the walker is given by $p -q$:
 If $X_t$ is the (random) 
 position  of the walker at time $t$ we have for $t \to \infty$
 $$   \frac{  X_t }{t} \to p -q   \quad  \hbox{(almost surely)} $$

  We can  define a large deviation function  $G(v)$  by the
 following relation: 
$$  { {\rm Proba}\left( \frac{X_t}{t} = v \right) \sim
 {\rm e}^{ - t\, G(v)}  }  \, $$ 
valid in the limit of large times.

This function can be calculated explicitly. It is given by:

 $$ { G(v) = q + \frac{v}{2} \log \frac{q}{p} - \sqrt{v^2 + 4pq}
 - |v| \log \frac{ 2 \sqrt{pq}}{|v|+ \sqrt{v^2 + 4pq}}  }$$

Note that 
\begin{itemize}
\item $G(v)$ is a { positive} function that {  vanishes } at $v = p-q$.
\item $G(v)$ is {convex.}
\item  $ \ca{ G(v) - G(-v) = v \log \frac{q}{p} \, .}$
\item Using the definition of the large deviation
 function, we observe that,  in the long time limit,
 the  previous identity implies 
  $$ {   \frac{  {\rm Proba}\left( \frac{X_t}{t} = v \right)}
 {  {\rm Proba}\left( \frac{X_t}{t} = - v \right)}
  =   {\rm e}^{ t \,   v \log \frac{p}{q}} } $$
\end{itemize}

\begin{figure}[ht]
 \begin{center}
  \includegraphics[height=6.0cm]{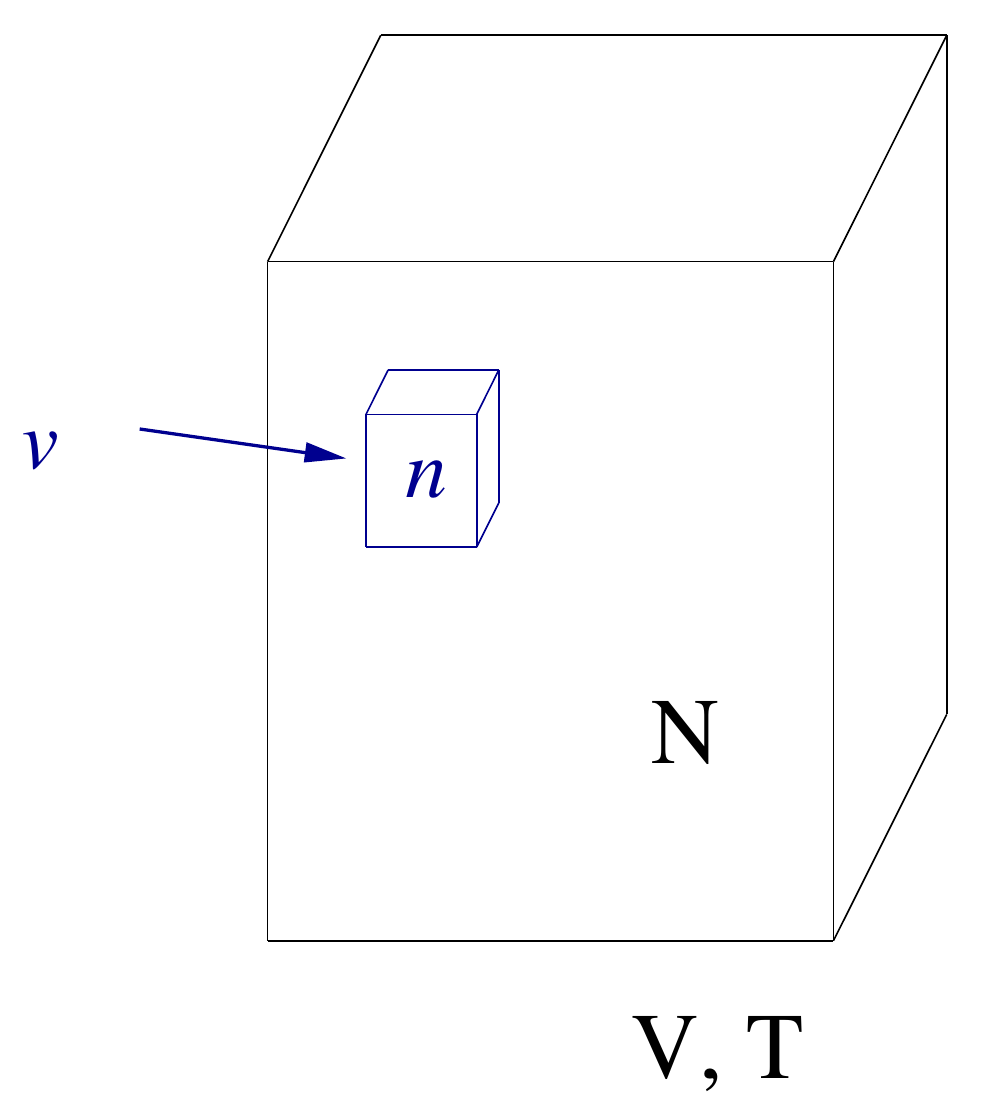}
  \caption{Equilibrium Fluctuations of density in a gas vessel.}
\label{fig:GasLDF}
 \end{center}
\end{figure}

 Our third example is closer to physics. Let us consider a gas,
 at thermodynamic equilibrium at temperature T,  consisting of  $N$
 molecules  enclosed in a vessel of total volume $V$. The average
 density is $\rho_0 = N/V$.
 We wish to
 probe local density fluctuations. We  consider an imaginary 
 volume $\mathrm{v}$, containing a large number of molecules
 but remaining  much smaller than the total
 volume, i.e.   such that $\rho_0^{-1} \ll \mathrm{v} \ll V$.
 Counting the number $n$ of molecules in  $\mathrm{v}$ will give us
 an empirical density $\rho = n/\mathrm{v}$ (see Figure~\ref{fig:GasLDF}).
 Clearly,  for $\mathrm{v}$ 
  large enough  the empirical  density  $\rho$ will be very close to 
  $\rho_0$ and typical fluctuations will scale as    $\sqrt{\mathrm{v}/V}$.
 What is the probability that   $\rho$ significantly deviates from 
 $\rho_0$?

The probability of observing
 large   fluctuations again satisfies a large deviation behaviour:
$$ {\rm Proba}\left( \frac{n}{\mathrm{v}} = \rho \right) \sim
 {\rm e}^{ - \mathrm{v} \, \Phi(\rho)}  \,\,\, \hbox{with   }   \,\,\,
 \Phi(\rho_0) = 0  \, .$$ 

In order to determine $\Phi(\rho)$, we 
 must count the fraction of
 the configurations of the gas  in which there are ${n = \rho v}$
 particles in the small volume ${\mathrm{v}}$ and ${N-n}$ particles in the rest
 of the volume ${(V -\mathrm{v})}$.
   Suppose that the interactions of the gas molecules are local.
 Then, neglecting surface effects,  this  number is given by
$${ {\rm Proba}\left( \frac{n}{\mathrm{v}} = \rho \right) \simeq
 \frac{ Z(\mathrm{v},n,T) Z(V-\mathrm{v},N-n,T)}{Z(V,N,T)} } $$ 
  Finally, we use that by definition, 
$ Z(\mathrm{v},n,T) = {\rm e}^{ -  v \beta  f( \rho , T) }$
 where $\beta =1/k_B T$ is the inverse temperature and $ f( \rho , T)$
 is the free energy per unit volume  and perform an expansion
 for $ 1 \ll \mathrm{v} \ll V$. This leads to 
 $$ \ca{  \Phi(\rho) = \beta \left( f(\rho, T)-  f(\rho_0, T) - (\rho -\rho_0)
 \frac{\partial f} {\partial \rho_0}   \right)   \, .}  $$

 We emphasize  that this large deviation function is very closely
 related to the thermodynamic Free Energy.

 A more general question would be the large deviation
 of a density profile. Suppose we  fully  cover the large box with
 $K = V/\mathrm{v}$ small boxes. What is the probability
 of observing an empirical density  $\rho_1$ in the first box, 
  $\rho_2$ in the second box etc...? Here again,
 we can show that a  large deviation principle is satisfied:
$$ {\rm Proba}\left(\rho_1,\rho_2,\ldots \rho_K  \right) \simeq
 {\rm e}^{ - V \,{ \mathcal F}(\rho_1, \rho_2,\ldots \rho_K )}   $$ 
 where the   large deviation function ${ \mathcal F}$
  depends on $K$ variables.
A reasoning similar to the one above allows us to show that
 $$ {\rm Proba}\left(\rho_1,\rho_2,\ldots \rho_K  \right) \simeq  
 \frac{ \prod_k Z(n_k, v, T)}{Z(V,N,T)} $$
Taking the infinite volume limit, we obtain
$$  \ca{ { \mathcal F}(\rho_1, \rho_2,\ldots \rho_K ) = \frac{\beta}{K}
 \sum_{k=1}^K \left(  f(\rho_i, T)-  f(\rho_0, T)\right) }$$

If now, we let the number $K$ of boxes go to infinity,
 then the question we are asking is the { probability of observing
 a given density profile } $\rho(x)$ in the big  volume $V$.
 For $K \to \infty$,
  the large deviation function $ { \mathcal F}$ becomes
 a  {\it functional of the density profile:}
$$  \ca{ { \mathcal F}[\rho(x)] = \beta \int dx 
 \left(  f(\rho(x), T)-  f(\rho_0, T)\right) } $$
  $f = -\log Z(\rho,T)$  being, as above,
   the  {\it free energy per unit  volume}.
  We conclude that  the  Free Energy of Thermodynamics could have been  defined
 {\it  ab initio}   as a 
  large deviation function.  More generally, all  thermodynamic potentials
  can be realized as large deviation functions. 

  However, the concept of  large deviations does not pertain to equilibrium.
  Large deviation functions can be introduced
  for very general processes, even far
  from equilibrium. They are positive functions that
  attain their minimum for when their argument
   takes the typical stationary variable.  These remarks suggest
 that these functions {\it may } be used  as 
    potentials in non-equilibrium statistical mechanics.

\subsection{Large Deviations and Cumulants}
\label{SubSec:Cumulants}

Let $X_t$ be a variable that satisfies a large deviation principle in 
 the limit  $t \to \infty$:
$$P\left(\frac{X_{t}}{t}=j\right) {\sim}e^{-t \Phi(j)}$$ 
where the large deviation function $\Phi(j)$ is 
positive and vanishes at $j =J$.

Another way to encode the statistics of $X_t$ is to study its 
moment-generating function,  defined as the average value 
$ {  \left\langle  {\rm e}^{\mu X_t}  \right\rangle }\,.$
Expanding   with respect of $\mu$, we get 
 $$  \log  \left\langle  {\rm e}^{\mu X_t}   \right\rangle
  = \sum_k \frac{\mu^k}{k!}  \langle \langle  X^k \rangle \rangle_c$$  
 where
$  \langle \langle  X^k \rangle \rangle_c$ denotes 
 the $k$-th cumulant of $X_t$.

 From  the  large deviation principle, we can show 
the  following behaviour:
\begin{equation}
  \left\langle  {\rm e}^{\mu X_t}   \right\rangle \simeq 
    {\rm e}^{ E(\mu) t}   \quad \quad  \hbox{ when} 
  \quad  t \rightarrow \infty  
\label{eq:defEmu}
\end{equation}
This implies that all  all cumulants of $X_t$ grow 
linearly with time and their values
 are given by the successive derivatives of ${E(\mu) }$.
 Moreover,  the cumulant generating function $E(\mu)$
 and  the large deviation function $\Phi(j)$
 are related by   Legendre transform. This  can be seen by using the 
 saddle-point method: 
$$  \left\langle  {\rm e}^{\mu X_t} \right\rangle
 = \int {\rm Pr}(X_t){\rm e}^{\mu X_t} dX_t = 
 t\int {\rm Pr}\left(\frac{X_t}{t}=j \right){\rm e}^{\mu t j } dj \sim 
 \int {\rm e}^{\mu t j -t \Phi(j)} $$ 
 From this relation we conclude that,
\begin{equation}
  \textcolor{red} 
 {\fbox{$   \ca{ E(\mu) = \max_j \left(
 \mu j -  \Phi(j)    \right)} $}}
\label{eq:LegendreEPhi}
\end{equation}
 
In the following examples, we shall often calculate $E(\mu)$
 first and then determine  $\Phi$ by a Legendre transformation.

\subsection{Generalized Detailed Balance}  
\label{SubSec:GDB}

  We have shown that detailed balance is a fingerprint of equilibrium.
  Conversely, out of  equilibrium, time
 reversibility and  detailed balance are  broken. A priori, one could
 imagine that any arbitrary Markov operator could  represent
 a physical system  far from equilibrium. This is not the case: the fundamental
 laws of physics are time-reversible (leaving apart  some aspects of
 weak-interaction). It is only after a coarse-graining procedure,
 when non-relevant degrees of freedom are integrated out, that
 the resulting effective dynamics  appears to be irreversible 
 for the restricted degrees of freedom we are interested in
  and for the space and time
 scales that we are considering.
 Nevertheless, whatever coarse-grained description is chosen
 at a  macroscopic scale,  a signature of this fundamental time-reversibility
  of physics   must
 remain. In other words, in order to have a sound physical model, even 
  very  far from equilibrium, detailed balance can not
 be violated  in an arbitrary manner: there   is  a  `natural way'
  of generalizing  detailed balance. We shall investigate here 
 what happens  to detailed balance for  a system connected
 to  unbalanced reservoirs (as in the pipe paradigm).

We first reformulate the equilibrium  detailed balance,  to
 make generalizations more transparent.

 \vskip 0.3cm

 {\bf The Equilibrium Case:}
\vskip 0.3cm

A  system is a thermal equilibrium with a reservoir at $T$ satisfies 
 the detailed balance   with respect to the Boltzmann weights.
 Equation~(\ref{eq:DetBal}) becomes 
\begin{equation}
   M({\mathcal C'}, {\mathcal C})  {\rm e}^{ -\beta E({\mathcal C})}
 =  M({\mathcal C}, {\mathcal C'}) \, 
\,    {\rm e}^{ -\beta E({\mathcal C'})} 
\end{equation}
 
Equivalently, defining $ { \Delta E =  E({\mathcal C'}) -  E({\mathcal C})},$
which represents the energy exchanged between the system and the reservoir
at a transition from ${\mathcal C}$ to ${\mathcal C'}$, the above equation
 becomes
\begin{equation}
   M_{\ca{+\Delta E}}({\mathcal C} \to {\mathcal C'})  
 =  M_{\ca{-\Delta E}}({\mathcal C'} \to {\mathcal C}) \, 
\,    {\rm e}^{ -\beta  \ca{\Delta E}} 
\end{equation}
where we have added an index to keep track of the exchanges of energy.

\vskip 0.3cm

{\bf The Non-Equilibrium Case:}
\vskip 0.3cm

Consider now  a system $S$ in contact with {\it two} reservoirs  
 $R_1$ and $R_2$ at $T_1$ and $T_2$. Suppose that during an elementary
step of the process, the system can exchange  energy (or matter...)
 ${\Delta E_1}$  with the first reservoir and  ${\Delta E_2}$ 
with the second one. Then,  generalized  detailed  balance is given by 
\begin{equation}
\textcolor{red}
{\fbox{$ \,\,\,\,
 {M}_{\ca{\Delta E_1, \Delta E_2}}{({\mathcal C} \to {\mathcal C'})} 
 =  {M}_{\ca{- \Delta E_1,  - \Delta E_2} } {({\mathcal C'} \to {\mathcal C})} \, 
  \ca{ {\rm e}^{ -\frac{\Delta E_1}{kT_1}-\frac{\Delta E_2}{kT_2} }} 
\,\,\,\,   $} }
\label{eq:GDB}
\end{equation}
where $\Delta E_i =  E_i({\mathcal C'}) -  E_i({\mathcal C})$ for $i=1,2$.
 This equation can be obtained by the following physical reasoning.
 Consider the global system $S + R_1 + R_2$: this is an isolated system,
 its total energy  $E({\mathcal C}) + E_1 + E_2$
 is conserved by the dynamics. Considered as a whole, the global  system 
  is governed by a reversible dynamics and in the infinite time limit
 it will reach  the  microcanonical measure. Besides,
 its dynamics must satisfy
detailed balance with respect to this microcanonical measure.
 This condition is expressed as 
\begin{eqnarray}
 \ca{ {\rm e}^{\frac{S_1(E_1) + S_2(E_2)}{k}} } \,\, 
  M( \{ {\mathcal C}, E_1,  E_2  \} \to  \{ {\mathcal C'}, E_1',  E_2'  \}) 
 \quad 
  =  M( \{ {\mathcal C'}, E_1',  E_2'  \} \to  \{ {\mathcal C}, E_1,  E_2  \})
 \,\,   \ca{ {\rm e}^{\frac{S_1(E_1') + S_2(E_2')}{k}} }   \nonumber 
 \end{eqnarray}
where $S_1$  and $S_2$  are the entropies   of the reservoirs (i..e the logarithms
 of the phase-space volumes). Using the fact that the
  reservoirs are at well-defined temperatures  and that energy exchanges
  with the system are small (i.e. $E_i' - E_i \ll E_i$), we can expand
 the entropy variations of each reservoir and this leads us to 
 Equation~(\ref{eq:GDB}). 

  We give now
 a  more abstract formulation of generalized  detailed  balance in which
  energy exchanges with the reservoirs are replaced by the flux of 
 an arbitrary quantity $Y_t$ (it can be a mass,  a charge, an  entropy...).
 Let us suppose  that during an elementary transition from
  ${\mathcal C}$ to ${\mathcal C'}$ between 
   $t$ and $t+dt$, the  observable    $Y_t$,  is incremented by $y$:
$$   { {\mathcal C}  \rightarrow  {\mathcal C'}} 
 \hbox{ and }  
  \,\, {  Y_t  \rightarrow Y_t + y }   \quad \hbox{  with probability }
 { M_y({\mathcal C'}, {\mathcal C})dt}  
 $$

 We shall  also assume that $Y_t$ is odd with respect to time-reversal,
 i.e. by  time reversal, the increment $y$  changes  its sign:
 $  { {\mathcal C'}  \rightarrow  {\mathcal C} \hbox{ and }  
  \,\, Y_t  \rightarrow Y_t - y } \, .$

A generalized detailed balance relation with respect to  $Y_t$
 will be satisfied if there exists  a constant $\mu_0$
 such that  the transition rates satisfy
\begin{equation}
 M_{\ca{+y}}({\mathcal C'}, {\mathcal C})
 =  M_{\ca{-y}}({\mathcal C}, {\mathcal C'}) \,  \ca{ {\rm e}^{\mu_0 y}}
 \label{eq:YGDpourY}
\end{equation}

This formula can be further extended by considering 
multiple  exchanges of various quantities between
 different reservoirs: the  statement of generalized detailed balance
 becomes 
$$
{ M}_{\ca{y_1,y_2,\ldots y_k}}  ( {\mathcal C}  \to {\mathcal C'}) 
 =  {M}_{\ca{-y_1, -y_2,\ldots -y_k}}   ( {\mathcal C'}  \to {\mathcal C}) 
  \,  \ca{ {\rm e}^{\mu_1^0 y_1 + \ldots \mu_k^0 y_k}} $$

\subsection{The Fluctuation Theorem}
\label{SubSec:FT}

 We shall now discuss a very  important
 property of systems out of equilibrium,  known as the
  Fluctuation Theorem. This relation
  was derived by G. Gallavotti and E. D. G. Cohen \cite{Gallavotti}.
 Here, we follow the  proof  of the  Fluctuation Theorem for
 valid for  Markov processes that was given by Lebowitz and Spohn \cite{LeboSpohn}
 A crucial emphasis is put on the  generalized detailed balance relation.

 The idea is 
to investigate how the  generalized detailed balance 
 equation~(\ref{eq:YGDpourY}) modifies
 the reasoning of  in  section~\ref{SubSec:DetailedB}, where we 
showed  that  detailed balance
 and time reversibility are equivalent. 
 
\begin{figure}[ht]
 \begin{center}
  \includegraphics[height=5.5cm]{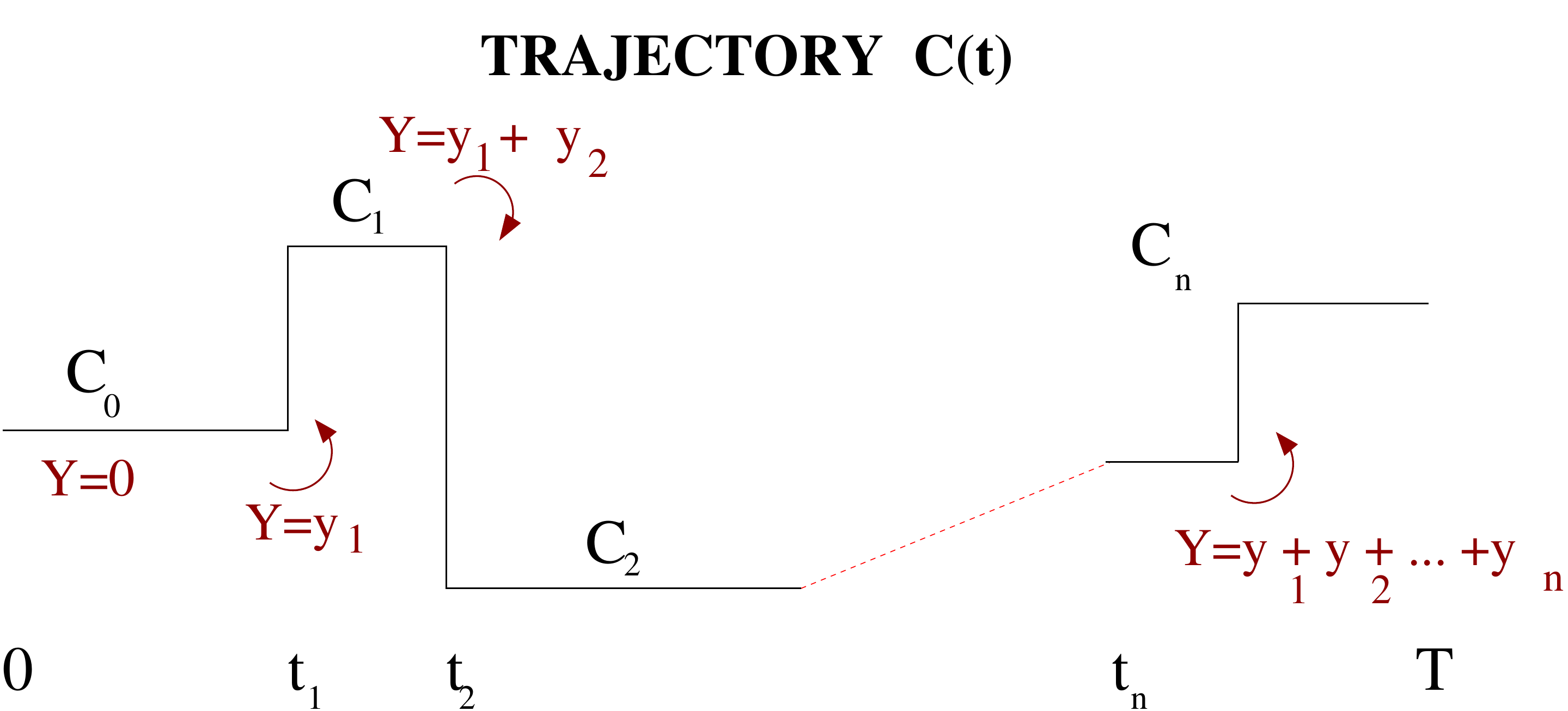}
  \caption{A trajectory of the system for $0 \le t \le T$. At each transition,
 the observable $Y$ is incremented by a quantity $y$.}
\label{fig:FT}
 \end{center}
\end{figure}

 As in section~\ref{SubSec:DetailedB}, we consider a trajectory
 (or history) of the system between time 0 and $T$. Now, for each jump
 between two configurations, we also keep track of the increment in the
 quantity $Y_t$ for which  the  generalized detailed balance 
 equation~(\ref{eq:YGDpourY}) is valid (see Figure~\ref{fig:FT}).

 As in  section~\ref{SubSec:DetailedB}, we compute 
 the  ratio between  the probabilities of forward and backward trajectories:
 equation~(\ref{eq:RatioFB}) is not modified (it is true for any Markov 
 process). But, now, we simplify this ratio by using
  the  generalized detailed balance 
 equation~(\ref{eq:YGDpourY}). We obtain 
\begin{eqnarray}
\ca{  \frac{ {\rm Pr} \{ { C}(t)\} }
{ {\rm Pr} \{\hat{C}(t)\}} = } 
\ca{ {\rm e}^{\mu_0 Y  \{ {C}(t)\}  }} \quad 
\ca{ \frac{ P_{{\rm stat}}({\mathcal C}_0)}{  P_{{\rm stat}}({\mathcal C}_n) }}  
\label{eq:GeneralRatio}
\end{eqnarray}
 where  $Y\{ C(t)\} = y_1 + y_2 +\ldots y_n$ represents the
 total  quantity  of $Y$ transferred  when the system
  follows  the trajectory $C(t)$ between 0 and $T$.

 Here, the  ratio between  the forward and backward  probabilities  
 is  different from unity. The dynamics is not reversible anymore
 and the  breaking of time-reversal is precisely  quantified by the total flux 
  of $Y$. Recall that $Y$ is odd, under time reversal. Hence, we have
$${ Y\{\hat{ C}(t)\} = -Y\{ { C}(t) \} }$$

It is  now useful to define the auxiliary quantity:
 $$ Z\{ { C}(t)\} = Y\{ { C}(t)\} + \frac{1}{\mu_0}
 \log \frac{ P_{{\rm stat}}({\mathcal C}_0)}{  P_{{\rm stat}}({\mathcal C}_n) } 
  $$

 The quantity $Z\{ {C}(t)\}$ is again odd w.r.t. time-reversal
 and it satisfies
 \begin{eqnarray}
 \frac{ {\rm Pr} \{ { C}(t)\} }
{ {\rm Pr} \{\hat {C}(t)\}} =  
{ {\rm e}^{\mu_0 Z \{ { C}(t)\}  }}   
\nonumber  
\end{eqnarray}
   or, equivalently, 
 \begin{eqnarray}
  {\rm e}^{(\mu-\mu_0) Z\{ { C}(t)\} }
   {\rm Pr} \{ { C}(t)  \} =
  {\rm e}^{ \mu Z\{ { C}(t)\} } {\rm Pr} \{\hat { C}(t)\}  = 
 {\rm e}^{-\mu{Z} \{\hat { C}(t)\} } {\rm Pr} \{\hat { C}(t)\} 
\end{eqnarray}
Summing over all possible histories
  between  time 0 and $t$, leads us to 
\begin{eqnarray}
\sum_{ \{ { C}(t) \}} {\rm e}^{(\mu-\mu_0) Z } \,\, 
   {\rm Pr} \{ { C}(t)  \} =
\sum_{ \{  {\hat{ C}(t) } \}}
 {\rm e}^{-\mu{Z} \{\hat { C}(t)\} }  \,\,  
   {\rm Pr} \{ \hat {  C}(t)  \}
 \nonumber
\end{eqnarray}
Interpreting both sides as average values, we obtain 
  \begin{eqnarray}
 { \left\langle  {\rm e}^{(\mu-\mu_0)  Z_t}   \right\rangle 
 =  \left\langle  {\rm e}^{-\mu Z_t}   \right\rangle  } 
\label{FTLaplace}
 \end{eqnarray}
 This is the statement of the Fluctuation Theorem in  Laplace
 space, for the auxiliary variable $Z$.  The quantities
 $Z_t$ and $Y_t$  grow with time, 
linearly in general.
 Their difference remains,  generically,   bounded (beware: this 
  could  be untrue  for some specific systems where `condensation' in some
 specific configurations  occurs. We assume here this does not happen). 
 This implies 
 that,  in the long time limit,  $Z_t$ and $Y_t$  have the
 same statistical behaviour and therefore 
$$ 
  \ca{ { \left\langle  {\rm e}^{(\mu-\mu_0)  Y_t}   \right\rangle 
 \simeq   \left\langle  {\rm e}^{-\mu Y_t}   \right\rangle  } }
 \quad    { {\rm when } \, \,  t \to \infty } $$ 

 Inserting now the typical behaviour~(\ref{eq:defEmu})
 derived in section~\ref{SubSec:Cumulants},
 $\left\langle  {\rm e}^{\mu Y_t}   \right\rangle \simeq 
    {\rm e}^{{  E}(\mu) t} \, ,$
 we conclude that
\begin{equation}
 \cc{ {  E}(\mu - \mu_0  ) =   {  E}(-\mu) }
\end{equation}

A Legendre transform yields  the 
 {\bf  Gallavotti-Cohen Fluctuation Theorem}  for the large deviation
 function
\begin{equation}
\textcolor{red}
{\fbox{$  \quad
  \ca{ \Phi(j) =   \Phi(-j) - \mu_0 j}
    \quad $} } 
\end{equation} 
 Using  the  large deviation principle,  this  identity
 implies 
\begin{equation}
\ca{ \frac{ {\rm Pr}\left(\frac{Y_{t}}{t}=j\right)}
 { {\rm Pr}\left(\frac{Y_{t}}{t}=-j\right)}
 {\simeq} \, {\rm e}^{\mu_0 j t} } 
\end{equation}

 This relation is the generic way of stating the Fluctuation Theorem.
 It compares the probability of occurrence of an event (for example,
 a total flux  of charge, or energy or entropy)  with that of the 
  opposite event. This relation   is true  
 {\bf far from equilibrium}. It   has been proved rigorously
 in various  contexts (chaotic systems,
 Markov/Langevin dynamics...).

\hfill\break
{\it Remark:} In the multiple variable case, we
 would obtain for the multi-cumulant generating function 
$$\cc{ {  E}(\mu_1 - \mu_1^0, \ldots,
 \mu_k - \mu_k^0  ) =   {  E}(-\mu_1, \ldots, -\mu_k) } $$ 
Or equivalently, for the large deviation function, 
$$
\textcolor{red}
{\fbox{$  \quad
  \ca{ \Phi(j_1,\ldots, j_k ) = 
   \Phi(-j_1, \ldots, -j_k) -\sum_{i=1}^k  \mu_i^0 j_i   }
    \quad $} }  $$

\section{The Exclusion Process}
\label{Sec:ASEP}

A fruitful strategy to gain insight into non-equilibrium physics  is 
 to extract as much information as possible from analytical
 studies and from  exact solutions of some special  models. 
Building a simple representation for  complex
   phenomena  is a  common  procedure  in    physics, leading
   to  the emergence of  paradigmatic systems: the harmonic oscillator,
   the random walker, the Ising model. All  these `beautiful models'
   often display  wonderful  mathematical structures
  \cite{Baxter}.

 In the field of
 non-equilibrium statistical mechanics, the Asymmetric Simple
 Exclusion Process (ASEP) is reaching  the  status of such  a
 paradigm. The ASEP consists of  particles   on a lattice, that
 hop  from a site  to its  immediate neighbours,   and
 satisfy the {\it exclusion condition} (there is at most 
 one particle per site). Therefore, a jump  is allowed only if the
 target site is empty. Physically, the   exclusion constraint mimics short-range
 interactions amongst particles.  Besides, in order to drive this
 lattice gas out of equilibrium, non-vanishing  currents must  be
 established in the system. This can be achieved by various means: by
 starting from  non-uniform initial conditions, by coupling  the system
 to external reservoirs  that drive   currents   through the system
 (transport of particles, energy, heat) or  by introducing some
 intrinsic bias in the dynamics that favours motion in a privileged
 direction.  Thus, each  particle is  an asymmetric random walker that
 interacts with the other and
  drifts  steadily along the direction of an external driving force.

\begin{figure}[ht]
 \begin{center}
  \includegraphics[height=4.0cm]{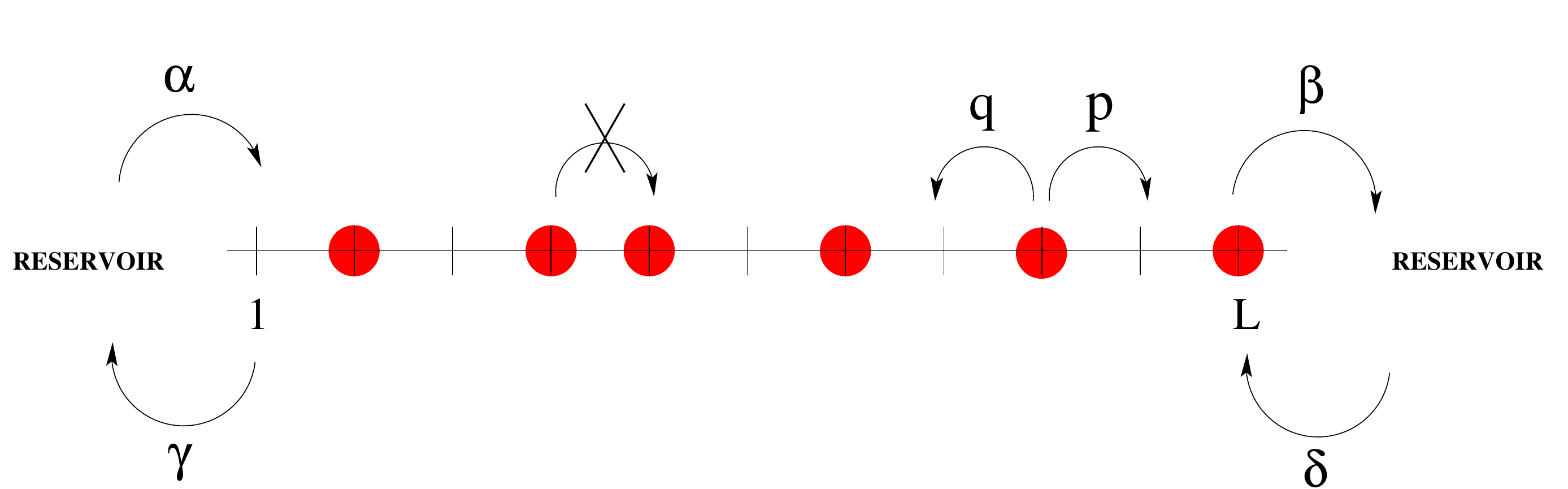}
  \caption{The Asymmetric Exclusion Process with Open Boundaries.}
\label{fig:ASEP}
 \end{center}
\end{figure}

  From Figure~\ref{fig:ASEP}, it can be seen that the ASEP on a finite lattice,
 in contact with two reservoirs, is an idealization of the paradigmatic pipe picture
 of Figure~\ref{fig:Courant}  that we have been constantly discussing.

  To summarize,
  the ASEP is a minimal model to study non-equilibrium behaviour. It is simple
 enough to allow  analytical studies, however  it contains the necessary
 ingredients for the emergence of a  non-trivial phenomenology:
\begin{itemize}
 \item ASYMMETRIC: The  external driving  breaks detailed-balance and
 creates a stationary current in the system.
  The model exhibits a non-equilibrium  stationary state. 

 \item EXCLUSION:  The hard core-interaction implies
 that there is  at most 1 particle per site. The ASEP is a genuine N-body problem.

\item PROCESS:   The dynamics is stochastic and  Markovian: there is  no
 underlying   Hamiltonian. 
\end{itemize}

\subsection{Definition of the  Exclusion Process}
\label{SubSec:DefASEP}

\begin{figure}[ht]
 \begin{center}
  \includegraphics[height=5.0cm]{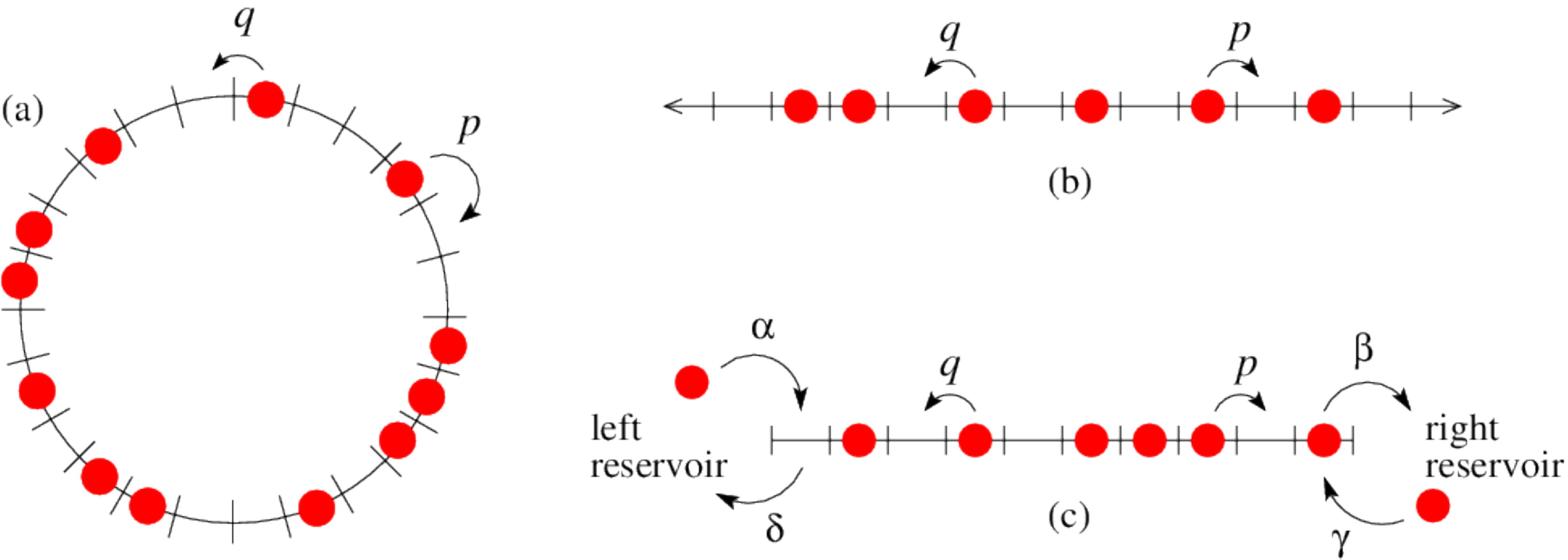}
  \caption{Different types of boundary conditions: The   ASEP can be studied  on a periodic chain (a), 
 on the infinite lattice (b)
 or on a finite lattice connected to two reservoirs (c).}
\label{fig:BC}
 \end{center}
\end{figure}

 The ASEP is a Markov process, consisting  of   particles
 located  on a discrete lattice that 
 evolves  in continuous time. We shall consider  only the case 
 when the underlying lattice is one dimensional. The  stochastic 
 evolution rules are the following:   at time $t$
 a particle located  at  a site $i$ in  the bulk of the system   jumps, in
the interval between  $t$ and $t+dt$, with probability $ p\ dt$ to the next
neighbouring site $i+1$ if this site is empty ({\it exclusion rule})
 and with  probability $q\ dt$ to the 
 site $i-1$ if this site is empty. The scalars $p$ and $q$ are parameters
 of the system; by rescaling time, one often takes $p=1$ and $q$ arbitrary.
 (Another commonly used rescaling is $p+q = 1$). 
 In   the totally asymmetric exclusion process (TASEP) 
 the jumps are totally biased in one direction ($q =0$ or $p =0$). 
  On the other hand, the {\it  symmetric} exclusion
  process (SEP) corresponds to the choice $ p = q$. The physics 
 and the phenomenology of the
 ASEP  are extremely sensitive to the boundary conditions. We shall
 mainly  discuss three types of  boundary conditions (see Figure~\ref{fig:BC}):

(i) The  periodic system: the exclusion process is defined 
 on a one dimensional lattice with $L$ sites (sites $i$ and $L + i$ are
identical) and $N$ particles.  Note that   the dynamics
 conserves the total number $N$ of particles

 (ii) The  finite one-dimensional
 lattice of $L$ sites  with open boundaries.
  Here, the site number $1$ (entrance site)
 and site number $L$ play a special role.
  Site 1 interacts   with the left reservoir as follows: 
 if    site 1  is empty, a particle can enter with rate $\alpha$ whereas 
 if it  is occupied it can become vacant  with rate $\gamma$.  Similarly,
 the interactions with the right  reservoir are as follows: 
  if site $L$  is empty, a particle can enter the system  with rate $\delta$
  and  if  $L$ is occupied, the  particle
   can leave the system with rate $\beta$. 
 The entrance and exit rates represent the coupling of the finite
 system with infinite reservoirs  which are  at different potentials 
 and  are located at the  boundaries. In 
  the special TASEP case, $q = \gamma = \delta = 0$:
  particles are injected by the left reservoir, they hop in  the right
 direction only and can  leave the system from the site number $L$ to the
 right reservoir.

(iii) The ASEP can also be defined on the infinite one-dimensional lattice.
 Here, the boundaries are sent to $\pm \infty$. Boundary conditions are
 here of a different kind:  the infinite  system remains always sensitive to  
  the  configuration it started from. Therefore, when studying the 
  ASEP on the  infinite lattice one must carefully specify the 
  initial configuration (or statistical set of configurations) the dynamics has
  begun with.

\subsection{Various Incarnations of the ASEP}
\label{SubSec:Avatars}

\begin{figure}[ht]
 \begin{center}
  \includegraphics[height=5.0cm]{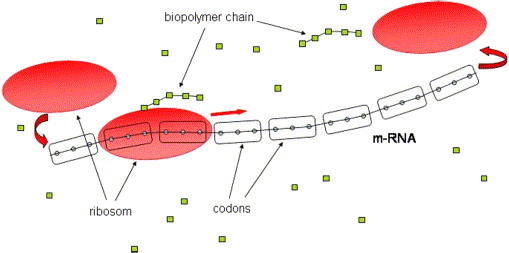}
  \caption{The ASEP was first defined and studied
 as a model of biopolymerization on nucleic acid templates \cite{MGP,Andreas1,Andreas2}.}
\label{fig:Ribosome}
 \end{center}
\end{figure}

 Due to its simplicity, the ASEP  has been introduced and used   in various
 contexts. It was  first proposed as a prototype to describe the
 dynamics of ribosomes along RNA \cite{MGP}  (see Figure~\ref{fig:Ribosome}). 
 In the mathematical literature,
 Brownian processes  with hard-core interactions were defined  by
 Spitzer   who coined the name  exclusion process. The ASEP also
 describes  transport in low-dimensional  systems with
 strong  geometrical  constraints  such as  macromolecules transiting
 through  capillary vessels,  anisotropic conductors,  or quantum
 dots  where electrons hop to  vacant locations  and repel
 each other via  Coulomb interaction.  Very popular  modern
 applications of the exclusion process include molecular motors that
 transport proteins  along  filaments inside the cells  and, of
 course, ASEP and its variants are ubiquitous in discrete models of
 traffic flow \cite{Andreas2}. 
  More realistic models that are  relevant for applications will not described  
 further: we refer the reader to the review paper \cite{CKZ}  that puts  emphasis on biophysical
 applications.

\begin{figure}[ht]
 \begin{center}
  \includegraphics[height=6.0cm]{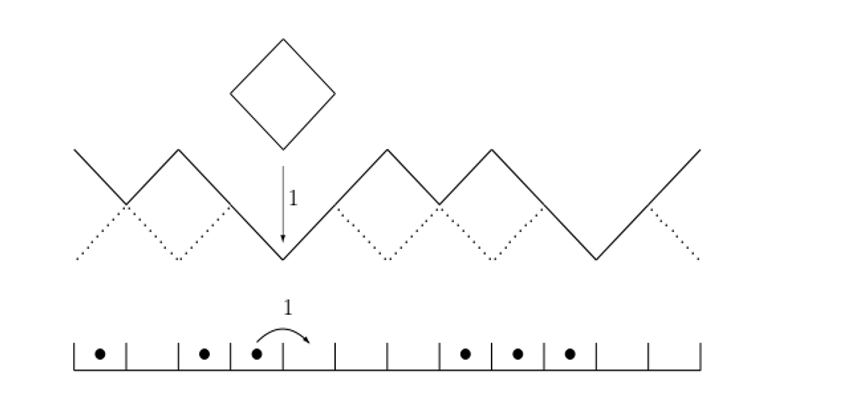}
  \caption{The one-dimensional ASEP is  a discrete version of the KPZ equation, satisfied by the height  $h(x,t)$
 of a  continuous interface:
$  \ca{ \frac{\partial h}{\partial t} = \nu \frac{\partial^2 h}{\partial x^2}
  + \frac{\lambda}{2}  \left(\frac{\partial h}{\partial x}\right)^2 + \xi(x,t) }  $
  where $\xi(x,t)$ is a Gaussian white noise. }
\label{fig:KPZ}
 \end{center}
\end{figure}

 Another feature  of ASEP is its relation with growth processes and in particular to the Kardar-Parisi-Zhang
 equation in one-dimension (see Figure~\ref{fig:KPZ}). A classic review on this subject is \cite{HHZ}.
 The  relation between the exclusion process and KPZ has lead  recently  to superb  mathematical developments:
 we refer  the readers to recent  reviews \cite{Sasamoto,Kriecherbauer} for details and references. 

 More generally, the ASEP belongs to the class of driven diffusive systems
 defined by Katz, Lebowitz and Spohn in 1984 \cite{KLS} (see \cite{Zia} for a review). 
   We emphasize that the ASEP is defined through dynamical rules: there
  is no energy associated with a microscopic configuration. More generally,
  the kinetic point of view seems to be a promising and
   fruitful approach to non-equilibrium systems.

 We emphasize that  there are  very many  variants of the fundamental ASEP:
 the dynamical rules can be modified (discrete-time dynamics, sequential 
or parallel updates, shuffle updates); 
 one can introduce local defects
 in the model  by modifying the dynamics on some specific bonds; 
 it is  possible to consider different types of particles with  different
 hopping rates; one can also consider  quenched or dynamical 
 disorder in the lattice; 
 the lattice geometry  itself can be changed (two-lane models, 
  ASEP defined on a stripe,
 on a network etc...). 
 All these alterations  drastically modify the outcome of the dynamics and require
 specific methods to be investigated. Literally hundreds of works have been devoted
 to the ASEP and its variants during the last fifteen years.
  Here, we shall  focus only  on the homogeneous case with the three ideal 
 types of boundary 
 conditions  discussed above and  present some of the 
  mathematical methods that have been developed for these three ideal cases.
  

\subsection{Basis  Properties of ASEP}
\label{SubSec:Ptes}

 The evolution of the ASEP  is encoded in the Markov operator
 $M$. For a finite-size system, the  Markov operator $M$
 is a matrix; for the infinite system  $M$ is an operator and its  precise 
 definition needs more elaborate mathematical tools \cite{Spohn91}.
 Unless stated otherwise, we shall focus here on the
  technically simpler case of a finite
 configuration space of size $L$  and the infinite system
 limit is obtained formally by taking  $L \to \infty$.
 An important feature of the ASEP on a finite lattice is ergodicity: any configuration
 can evolve to any other one in a finite number of steps. This property insures
 that  the Perron-Frobenius theorem holds true
 (see, for example  \cite{VanKampen}). This implies that the Markov matrix $M$
  contains the value  0 as  a non-degenerate eigenvalue and that all other 
eigenvalues $E$  have a strictly negative real
part. The physical interpretation of the spectrum of $M$ is the following: 
 the  right eigenvector associated with  the eigenvalue 0  corresponds
 to  the stationary state (or steady-state) of the dynamics.  Because  all non-zero 
eigenvalues $E$  have a strictly negative real
part,   the corresponding  eigenmodes of $M$ 
 are relaxation states: 
 the relaxation time is given by  $\tau =
-1/\mathrm{Re}(E)$ and   the imaginary part of $E$ leads  to  oscillations.

  We emphasize that   from the mathematical point of view,
 the  operator $M$ encodes all the required data of the dynamics and any 
 `physical' questions that one may ask about the system ultimately refers  to some
  property of $M$.  We shall now list some  fundamental issues that  may arise:
\begin{itemize}

\item Once the dynamics is properly defined, the basic question is to determine
 the  steady-state $P_{{\rm stat}}$  of the system
  {\it i.e.,} the eigenvector of $M$ with eigenvalue 0.
 Given a  configuration  ${\mathcal C}$, the value of  the component 
  $P_{{\rm stat}}({\mathcal C})$  is 
 the stationary weight (or measure) of  ${\mathcal C}$ in  the  steady-state, 
 {\it i.e.,} it  represents the  frequency of occurrence  of  ${\mathcal C}$
in the  stationary state.

\item  The knowledge of  the vector   $P_{{\rm stat}}$   is similar to knowing the
 Gibbs-Boltzmann canonical law in equilibrium statistical mechanics. 
 From  $P_{{\rm stat}},$  one can determine steady-state properties
 and all equal-time steady-state correlations. Some important questions
 are: what is the mean occupation $\rho_i$ of a given site $i$?
 What does the most likely density profile, given by the function $ i \to \rho_i$,
 look like? Can one calculate density-density correlation functions between
 different sites?   What is the probability of occurrence
 of a  density profile that differs significantly from the  most likely one
 (this probability is called the large deviation of the density profile)?

\item The ASEP, being a non-equilibrium system, carries a  finite, non-zero,  steady-state
 current $J$. The value of this current is an important physical observable of the model.
 The dependence of  $J$ on the external parameters of the system can allow to define
 different phases of the system. 

\item Fluctuations in the steady-state: the  stationary state  is a dynamical
 state in which the  system constantly evolves from one micro-state to another.
 This  microscopic evolution induces macroscopic fluctuations (which are the equivalent
 of the Gaussian Brownian fluctuations at equilibrium). How can one characterize
 steady-state  fluctuations? Are they necessarily Gaussian? How are  they related
 to the linear response of the system to small perturbations in the vicinity
 of  the steady-state? These issues can be tackled  by considering 
  tagged-particle dynamics, anomalous diffusion,
 time-dependent  perturbations of  the dynamical  rules  etc...

\item The existence of a  current $J$  in the stationary state
  corresponds to the physical transport
 of some extensive quantity $Q$ (mass, charge, energy) through the system. The total
 quantity  $Q_t$  transported during a (long)  period of time $t$ in the steady-state
 is a random quantity. We know that the mean value of  $Q_t$ is given by $J t$
 but there are fluctuations.  
 More specifically,  in the long time limit,
 the distribution of the random variable $(Q_t/t - J)$ represents exceptional
 fluctuations of the mean-current (known as  large deviations): this 
 is an important observable that quantifies the transport properties of the system.

\item The way a system relaxes to its  stationary state is also an important
 characteristic  of the system. The typical relaxation time $T$
 of the ASEP scales with
 the size $L$ of the system as $ T \sim L^z$, where $z$ is the  dynamical exponent. 
 The value of $z$ is related to the spectral gap of the Markov matrix $M$, {\it i.e.}, to
 the real-part of its largest non-vanishing eigenvalue. 
 For a diffusive system, one has $z =2$. For the ASEP with periodic
 boundary condition, an exact calculation leads to $z =3/2$. 
 More generally,   the transitory state 
 of the model can be probed using   correlation functions at different  times.

\item    The   matrix $M$ is generally a  non-symmetric matrix and, therefore, 
 its right eigenvectors differ from its  left eigenvectors. For instance, 
 a  right eigenvector $\psi_E$ corresponding to the eigenvalue $E$ is defined as 
\begin{equation}
  M \psi_E = E\psi_E      \, . 
  \label{eq:mpsi=epsi}
\end{equation}
 Knowing the spectrum of  $M$ conveys a lot of  information about the dynamics. There
 are analytical techniques, such as the Bethe Ansatz,  that allow us
 to diagonalize  $M$  in some specific cases.
 Because $M$ is a real matrix, its 
 eigenvalues (and eigenvectors) are either real numbers or complex
 conjugate pairs.

\item  Solving analytically  the master equation 
 would allow  us to calculate exactly the  evolution
 of the system.  A challenging goal is to determine the   finite-time Green function 
 (or transition probability)
 $P_{t}({\mathcal C}|{\mathcal C}_0)$, the probability for the system to be
 in configuration ${\mathcal C}$  at time $t$, knowing that  the
 initial configuration at time $t=0$ was  ${\mathcal C}_0$.  The knowledge of 
 the transition probability, together with the Markov property, allows us in principle
 to calculate all the  correlation functions of  the system.

\end{itemize}

 The following sections are devoted to explaining some  analytical
 techniques that have been  developed 
 to  answer some of these issues for the ASEP.

\subsection{Mean-Field analysis of the ASEP}
\label{SubSec:MF}

    Before discussing exact techniques to solve the dynamics of the ASEP, we want to explain
 the   mean-field approach. In many physically important  cases with  inhomogeneities or more complex
 dynamical rules,  mean-field calculations are the only available technique and they often lead to
 sound results that can be  checked and compared with numerical experiments.

\subsubsection{Burgers Equation in the Hydrodynamic Limit.}

   In  the limit of large systems, it is natural to look for a continuous description
 of the model.  Finding an accurate hydrodynamic model for interacting particle processes
  is a difficult and important problem.  Here, we present a naive approach that reveals the relation
 between the ASEP and Burgers equation.

  We recall that the  binary variable $\tau_i=0,1$ characterizes 
 if site $i$ is empty or occupied.
 The average value  $\langle \tau_i(t) \rangle$ satisfies the following 
 equation:
 \begin{eqnarray}   \frac{ d \langle \tau_i \rangle}{ d t}   & = & 
 p [\langle \tau_{i-1} ( 1 - \tau_i) \rangle - 
 \langle \tau_{i} ( 1 - \tau_{i+1}) \rangle]
 + q [\langle \tau_{i+1} ( 1 - \tau_i) \rangle 
 -  \langle \tau_{i} ( 1 - \tau_{i-1}) \rangle ] \nonumber \\
  & = & p \langle \tau_{i-1} \rangle + q \langle \tau_{i+1} \rangle 
 -(p + q) \langle \tau_i \rangle + (p - q)
 \langle \tau_{i} (\tau_{i+1} - \tau_{i-1}) \rangle  \nonumber
 \end{eqnarray}

For  $p \neq q$:   1-point averages couple to 2-points averages
 etc... A   hierarchy of differential equations  is generated
 ({\it cf} BBGKY). This set of coupled equations  can not be solved in general.
 The mean-field  approach can be viewed as a technique for  closing 
 the hierarchy. For the ASEP, the procedure is quite simple. We
 sketch it below: 
\vskip0.3cm
${\bullet}$ Define the  continuous space variable ${ x = \frac{i}{L}}$.
 and  the limit  $L \gg 1 .$

\vskip0.3cm
  ${\bullet}$ Define a smooth local density by 
 $  {\langle \tau_i(t) \rangle = \rho(x,t)}$.

\vskip0.3cm
 ${\bullet}$  Rescale  the   rates:
  $ {p =  1  +  \frac{\nu}{L}}$ and
  $ {q =  1  - \frac{ \nu}{L}}$
 
\vskip0.3cm
 ${\bullet}$  {Mean-field assumption:}  write  the 
 2-points averages as products of 1-point averages.

 After carrying out this program
 leads to,
 after a  {diffusive rescaling of time }$ \ca{t \to t/L^2,}$ to
 the following partial differential equation
\begin{equation}
 \ca{  \frac{\partial \rho}{\partial t} = 
   \frac{\partial^2 \rho }{\partial x^2}    - 2  \nu 
 \frac{\partial \rho (1 -\rho )}{\partial x}  } 
\label{eq:Burgers}  
\end{equation}
 This is  the { Burgers equation with viscosity.}

\begin{figure}[ht]
 \begin{center}
 \includegraphics[height=3.0cm]{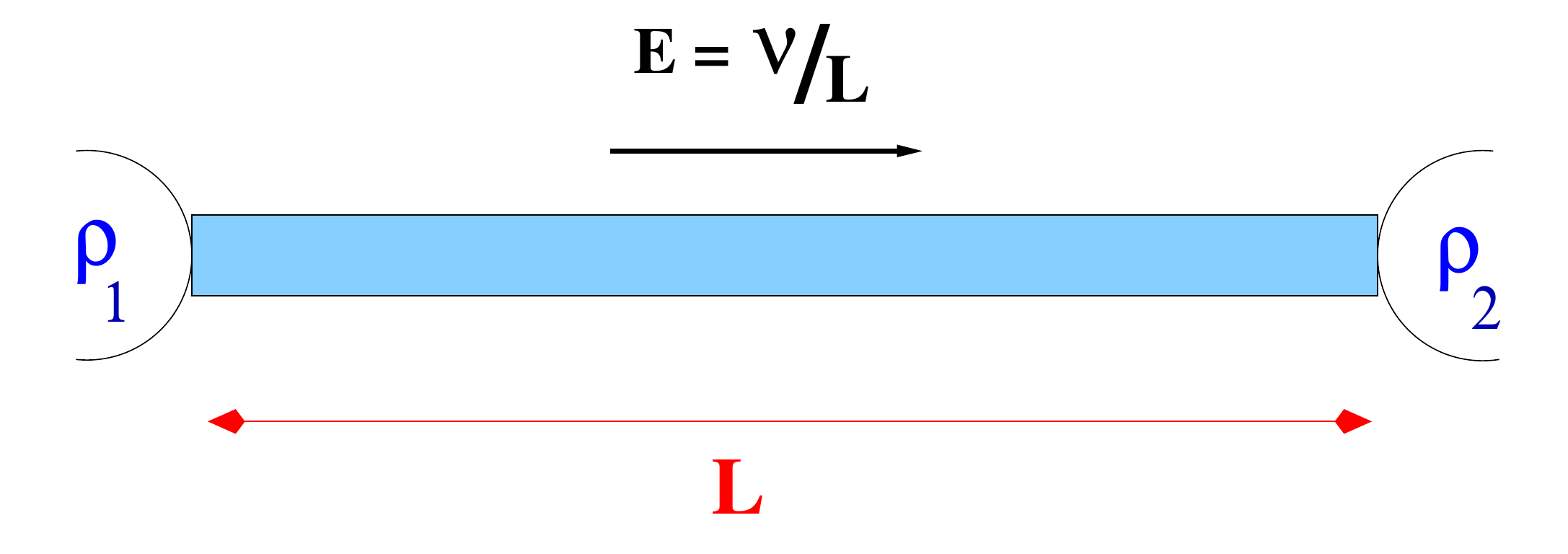}
\caption{Hydrodynamic description of the ASEP with open boundaries: the pipe model is retrieved.
A weak asymmetry in the jumping rates can be interpreted as a weak
bulk electric field, inversely proportional to the size of the system.}
\label{fig:Courant2}
 \end{center}
\end{figure}

To resume,  starting from the microscopic level,  we have  defined a  local density
 $\rho(x, t)$ and a local  current  $j(x, t)$ that depend on   macroscopic
 space-time variables   $x = i/L, t = s/L^2$ (diffusive scaling)
 in the limit of   {\it weak asymmetry} $ {p-q = \nu/L}$ 
 (see Figure~\ref{fig:Courant2}).  Then,  we have found
that the typical  evolution of the system  is given  by 
 the   hydrodynamic behaviour:
 $$ \cc{ \partial_t \rho =  \nabla^2  \rho  - \nu 
 \nabla\sigma(\rho ) } \quad 
 \hbox{ with } \quad \cc{  \sigma(\rho ) =  2 \rho (1 -\rho ) }  $$
 We have explained how to obtain this equation in a  `hand-waving' manner. However,
 the result is mathematically correct: it can be proved rigorously that the continuous
 limit of the weakly-asymmetric ASEP is described, on average, by this equation.
 The   proof is  a major  mathematical achievement.

 Had  we kept  a finite asymmetry:  $ p - q = {\mathcal O}(1)$, 
 the same procedure (with ballistic time-rescaling) would have led us 
 to the   inviscid limit of  Burgers equation:
$$   \frac{\partial \rho}{\partial t} = \frac{{\bf\ca{1}}}{{\bf\ca{L}}}
   \frac{\partial^2 \rho }{\partial x^2}    - 2\nu 
 \frac{\partial \rho (1 -\rho )}{\partial x}     $$
 {This equation is a textbook example of a PDF that generates shocks, even if the initial
condition is smooth.} A natural question that arises  is whether these shocks
  an artefact of the hydrodynamic limit or do they
 genuinely exist at the microscopic level. The following model, invented by J. L. Lebowitz
 and S. A. Janowsky sheds light on this issue.

  \subsubsection{ A worked-out example: The Lebowitz-Janowsky model}

The Lebowitz-Janowsky model describes  the formation of shocks at the microscopic scale \cite{JL1}.
 This is very simple model, but it has not been solved exactly. It  will provide  us with a 
 good illustration  of mean-field methods. 

The simplest version of Lebowitz-Janowsky model is  a TASEP on a periodic ring with
 one defective bound: through that bound (say the bond between site $L$ and site 1) the jump
 rate is given by $r$ whereas through all the other bonds, the  jump rates are equal to 1.
 For $r <1$ (which is the most interesting case), we have a slow bond, i.e. a constriction,
 that may prevent the flow of particles through and generate a 'traffic-jam'. This is indeed
 what is going to happen: for any given density $\rho$, there is a critical value $r_c(\rho)$ such that
 for $r \le r_c$ a  separation will occur  between a dense phase before the slow bond
 and a sparse phase after that bond. 

 \begin{figure}[ht]
 \begin{center}
  \includegraphics[height=4.0cm]{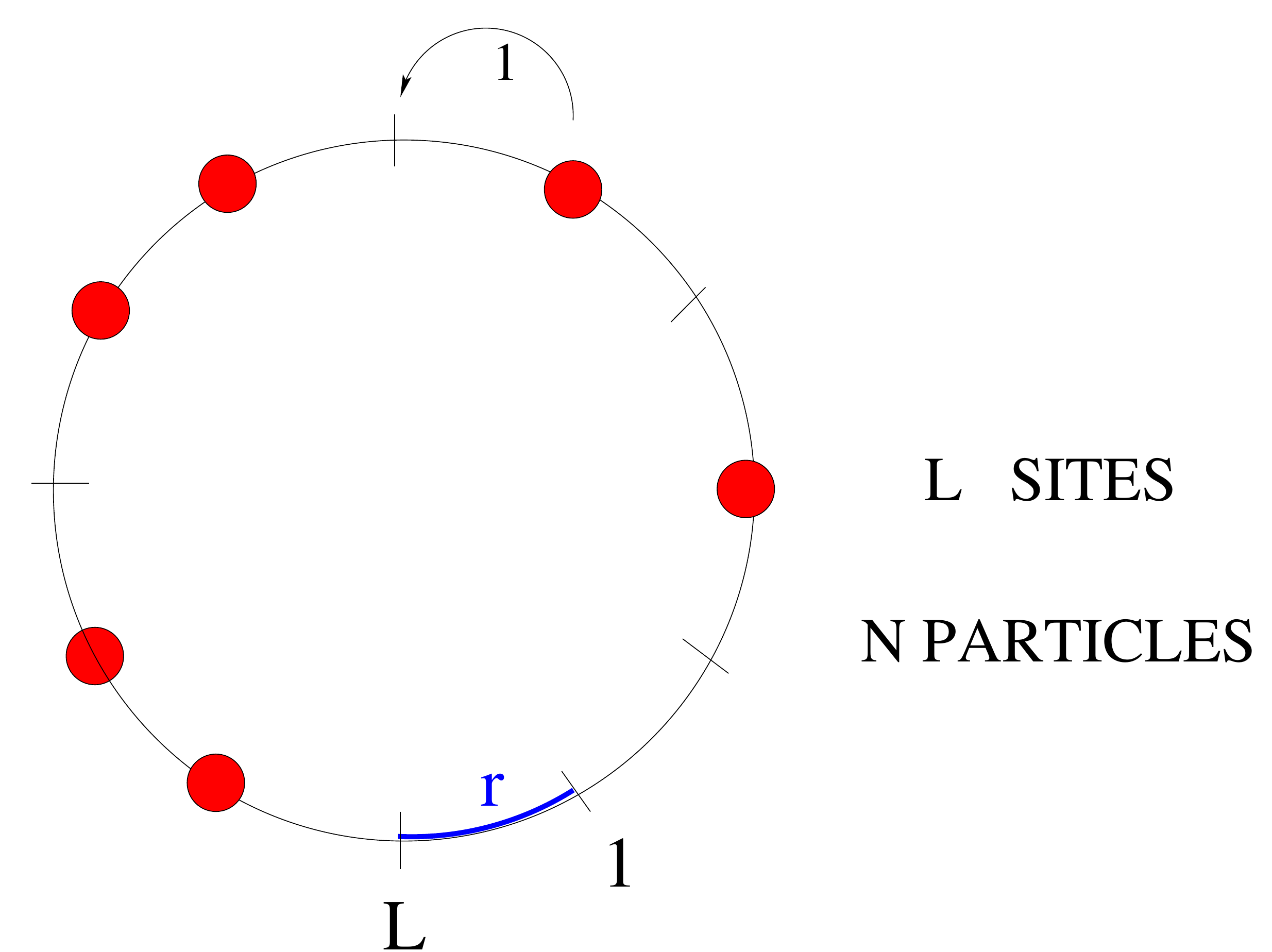}
  \caption{The TASEP on a ring with an inhomogeneous bond with jump rate $r$.}
\label{fig:JLeb} 
 \end{center}
\end{figure}

 The blockage model can be analysed by elementary mean-field
 considerations.  Through a `normal'  bond $(i,i+1)$ the current 
 is exactly  given by 
 ${J_{i,i+1} =  \langle \tau_{i} ( 1 - \tau_{i+1}) \rangle\,.} $
 In the stationary state, this current is uniform $J_{i,i+1} = J$.
 Far from the blockage and from the shock region,the density 
  is approximately  uniform (as numerical  simulations show, see Figure~\ref{fig:JLeb}). Thus, 
 using  a  mean-field assumption we can write
 $$ { J = \rho_{low} ( 1 - \rho_{low}) = \rho_{high} ( 1 - \rho_{high})} $$
 where  $\rho_{low}$  and $ \rho_{high}$  are the values of the density plateaux on both sides
  of the slow bond. This relation leads to two  possible solutions:

${\bullet}$ Either the  density  is uniform   everywhere, i.e.,  
  ${\rho_{low} = \rho_{high}= \rho_0}$

${\bullet}$  Or we have  different densities on 
the sides that make a  shock. Then, necessarily:     ${\rho_{low} =    1 -\rho_{high}} $

 To find the values of the density plateaux, we calculate the mean-field current 
 right at the defective bond:
 $${ r \rho_{L}( 1 -  \rho_{1}) = r \rho_{high}  ( 1 - \rho_{low}) = J}$$
(This is a strong approximation that neglects correlations through the slow bond.)
 We now have enough  equations to  obtain
$$ \rho_{low} =   \frac{r}{1 + r} 
 \quad \quad   \rho_{high} =   \frac{1}{1 + r} 
 \quad \hbox{ and }  \quad    J = \frac{r}{(1 + r)^2}  $$

 To conclude the analysis, we must find the condition for the existence of the shock: when do we have
 $\rho_{low} = \rho_{high}$ and when  $ \rho_{low} < \rho_{high}$?
 We must  use the conservation of the number
 of particles. Let  $1 \le S \le L$ be the position of the shock,  then we have 
 ${ N = S \rho_{low} + (L -S) \rho_{high}}$ i.e., dividing by $L$:
$$ {\rho_0 = \frac{ s \, r + (1 -s)}{ r + 1} }  \quad \quad \hbox{ with }
  \quad \quad  { 0 \le\,\,  s = \frac{S}{L} \,\, \le 1} $$
This relation defines the phase boundary between the uniform and the shock phases. It can be rewritten
 in a more elegant manner as follows:
$$    \ca{    \left| \rho_0 -\frac{1}{2} \right|  \le
 \frac{1 - r}{2( r + 1)}  }$$
    
\begin{figure}[ht]
 \begin{center}
\begin{tabular}{ccc}
  \includegraphics[height=5.0cm]{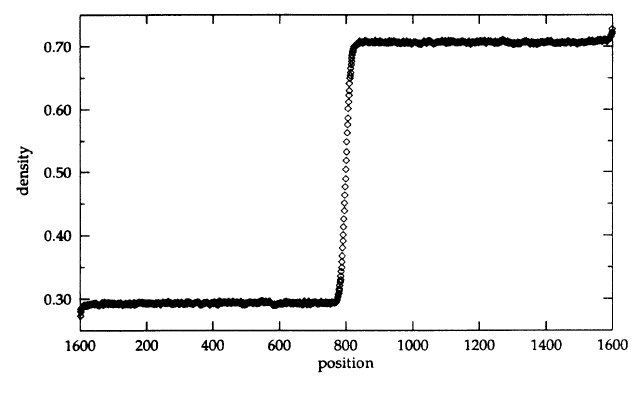}
 &\quad\quad\quad\quad  & 
 \includegraphics[height=5cm]{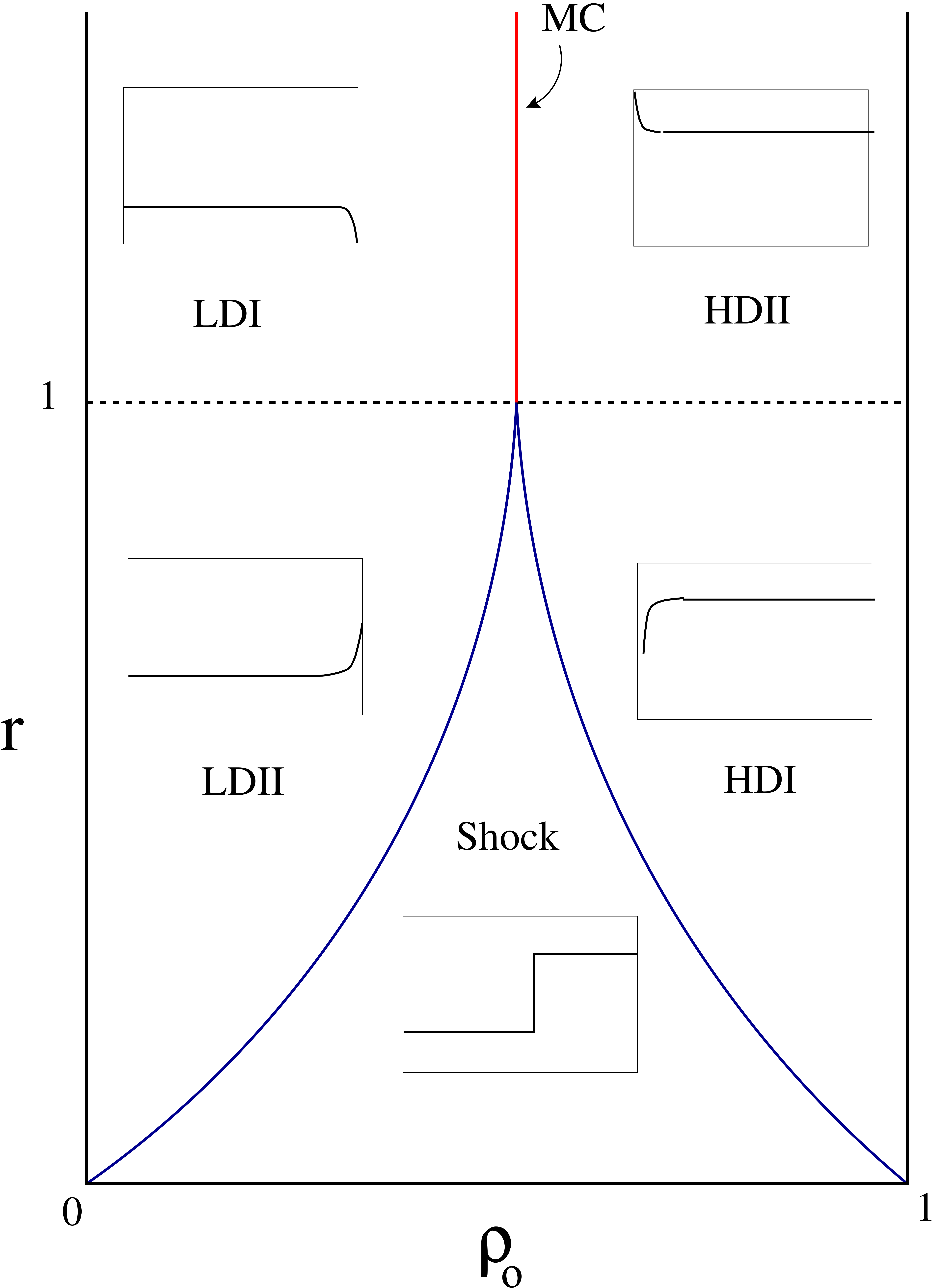}
 \end{tabular} 
\caption{Numerical simulations of the  Lebowitz-Janowsky model in the shock phase.
 Picture of the  phase diagram derived from the mean-field for all values of
 $r$ and average  density $\rho_0$.}
\label{fig:JLDiag}
 \end{center}
\end{figure}

From this equation, we observe, in particular,  that  a shock will always appear for $\rho_0 = 1/2$
 as soon as $ r < 1$. The full phase diagram of the system is drawn in Figure~\ref{fig:JLDiag}.
Numerical simulations seem to support this phase diagram. However, no exact proof is available.
 Besides, using an improved mean field analysis, the form of the shock
 can be calculated. However,  the results do not coincide with  simulations.
\vskip 0.3cm
The exact solution of the Lebowitz-Janowsky model is a celebrated open problem in this field (see \cite{JL1} for references
 and some recent progress).

\subsection{The Steady State of the open ASEP: The Matrix Ansatz}
\label{SubSec:Steady}

 There does not exist a  method
 to calculate the stationary measure for a non-equilibrium interacting model.
 After briefly describing the stationary state of 
 the ASEP with periodic boundary
 conditions  and of the ASEP on the infinite line, we shall focus
 on the ASEP with open boundaries.

  For the ASEP
 on a ring the steady state  is uniform:  all configurations
 have the same probability. The proof is elementary: for any configuration
 ${\mathcal C}$
the number of states it can reach to  by an elementary move is equal to 
the total  number of  configurations that can evolve into it.

For the exclusion process on an  infinite line, the stationary measures
 have been fully  studied and classified \cite{Spohn91}.  There are two 
 one-parameter families of invariant measures: one family,  denoted by $\nu_\rho$, 
 is   a  product of local Bernoulli measures of constant  density $\rho$, where each site
 is occupied with probability $\rho$; the other family is discrete
 and  is concentrated on a countable subset of configurations. For  the TASEP, this second family
 corresponds  to {\it blocking measures}, which are point-mass measures concentrated on
 step-like  configurations ({\it i.e.},
  configurations where all sites to  the left of a given site $n$
 are empty and  all sites to  the right of $n$ are occupied).

 We  now consider  the case of the ASEP on a finite and open  lattice  with open 
 boundaries. For convenience, we rescale the time so that forward jumps occur
 with rate 1 and backward jumps with rate $x = q/p$. This scalar  $x$ is called
 the asymmetry parameter.

 Here, the  mean-field method  gives 
  a reasonably good approximation. However, it is not exact
  for a finite system. There are notable deviations in the density profile
(i.e. the function  $i \to \rho_i$ where  $\rho_i$ is
  the average density at site $i$). Moreover, 
 fluctuations and rare events  are not well accounted for by  mean-field.
 We shall now explain a method to obtain the exact 
 stationary measure for  the ASEP  with open boundaries. This technique,
 known as the Matrix Representation Method, was developed in \cite{DEHP}.
 Since that seminal paper,  it  has become a  very important  tool
 to investigate  non-equilibrium models. The  review  written by R. A.  Blythe and M. R. Evans
 allows one to  learn the 
 Matrix Representation Method for the ASEP and other models  and also to find many  references \cite{MartinRev2}.

 A  configuration
 ${\mathcal C}$  can be represented by a  string of length $L$,
 $(\tau_1, \ldots, \tau_L)$,  where $\tau_i$ is the binary occupation
 variable of  the site $i$  (i.e. $\tau_i = 1$ or 0
  if the site $i$ is occupied or empty). The idea is to  associate
  with each configuration
${\mathcal C}$, the following matrix element: 
\begin{equation}
  P(\mathcal C) = \frac{1}{Z_L} \langle W | \prod_{i=1}^L 
 \left( \tau_i D + (1 - \tau_i) E \right) | V  \rangle \, .
\label{MPA}
\end{equation}

 The operators $\ca{D}$ and  $\cc{E}$, the vectors  $\cb{ \langle W  |} $
 and  $\cb{ | V  \rangle} $  satisfy 
\begin{eqnarray}
      \quad  \quad  \quad  \quad   \quad  \quad   
        \ca{D} \,  \cc{E}  -  x \cc{E}  \ca{D} 
  &=& (1 -x) (\ca{D} \,  +  \,  \cc{E})   \nonumber \\
            ( \beta \,  \ca{D} - \delta\,  \cc{E}  )
 \, \cb{| V  \rangle}       &=& 
  \cb{| V  \rangle}    \nonumber \\  
            \quad  
            \cb{ \langle W |}  (\alpha \, \cc{E} -  \gamma \, \ca{D} )
     &=&   \cb{ \langle W |}  \,
 \label{DEHPASEP}
\end{eqnarray}

 The claim is that if the algebraic relations~(\ref{DEHPASEP}) are satisfied 
 then, the matrix element~(\ref{MPA}), duly normalized, is 
 the stationary weight of the configuration
${\mathcal C}$. Note that the  normalization constant is   
 $Z_L =  {\langle W | }\left( \ca{D}  + \cc{E} \right)^L
  { | V \rangle} =   {\langle W | } C^L
  { | V \rangle}$ where $ C = D + E $

This algebra encodes combinatorial recursion relations between systems
 of different sizes. Generically, the representations of this quadratic algebra
   are  infinite dimensional
 ($q$-deformed oscillators).

\begin{eqnarray}
  D  = \left( \begin{array}{ccccc}
                  1&\sqrt{1 -x}&0&0& \dots\\
                  0&1&\sqrt{1 -x^2}&0& \dots\\
                  0&0&1&\sqrt{1 -x^3}&\dots\\
                  &&&\ddots&\ddots 
                   \end{array}
                   \right)  
\quad \hbox{and}  \quad E  =  D^{\dagger}
 \,\, 
  \nonumber 
 \end{eqnarray}


The matrix Ansatz allows one   to calculate
 {Stationary State Properties}
  (currents, correlations, fluctuations) and to derive the  {Phase Diagram}
 in the infinite size limit.  There are three phases, 
  determined by
 the values of $\rho_a$ and $\rho_b$ that represent effective densities of the left 
 and the right reservoir, respectively. 
  The precise formulae for these effective densities for the general ASEP model  were found by
  T. Sasamoto:
\begin{eqnarray}
 \rho_a =\frac{ 1}{a_{+} + 1}  \quad  &\hbox{ where }&  \quad 
  a_{\pm}  =   \frac{ (1-x-\alpha+\gamma)\pm
  \sqrt{(1-x-\alpha+\gamma)^2+4\alpha\gamma}}{2\alpha} \, ,\nonumber \\
 \rho_b = \frac{b_{+}}{b_{+} + 1}  \quad    &\hbox{ where }&  \quad 
  b_{\pm} =  \frac{(1-x-\beta+\delta) \pm
  \sqrt{(1-x-\beta+\delta)^2+4\beta\delta}}{2\beta} 
\label{def:abpm}
\end{eqnarray}

 \begin{figure}[ht]
 \begin{center}
  \includegraphics[height=5.0cm]{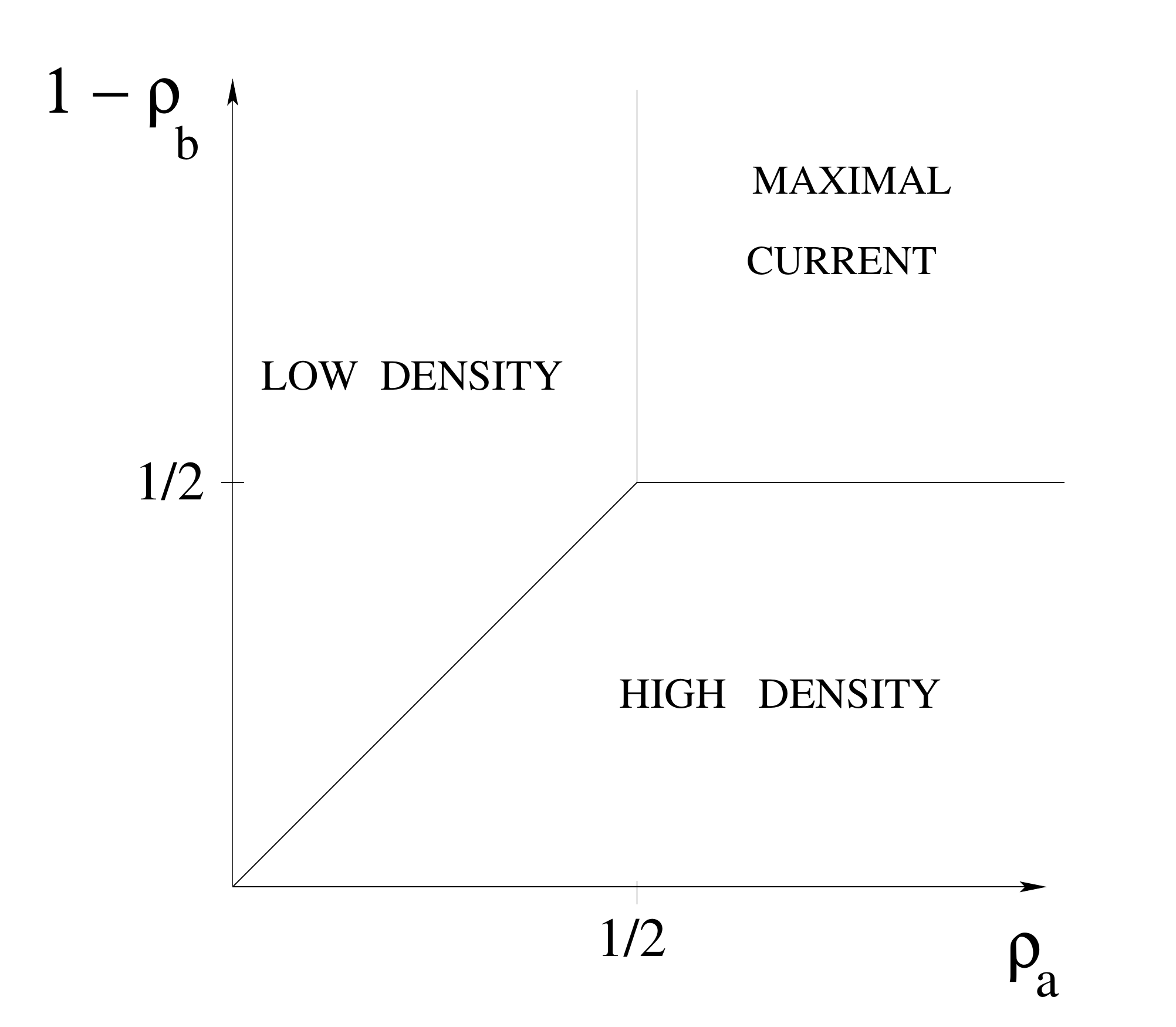}
  \caption{The Phase Diagram of the open ASEP. The phase in which the system is found
 depends on the effective densities $\rho_a$ and $\rho_b$.}
\label{fig:PhaseDiagASEP}
 \end{center}
\end{figure}

 In the TASEP case $q = \gamma = \delta =0$ and $p=1$,  the algebra simplifies  and reduces to 
\begin{eqnarray}
      \quad  \quad  \quad  \quad   \quad  \quad   
        \ca{D} \,  \cc{E} &=& \ca{D} \,  +  \,  \cc{E}   \nonumber \\
              \ca{D}  \, \cb{| \beta  \rangle}   & = & 
 \ca{  \frac{1}{\beta} } \, \cb{ | \beta  \rangle}     \nonumber   \\
             \cb{ \langle \alpha |} \, \cc{E} & = & 
  \cc{ \frac{1}{\alpha}}  \cb{ \langle \alpha | } \,\,   \nonumber
 \label{DEHPAlgebra}
\end{eqnarray}
 For this case, many  calculations can be performed in rather simple manner.
 In particular,  the average stationary current is found to be 
 $$  { J =  \langle \tau_{i} ( 1 - \tau_{i+1}) \rangle = 
\frac{\langle \alpha | C^{i-1}\, D\, E\,C^{L-i-1}  | \beta  \rangle}
{\langle \alpha | C^L  | \beta  \rangle}
= \frac{\langle \alpha | C^{L-1} | \beta  \rangle}
{\langle \alpha | C^L  | \beta  \rangle} = \frac{Z_{L-1}}{Z_L} } $$
 In fact, the Matrix Ansatz gives access to all equal
 time correlations in the steady-state. For example, the density profile:
 $$  {\rho_i = \langle \tau_{i}  \rangle = 
\frac{\langle \alpha | C^{i-1}\, D\,C^{L-i}  | \beta  \rangle}
{\langle \alpha | C^L  | \beta  \rangle} }$$
 or even  Multi-body correlations:
 $$ {  \langle \tau_{i_1} \tau_{i_2} \ldots \tau_{i_k} \rangle =
  \frac{\langle \alpha | C^{i_1-1}\, D\,C^{i_2-i_1-1}\, D\,\ldots  D\,C^{L-i_k} 
  | \beta  \rangle}
{\langle \alpha | C^L  | \beta  \rangle} } $$

The expressions look formal but it is possible to derive explicit formulae:
 either by using purely combinatorial/algebraic techniques or via a
 specific representation (e.g., $C$ can be chosen as a discrete Laplacian).
 For example,  we have
 $$ \langle \alpha | C^L  | \beta  \rangle
 \displaystyle =\sum_{p=1}^{L}\frac{p\,(2L-1-p)!}{L!\,(L-p)!}
\frac{\beta^{-p-1}-\alpha ^{-p-1}}{\beta ^{-1}-\alpha ^{-1}} $$

\begin{figure}[ht]
 \begin{center}
  \includegraphics[height=5.0cm]{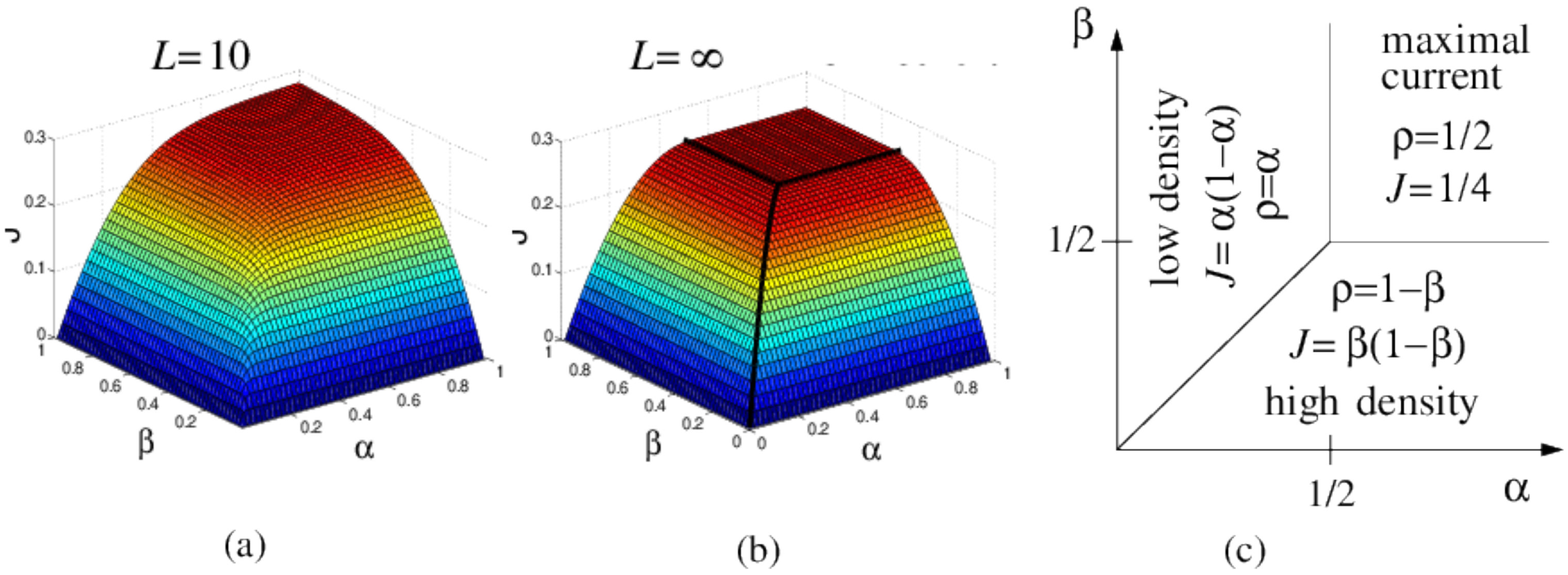}
  \caption{The Phase Diagram of the open ASEP}
\label{fig:PhaseDiagTom}
 \end{center}
\end{figure}

The TASEP phase diagram is shown in Figure~\ref{fig:PhaseDiagTom}.

 For the general ASEP  case, results can be derived through more elaborate methods involving
 orthogonal polynomials, as shown by T. Sasamoto (1999).  More details
 and precise references can again  be found in the  review of  R. Blythe and
  M. R. Evans \cite{MartinRev2}. 

  The Matrix Ansatz is an efficient tool to investigate the stationary state.
 However, it does not allow us to access to time-depend properties: how does the system
 relax to its stationary state? Can we calculate fluctuations of history-dependent
 observables (such as the total number of particles exchanged between the two reservoirs
 during a certain amount of time)?

\section{The Bethe Ansatz for the Exclusion Process}
\label{Sec:Bethe}

 In order to investigate the behaviour of the system which is not stationary,
 the spectrum of the Markov matrix is needed.  For an arbitrary stochastic system, the 
 evolution operator can not be diagonalized. However, the ASEP belongs to a very special
 class of models: it is {\it integrable} and it can be solved using  the Bethe Ansatz
 as first noticed by D. Dhar in 1987.  Indeed, 
 the Markov
 matrix that encodes the stochastic dynamics of the ASEP
 can be rewritten  in terms  of Pauli matrices;   in 
 the absence of a driving field,
the {\it symmetric} exclusion process can be mapped exactly  into the
Heisenberg spin chain.  The asymmetry due to a non-zero external
driving field breaks the left/right symmetry and the ASEP becomes 
equivalent to a non-Hermitian spin chain of the XXZ type
 with boundary terms that preserve the integrable 
 character of the model.  The ASEP can also be mapped
 into a six  vertex model.
 These mappings  suggest the  use of  the Bethe Ansatz
 to derive spectral information about 
  the evolution operator, such as the spectral
 gap \cite{Gwa,ogkmrev}
 and large deviation function. 

  Let us  apply  the  Bethe Ansatz to  the ASEP on a ring.
 A configuration can   be characterized
  by the positions of the $N$ particles
on the ring, $(x_1, x_2, \dots, x_N)$ with $1 \le x_1 < x_2 < \dots < x_N
\le L$. With  this representation, the eigenvalue equation
 (\ref{eq:mpsi=epsi})   becomes
\begin{eqnarray}
  \ca{  E  \, \psi_E(x_1,\dots, x_N) = }  &&  
\ca{  \sum_i{'}   \,   p \left[ 
             \psi_E(x_1, \dots, x_{i-1},\ x_i-1,\ x_{i+1}, \dots, x_N) 
 - \psi_E(x_1,\dots, x_n) \right] + }
    \nonumber \\      
  &&  \ca{  \sum_j{'}   \,   q \left[ \psi_E(x_1, \dots, x_{j-1},\ x_j+1,\ x_{j+1}, \dots, x_N)
              - \psi_E(x_1,\dots, x_N) \right]  }
 \label{Eq:VPASEP}
\end{eqnarray}
where the sum  {\it are restricted}  over   the indexes $i$ such that $ x_{i-1} < x_i-1$
 and   over the indexes $j$ such that $  x_j + 1 < x_{j+1} \,;$   these conditions 
 ensure  that the corresponding jumps  are  allowed.

    We observe that equation~(\ref{Eq:VPASEP}) is akin to  a discrete Laplacian
 on a $N$-dimensional lattice: the major difference is that the  terms corresponding
 to  forbidden  jumps are absent. Nevertheless, this
 suggests that a trial solution (Ansatz in German) in the form of plane waves may be useful.
 This is precisely the idea underlying the Bethe Ansatz. In the following
 we shall give an elementary introduction to this technique which is used in very many different
 areas of theoretical physics. Originally, H. Bethe  developed it to study the Heisenberg
 spin chain  model of quantum magnetism.  The ASEP is the one of the
 simplest systems  to learn the Bethe Ansatz.

\subsection{Bethe Ansatz for ASEP: a crash-course}
\label{SubSec:crashcourse}

Our aim is to solve the linear eigenvalue problem~(\ref{Eq:VPASEP}) which corresponds
 to the relaxation modes of the ASEP  with $N$  particles  on  a ring of $L$ sites. 
 We shall study some special cases with small values of $N$ in order to unveil
 the  general structure of the solution.

\hfill\break
 {\bf The 1 particle case:} 
\vskip 0.3cm 

 For  $N =1$, equation~(\ref{Eq:VPASEP})  reads 
\begin{equation}
 E \psi_E(x) = p  \psi_E(x -1)  +  q  \psi_E(x + 1) - (p+q) \psi_E(x) \, , 
 \label{N=1PASEP}
\end{equation}
 with $ 1 \le x \le L$ and where periodicity is assumed
\begin{equation}
  \psi_E(x + L ) =  \psi_E(x)  \, .
\end{equation}
  Equation~(\ref{N=1PASEP}) is simply a linear recursion of order 2 that is solved as
 \begin{eqnarray}
  \psi_E(x) = A z_{+}^x + B z_{-}^x \,  ,
 \end{eqnarray}
 where $r=z_{\pm}$ are the two  roots of the characteristic equation
\begin{eqnarray}
    q r^2 -(E +p +q) r + p = 0  \, .
 \end{eqnarray}
 The periodicity condition imposes that at least  one of the two characteristic values
 is a $L$-th root of unity (Note that because $z_{+} z_{-} = p/q$ both of them can not
 be roots of unity as soon as $ p \neq q$). The general solution is, finally,
 \begin{eqnarray}
  \psi_E(x) = A z^x  \quad \quad \hbox{ with }  \quad z^L = 1  \,  ,
\label{1bodyEigenP}
 \end{eqnarray}
   The solution is therefore a simple {\it plane wave} with  momentum  
 $2 k \pi /L$ and with   eigenvalue 
\begin{equation}
  E  =  \frac{p}{z} + q z - (p+q)  \, .
\label{1bodyEnergy}
\end{equation}

\hfill\break
 {\bf The 2 particles  case:} 
\vskip 0.3cm

    The case $N=2$ where two particles are present  is more interesting because when
 the particles are located on adjacent sites the exclusion effect  plays a role.
 Indeed, the general eigenvalue equation~(\ref{Eq:VPASEP}) can be split into
 two different cases: 
 
\hfill\break
  $\bullet$  The  Generic case: here  $x_1$ and $x_2$ are separated by at least one empty site:
  \begin{eqnarray}
 E \psi_E(x_1,x_2) = &&  p  \left[ \psi_E(x_1 -1 ,x_2) + \psi_E(x_1,x_2 -1)   \right]
    +  q  \left[ \psi_E(x_1 +1,x_2) + \psi_E(x_1,x_2 +1)   \right]  
 \nonumber \\
&&  - 2(p+q) \psi_E(x_1,x_2) \, .
\label{2-generic}
\end{eqnarray}

\hfill\break
  $\bullet$   The (special) adjacency case:  $x_2 = x_1 +1$, some jumps are forbidden
 and the eigenvalue equation reduces to:
 \begin{equation}
 E \psi_E(x_1,x_1+1) = p   \psi_E(x_1 -1 ,x_1)    +  q  \psi_E(x_1 ,x_1+ 2)   - (p+q) \psi_E(x_1,x_1+1) \, .
\label{2-special}
\end{equation}
  This equation differs from  the generic equation~(\ref{2-generic}) in which
 we substitute  $x_2 = x_1 +1$: {\it there are missing terms}. An equivalent   way to take into account
  the  adjacency case is to impose that the generic equation~(\ref{2-generic}) is valid
 {\it for all values} of  $x_1$ and $x_2$ and  add to it the following  {\it  cancellation boundary 
 condition:}
 \begin{equation}
   p \psi_E(x_1 ,x_1)    +  q  \psi_E(x_1+1 ,x_1+ 1)   - (p+q) \psi_E(x_1,x_1+1) = 0 \, .
\label{2-annulation}
\end{equation}

 We now examine how these equations can be solved. 
 In the generic case particles behave  totally independently ({\it i.e.},  they do not interact).
  The solution of the  generic equation~(\ref{2-generic}) can therefore be written
 as a product of plane  waves  $\psi_E(x_1,x_2) = A z_1^{x_1} z_2^{x_2} $, with the eigenvalue
 \begin{equation}
  E  =  p \left( \frac{1}{z_1} +  \frac{1}{z_2} \right)
  + q \left(  z_1 + z_2 \right)  -2 (p+q)  \, .
\label{2-Energy}
\end{equation}
 However, the simple  product solution can not be the full answer: indeed the
 cancellation condition  for the  adjacency case~(\ref{2-annulation}) has to be satisfied also. 
 The first crucial  observation, following H. Bethe,
  is that  the eigenvalue $E$, 
 given in~(\ref{2-Energy}) is invariant by the permutation $z_1 \leftrightarrow z_2$.  In other
 words, there are two plane waves  $ A z_1^{x_1} z_2^{x_2} $ and  $ B z_2^{x_1} z_1^{x_2} $ 
 with  the same eigenvalue $E$ which has a two-fold degeneracy; the full eigenfunction
 corresponding to $E$ can thus be written as
 \begin{equation}
\psi_E(x_1,x_2) = A_{12} z_1^{x_1} z_2^{x_2} + A_{21} z_2^{x_1} z_1^{x_2} \, ,
\label{2-Bethe}
\end{equation}
 where the amplitudes  $A_{12} $  and   $A_{21} $  are yet arbitrary.
  The second key step is to understand that  these amplitudes  can  now be chosen
  to fulfil  the  adjacency cancellation condition: if we substitute the
 expression~(\ref{2-Bethe}) in equation~(\ref{2-annulation}), we obtain the relation 
\begin{equation}
   \frac{  A_{21} } {  A_{12} } =  
 - \frac{ q z_1  z_2 - (p +q)  z_2  + p}{ q z_1  z_2 - (p +q)  z_1  + p} \, .
\label{2-BetheAmplitudes}
\end{equation}
 The eigenfunction~(\ref{2-Bethe}) is therefore determined, but for
  an overall multiplicative constant.
 We now implement the periodicity condition that takes into account the fact that the system
 is defined on a ring. This constraint can be written as follows for $ 1 \le x_1 < x_2 \le L $
\begin{equation}
  \psi_E ( x_1, x_2) =  \psi_E ( x_2, x_1 + L) \, .
\label{2-Period}
\end{equation}
 This relation plays the role of a quantification condition for the scalars $z_1$ and $z_2$
 that we shall call hereafter the {\it Bethe roots}.  Indeed,  if we impose the condition 
 that   the  expression~(\ref{2-Bethe}) satisfies   equation~(\ref{2-Period}) 
 for all generic   values of the positions  $x_1$ and $ x_2$ we obtain  new relations
 between the amplitudes:
 \begin{equation}
   \frac{  A_{21} } {  A_{12} } = z_2^L = \frac{1}{z_1^L} \, . 
\label{2-BetheAmpliBIS}
\end{equation}
 Comparing equations~(\ref{2-BetheAmplitudes}) and~(\ref{2-BetheAmpliBIS}) leads to a set of
 algebraic equations obeyed by the   Bethe roots $z_1$ and $z_2$:
 \begin{eqnarray}
        z_1^L &=& -  \frac{ q z_1  z_2 - (p +q)  z_1  + p}{ q z_1  z_2 - (p +q)  z_2  + p} \, \\
       z_2^L &=& -   \frac{ q z_1  z_2 - (p +q)  z_2  + p}{ q z_1  z_2 - (p +q)  z_1  + p}
\end{eqnarray}
 These equations are known as the {\it Bethe Ansatz Equations}.
 Finding the spectrum of the Matrix $M$ for two particles on a ring of size $L$ is
 reduced to solving these  two coupled polynomial equations of degree of order $L$
 with unknowns  $z_1$ and $z_2$. Surely, this still remains a very challenging task
 but  the Bethe equations are explicit and very symmetric.
 Besides, we emphasize that the size of the matrix $M$ (and the degree of its
 characteristic polynomial) is of order $L^2$. 

\hfill\break
 {\bf The 3 particles case:} 
\vskip 0.3cm

  We are now ready to  consider the case  $N=3$.  For a system
 containing  three particles, located at $x_1 \le x_2 \le x_3$, the generic equation,
 valid when the particles are well separated, 
 can readily be written using equation~(\ref{Eq:VPASEP}).  But now, the special
 adjacency cases are more complicated:

  (i) Two particles are  next to each other
 and the  third one is  far apart; such a  setting is called  a 2-body collision
 and the boundary condition that results is identical to the one obtained for
 the case $N=2$.  There are now two equations that correspond to the cases
 $x_1=x \le  x_2 = x +1 \ll x_3 $ and  $x_1 \ll  x_2 =x  \le x_3=  x  +1$:
  \begin{eqnarray}
   p \psi_E(x ,x , x_3)    +  q  \psi_E(x+1 ,x + 1, x_3 )   - (p+q) \psi_E(x ,x +1 , x_3 ) &=&  0
\label{3-annulation}  \, .\\
    p \psi_E(x_1, x ,x )    +  q  \psi_E(x_1, x+1 ,x + 1)   - (p+q) \psi_E(x_1,x ,x +1  ) &=&  0 
\label{3-annulationbis}
\end{eqnarray}
 We emphasize again that these equations are identical to equation~(\ref{2-annulation}) 
 because the third particle, located far apart, is simply a {\it spectator} ($x_3$
 is a spectator in the first equation; $x_1$ in the second one).

 (ii) There can be 3-body collisions, in which the three particles are adjacent, with 
 $x_1 =x,  x_2 = x +1,  x_3 = x +2$.
 The resulting  boundary condition is then given by
  \begin{eqnarray}
   p \left\{ \psi_E(x ,x , x+2) +  \psi_E(x , x+1 ,x + 1)   \right\}
   +  q  \left\{   \psi_E(x+1 ,x + 1, x +2 ) +   \psi_E(x , x + 2 ,x + 2)   \right\} \nonumber \\
  - 2 (p+q) \psi_E(x ,x +1 , x +2 )  =  0  \, .
\label{3-bodyCollision}
\end{eqnarray}
   The {\bf fundamental  remark is that 
   3-body collisions  do not lead to  an independent  new  constraint}.
 Indeed, equation~(\ref{3-bodyCollision}) is simply a linear combination of the
constraints~(\ref{3-annulation}) and~(\ref{3-annulationbis})    imposed by 
 the   2-body collisions. To be precise:  equation~(\ref{3-bodyCollision}) is the
 sum of  equation~(\ref{3-annulation}),  with the substitutions
  $x \to x$ and $x_3 \to x+2$,  and of 
   equation~(\ref{3-annulationbis}) with $x_1 \to x$  and   $x \to x+1$. Therefore,
{\it  it is sufficient to fulfil  the  2-body constraints  because  then the 
 3-body conditions are automatically satisfied.}  The fact that 3-body
 collisions  decompose or 'factorize' into  2-body  collisions  is the {\bf crucial property}
 that lies at the very heart of the Bethe Ansatz. If it were not true, the ASEP would
 not be exactly solvable or 'integrable'.

 For $N=3$, the plane wave $\psi_E(x_1,x_2,x_3) =  A  z_1^{x_1} z_2^{x_2} z_3^{x_3}$
 is a solution of the generic equation with  the  eigenvalue 
 \begin{equation}
  E  =  p \left( \frac{1}{z_1} +  \frac{1}{z_2} +  \frac{1}{z_3} \right)
  + q \left(  z_1 + z_2 + z_3 \right)  -3 (p+q)  \, .
\label{3-Energy}
\end{equation}
 However, such a  single  plane wave does not satisfy the boundary conditions
 (\ref{3-annulation}) and~(\ref{3-annulationbis}). Here again we note that the eigenvalue $E$
 is invariant under the permutations of  $z_1, z_2$ and $z_3$. There are 6 such
 permutations, that belong to   $\Sigma_3$  the permutation group of 3 objects.
 The  Bethe wave-function is therefore  written as a sum of the  6  plane waves,
  corresponding to the same eigenvalue $E$, with unknown amplitudes: 
 \begin{eqnarray}
\psi_E(x_1,x_2,x_3) =  &\,\,&    A_{123} \, z_1^{x_1} z_2^{x_2} z_3^{x_3}
   +  A_{132}\, z_1^{x_1} z_3^{x_2} z_2^{x_3}
 +     A_{213} \, z_2^{x_1} z_1^{x_2} z_3^{x_3}
 \nonumber  \\    &\,\,&  +   A_{231}\,  z_2^{x_1}  z_3^{x_2} z_1^{x_3}  +
   A_{312}\,  z_3^{x_1} z_1^{x_2} z_2^{x_3}  + A_{321}\,  z_3^{x_1} z_2^{x_2}  z_1^{x_3} 
 \\
    &=&  \sum_{\sigma \in \Sigma_3} A_{\sigma}  \,
  z_{\sigma(1)}^{x_1} z_{\sigma(2)}^{x_2} z_{\sigma(3)}^{x_3}  \, .
\label{3-Bethe}
\end{eqnarray}
 The 6  amplitudes  $A_{\sigma}$  are uniquely and unambiguously determined
 (up to an overall multiplicative constant) by the  2-body  collision  constraints.
 It is therefore  absolutely crucial that  3-body  collisions 
 do not bring  additional  independent constraints that   the Bethe wave function
 could not satisfy.  We strongly encourage the reader to perform the calculations
 (which are very similar to the $N=2$ case) of the  amplitude-ratios.

 Finally, the Bethe roots $z_1$, $z_2$ and $z_3$ are quantized through 
  the periodicity condition 
\begin{equation}
  \psi_E ( x_1, x_2, x_3) =  \psi_E ( x_2, x_3,  x_1 + L) \, , 
\label{3-Period}
\end{equation}
 for $ 1 \le x_1 < x_2 < x_3 \le  L $.  This condition leads to the Bethe Ansatz
 equations (the equations for general $N$ are given below).

 \hfill\break
 {\bf The general  case:} 
\vskip 0.3cm

   Finally, we briefly  discuss the general case $ N > 3$.  Here one can have
$k$-body collisions with $k=2,3,\ldots N$. However,  all multi-body
 collisions  'factorize' into  2-body  collisions  and ASEP
 can be diagonalized using  the Bethe Wave Function
 \begin{eqnarray}
\psi_E(x_1,x_2, \ldots, x_N) =   \sum_{\sigma \in \Sigma_N} A_{\sigma}  \,
  z_{\sigma(1)}^{x_1} z_{\sigma(2)}^{x_2} \ldots  z_{\sigma(N)}^{x_N}  \, , 
\label{N-Bethe}
\end{eqnarray}
 where  $\Sigma_N$  is the permutation group of $N$ objects.
  The $N!$  amplitudes  $A_{\sigma}$  are fixed 
 (up to an overall multiplicative constant) by the  2-body  collision  constraints.
 The corresponding  eigenvalue is given by
 \begin{equation}
  E  =  p  \sum_{i=1}^N  \frac{1}{z_i} 
  + q   \sum_{i=1}^N z_i   -N (p+q)  \, .
\label{N-Energy}
\end{equation}
The periodicity  condition 
 \begin{equation}
  \psi_E ( x_1, x_2,\ldots, x_N) =  \psi_E ( x_2, x_3,\ldots, x_N,  x_1 + L) \, 
 \quad \hbox{ with } \quad   1 \le x_1 < x_2 <\ldots < x_N \le  L \, ,
\label{N-Period}
\end{equation}
 leads to a set of algebraic equations satisfied by  the Bethe roots 
 $z_1, z_2,\ldots,z_N$.   The  Bethe Ansatz equations are given by 
 \begin{equation}
        z_i^L =  (-1)^{N-1}
 \prod_{j \neq i}  \frac{ q z_i  z_j - (p +q)  z_i  + p}{ q z_i z_j - (p +q)  z_j  + p}   = (-1)^{N-1}
 \prod_{j \neq i}  \frac{ x z_i  z_j - (1 +x)  z_i  + 1}{ x z_i z_j - (1 +x)  z_j  + 1}   \, ,
 \label{EQ:BetheAnsatz}
\end{equation}
 for $i =1, \ldots N$. The last equation is obtained by using
  the asymmetry  parameter $x = q/p$.

 The Bethe Ansatz thus provides
  us with  a set of $N$ coupled algebraic equations of degree
 of order $L$  (Recall that the size of the matrix $M$ is  of order $2^L$,
 when $N \simeq L/2$).  Although the Bethe  equations are  highly non-linear,
 a huge variety of methods have been developed to analyse them.

 We remark that for $p = q = 1$  the Bethe equations are the same as 
 the ones  derived by H. Bethe in 1931. Indeed, the symmetric exclusion process
 is identical to the isotropic Heisenberg  spin chain.

\subsection{Analysis of the Bethe Equations for the TASEP}
\label{SubSec:BATASEP}

\begin{figure}[ht]
\begin{center}
  \includegraphics[height=5.5cm]{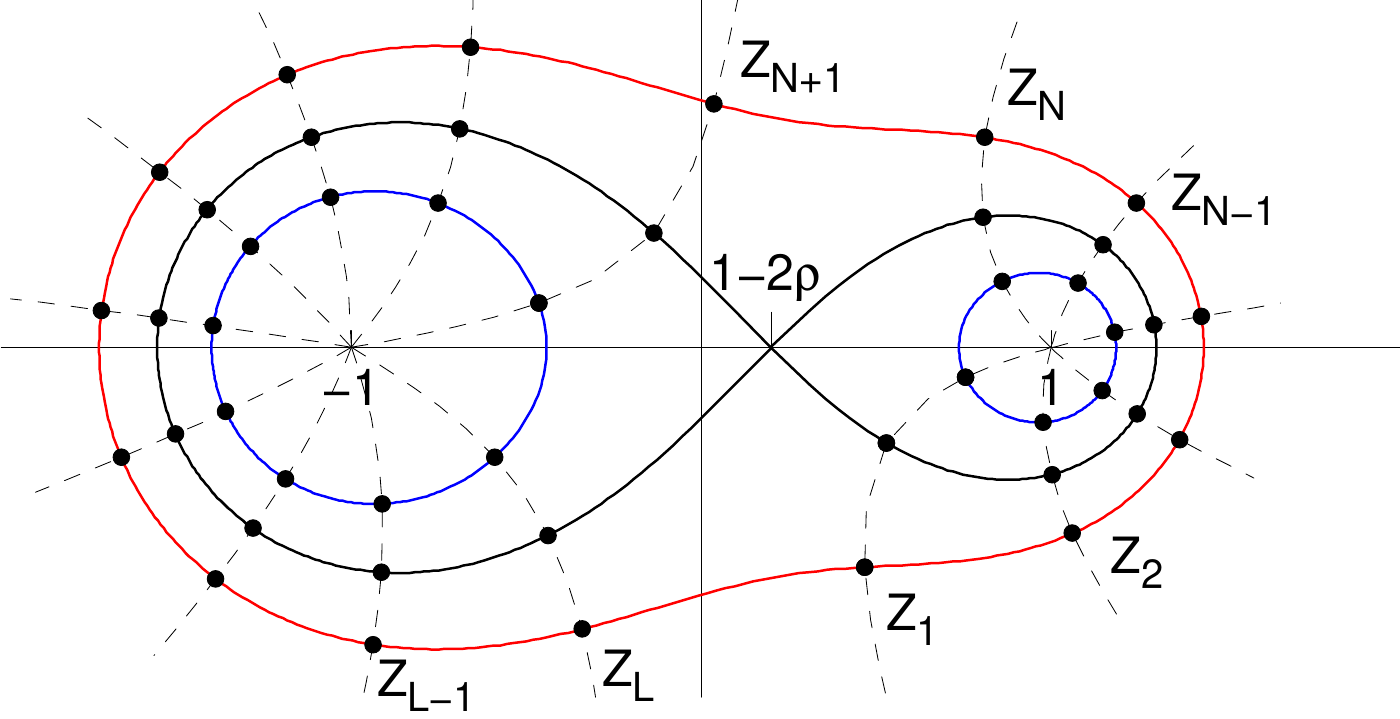}
 \caption{The loci of the roots of the Bethe Equations for TASEP
 are remarkable curves  known as the  Cassini Ovals.}
\label{fig:Ovales}
\end{center}
\end{figure}

 For TASEP, the Bethe equations take a simpler form:
  making the change of variable ${ \zeta_i}  = \frac{2}{z_i} -1$,
 these equations become

   $$  {{\bf   ( 1 -\zeta_i)^N   ( 1 +\zeta_i)^{L-N}  = 
  -2^L \prod_{j=1}^N   \frac{ \zeta_j -1}{ \zeta_j + 1} } }  
  \,\,\,\,  \hbox{ for  } \,\,\,\, {i =   1, \ldots  N  }    $$     
  {\it { Note that   the r.h.s.  is a constant independent of 
   $i$: There is  an effective  DECOUPLING.}}
 The corresponding eigenvalue is 
 $$ { E = \frac{1}{2}(-N + \sum_j \zeta_j) }$$
  For a fixed value of the r.h.s. the roots $\zeta_i$ lie on curves 
  that satisfy 
 $$ \left| 1 - \zeta \right|^\rho  \left| 1 + \zeta \right|^{1-\rho}
 = const $$
where $\rho = N/L$ is the density (see Figure~\ref{fig:Ovales}).

The fact that the Bethe equations can be reduced to an effective single-variable
polynomial suggests the following self-consistent 
procedure for solving them:

\begin{itemize}
\item[$\bullet$] For any given value of $Y$,  {\it SOLVE }  
        $ \,\,\, ( 1 -z_i)^N   ( 1 +z_i)^{L-N}  =  Y \, . $
   The roots are located on    {Cassini Ovals} 
\item[$\bullet$]   {\it CHOOSE   $N$ roots  }
  $z_{c(1)}, \ldots z_{c(N)}$ amongst  the $L$ available roots,
 with  a  {\it choice set} $c : \{ c(1),\ldots , c(N)\} \subset
 \{ 1,\ldots ,L\}  \, . $
\item[$\bullet$] {\it SOLVE }  the  {self-consistent}
 equation ${\bf   A_c(Y) = Y }$  where
          $$A_c(Y) =  -2^L \prod_{j=1}^N  
 \frac{ z_{c(j)} -1}{ z_{c(j)} + 1}  \, .$$
\item[$\bullet$]  {\it DEDUCE  }   from the value of $Y$,   the 
  $z_{c(j)}$'s and the energy corresponding to the choice 
  set  $c$~:
       $$ 2E_c(Y) = -N + \sum_{j=1}^N z_{c(j)} .$$
\end{itemize}

This program can be carried through to calculate 
 the spectral gap  of the Markov matrix $M$, which amounts to calculating 
 $E_1$ the  eigenvalue with largest strictly negative real part.   For a density $\rho = N/L$,
  one obtains for the TASEP
\begin{eqnarray}
    E_1  = 
 &&  { -2 \sqrt{\rho( 1 - \rho)} \frac{6.509189337\ldots}{L^{3/2}} }
      \pm  {\frac{ 2 i \pi (2 \rho -1)}{L}  \, .}  \nonumber \\
          &&   \,\,\,\,\,\,  \,\,\,\,\,\,  \,\,\,\,  {\rm {(RELAXATION)}}  
         \,\,\,\,  \,\,\,\, \,\,\,\, \,\,\,\, \,\,\,\, 
 {\rm  {(OSCILLATIONS)} }      \nonumber 
\end{eqnarray}
   The first excited state  consists
     of a  pair of conjugate  complex numbers when 
   $\rho$  is different from 1/2. The real part of $E_1$ describes the relaxation
    towards the stationary state: we find that   the largest relaxation time  
    scales as  $T \sim L^z$   with the dynamical exponent 
   $z=3/2$  \cite{Dhar,Gwa}.   This value
  agrees with the dynamical exponent of the   one-dimensional 
 Kardar-Parisi-Zhang equation   that belongs to the same universality
 class  as ASEP (see the review of Halpin-Healy and Zhang 1995 \cite{HHZ} and \cite{Sasamoto,Kriecherbauer}
 for recent developments.). The imaginary part of 
  $E_1$  represents the  relaxation oscillations and scales as $L^{-1}$;
 these  oscillations correspond to a kinematic  travelling   wave that propagates with
  the group velocity $2 \rho -1$.
  For the  partially asymmetric case
 ($x \neq 0$),  the Bethe equations do not decouple and analytical
 results are much harder to obtain (see   \cite{ogkmrev} for references).

\section{Large Deviation  far from  Equilibrium}
\label{Sec:LDFNonEq}

   We now return to  our  basic picture  of a conducting  pipe between two
   unbalanced  reservoirs (Figure~\ref{fig:Courant}) and formulate some questions
   involving the concept of large deviations.  Using the ASEP as a paradigm, we shall explain
   how the tools  developed above (Bethe Ansatz, Matrix Representation) can help us to give 
   mathematically precise answers to these problems. 

\begin{figure}[ht]
 \begin{center}
 \includegraphics[height=3.0cm]{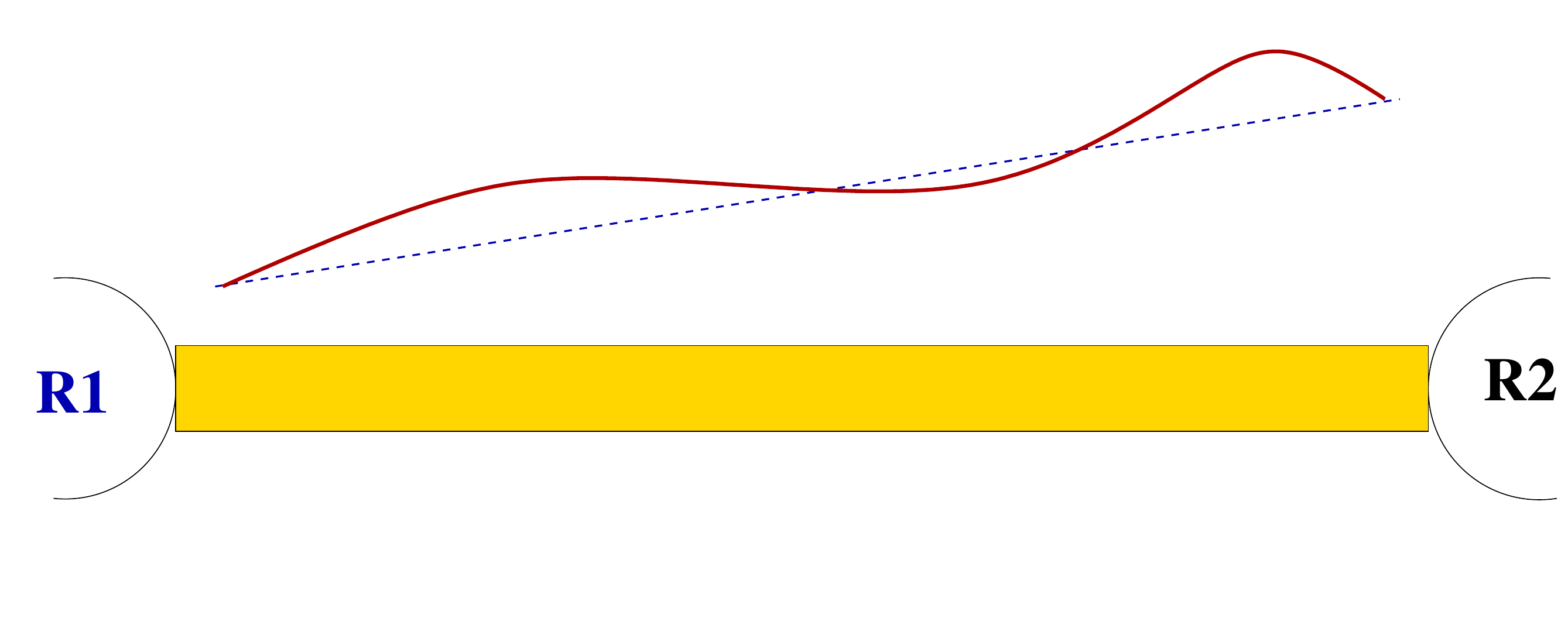}
\caption{Density Profiles in the pipe model: the average profile is linear. However,
non-typical density distributions may occur.}
\label{fig:ProfilAtypique}
 \end{center}
\end{figure}

  1.  In the pipe model, a steady-state is reached with a non-vanishing constant current $J$
   and a stationary density profile $\rho(x)$. Typically, this density profile will
   be linear (as seen  by using Fick's phenomenological law).
  We  have seen above  that for a system in equilibrium (i.e. when the two
   reservoirs are at the same potential) the large deviations of the density profile
   are determined by the free energy. A similar question can be raised in the 
 non-equilibrium stationary state: What is  the  probability of observing an
 atypical density profile in the steady state (see Figure~\ref{fig:ProfilAtypique})? 
More precisely, assuming a large deviation behaviour, 
$$  { {\rm Pr} \{ \rho(x) \} \sim   {\rm e}^{ - \beta L  \,   {\mathcal F}(\{\rho(x)\}) }   } \, , $$ 
 what  does  the functional 
 $ { {\mathcal F}(\{\rho(x)\}) }$ look like in  this  non-equilibrium 
 system?

  For the ASEP, the answer to this question was given by
 B. Derrida, J. Lebowitz  E. Speer in  2002.  These authors calculated 
the probability of observing an
 {atypical density profile}  in the steady state of the ASEP, 
  {starting from the exact microscopic  solution} of the exclusion
 process, with the help of the Matrix Ansatz.  For the symmetric exclusion process
(which corresponds to  a discrete version of Fick's law), the  large deviation functional 
 is given by
$$ { {\mathcal F}(\{\rho(x)\}) = \int_0^1 dx
  \left( B (\rho(x),F(x)) + \log \frac{F'(x)}{\rho_2 -\rho_1}  \right)} 
 $$  
where  $B(u,v) = (1 -u)  \log \frac{1 -u}{1 -v} + u\log \frac{u}{v}$  and
 $F(x)$ satisfies  
$$F \left( F'^2 + (1-F)F''  \right) = F'^2 \rho
 \quad \hbox{ with  } \quad   F(0) = \rho_1   \hbox{ and   } 
  F(1) = \rho_2 \, .$$ 
 It is important to note that this functional  is non-local as soon as 
$ \rho_1  \neq  \rho_2  $ and that it is 
  is NOT identical to the one given by  assuming local equilibrium.

 2. A similar problem can be raised about the current fluctuations.
 Let us  call  $ {Y_{t}}$ the total charge  (or time-integrated
 current)  transported through the system
between  time  0 and  time  ${t}$, then
$$\frac{Y_{t}}{t} \to J  \quad \quad  \hbox{ when} \quad \quad t \to \infty \,  $$ 
where $J$ is the average steady-state current.
 However, the observable $ {Y_{t}}$ is a random variable, that may take non-typical values.
 Its  fluctuations    obey  a  {large deviation principle:}
 $$P\left(\frac{Y_{t}}{t}=j\right) {\sim}e^{-t \Phi(j)}$$ 
 where  $ { \Phi(j) }$ is the large deviation function of the total current. 
 Note that $ { \Phi(j) }$  is positive, vanishes at $j =J$
 and is convex (in general). A natural question is to derive  a  mathematical formula 
$ { \Phi(j) }$. We shall describe below the exact  solution for the ASEP.

\subsection{Current Fluctuations far from equilibrium on a periodic ring}
\label{SubSec:periodic}

\begin{figure}[ht]
 \begin{center}
 \includegraphics[height=4.0cm]{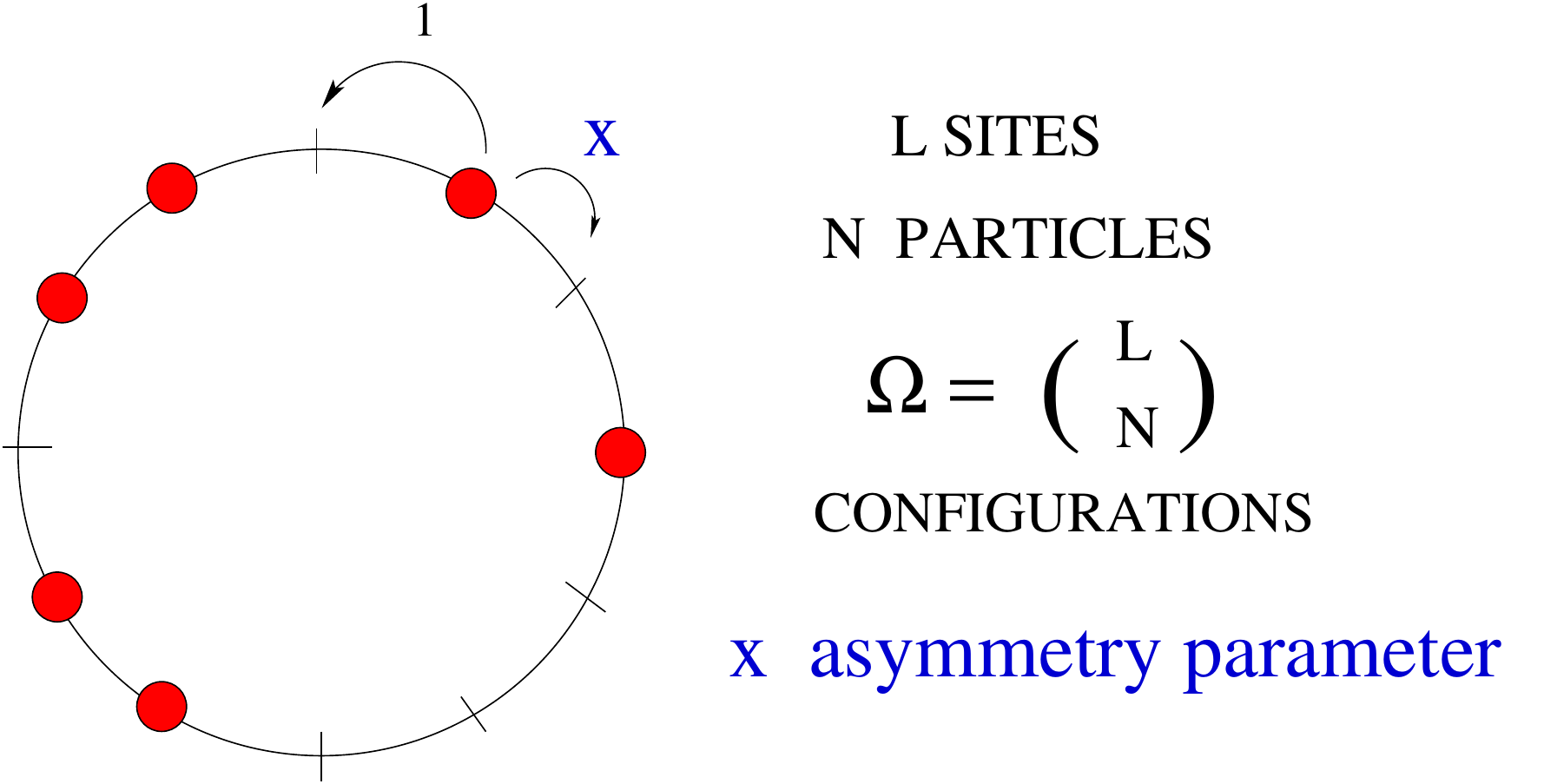}
\caption{The ASEP on a periodic ring. After a suitable rescaling of time, the asymmetry parameter
 is denoted by $x$.}
\label{fig:ASEPR2}
 \end{center}
\end{figure}

 We consider in this section the case of a periodic ring and study the statistics
 of the total displacement of all the particles in the system time 0 and time $t$.
 Because the system is finite, this total displacement is proportional to the time
 integrated current. The  ASEP on a periodic ring is drawn again in Figure~\ref{fig:ASEPR2},
 For convenience, time has been rescaled so that forward jumps occur with rate 1, whereas
 backward jumps occur with rate $x = q/p$.

      Thus, here,    $Y_t$  will represent the total distance
 covered by all the particles between time 0 and time $t$
 and   $P_t(\mathcal{C}, Y)$ is  the joint probability
 of being   in the configuration $\mathcal{C}$ at time $t$ and
 having $Y_t = Y$. The evolution equation of  $P_t(\mathcal{C}, Y)$ is:
\begin{equation}
    \frac{d}{dt} P_t(\mathcal{C}, Y)  = \sum_{\mathcal{C}'} \Big(
   M_0(\mathcal{C},\mathcal{C}') P_t(\mathcal{C}', Y)  
+   M_1(\mathcal{C},\mathcal{C}')  P_t(\mathcal{C}', Y -1)  
+   M_{-1} (\mathcal{C},\mathcal{C}')  P_t(\mathcal{C}', Y +1)  
        \Big)   \, . 
 \label{eq:Markov2}
\end{equation}
 Using  the generating function  $F_t(\mathcal{C})$
 \begin{equation}
  F_t(\mathcal{C}) =  \sum_{ Y =0}^\infty 
  {\rm e}^{\mu Y} P_t(\mathcal{C}, Y) \, ,
 \label{eq:defF}
\end{equation}
  the evolution equation becomes
\begin{equation}
    \frac{d}{dt} F_t(\mathcal{C})  = \sum_{\mathcal{C}'} \Big(
   M_0(\mathcal{C},\mathcal{C}') 
+ {\rm e}^\mu  M_1(\mathcal{C},\mathcal{C}')  
+   {\rm e}^{-\mu}  M_{-1} (\mathcal{C},\mathcal{C}')   
        \Big)  F_t(\mathcal{C}') =  \sum_{\mathcal{C}'} 
  M(\mu)(\mathcal{C},\mathcal{C}')   F_t(\mathcal{C}')
   \, . 
 \label{eq:Markov3}
\end{equation}
This equation is   similar to the original 
 Markov  equation for the probability
 distribution  but now  the original
 Markov matrix $M$ is deformed  by a  jump-counting   {\it  fugacity } ${\mu}$ into 
  $M(\mu)$  (which  is  not a Markov matrix in general),  given by
 \begin{equation}
  M(\mu) =   M_0 +  {\rm e}^\mu  M_1 +  {\rm e}^{-\mu}  M_{-1} \, . 
 \label{eq:defMmu}
\end{equation}
 In the long time limit, $ t \to \infty$, 
  the behaviour of   $F_t(\mathcal{C})$
 is dominated by the largest eigenvalue  $E(\mu)$  and one can write 
  \begin{equation}
  {  \left\langle  {\rm e}^{\mu Y_t}   \right\rangle \simeq 
    {\rm e}^{ E(\mu) t} }  \, . 
 \label{eq:limF}
\end{equation}
 Thus,  in the long time limit, the   function $E(\mu)$  is the generating function of 
 the  cumulants  of the total current $Y_t$.  But   $E(\mu)$  is also the
 dominant  eigenvalue of the   matrix   $M(\mu)$. Therefore, 
  the current statistics has been traded into  an eigenvalue problem. Fortunately,  the deformed matrix 
  $M(\mu)$  can still be diagonalized by the Bethe Ansatz.
 In fact, a  small modification of the calculations described in Section~\ref{Sec:Bethe}
 leads  to the following  Bethe Ansatz equations 
\begin{equation}
  z_i^L  = (-1)^{N-1} \prod_{j =1}^N
   \frac{x {\rm e}^{-\mu}  z_i z_j - (1+x) z_i + {\rm e}^{\mu}  }
  {x{\rm e}^{-\mu}  z_i z_j -(1+x) z_j +{\rm e}^{\mu} }  \, .
\label{eq:BAmu}
\end{equation}
where {the asymmetry parameter ${x= q/p}$} is the ratio of the rates
 of backward to forward jumps.
 The  eigenvalues of $M(\mu)$  are   given by 
\begin{equation}
 E(\mu; z_1, z_2 \ldots  z_N ) =  {\rm e}^{\mu}  \sum_{i =1}^N 
   \frac{1}{ z_i}  + 
        x {\rm e}^{-\mu}  \sum_{i =1}^N   z_i \, - N (1+x)\, .
\label{eq:Emu}
\end{equation}
 The cumulant generating function corresponds to the largest eigenvalue.

\vskip 0.5cm

\subsubsection{The  periodic TASEP Case}

  For the  TASEP,  $x=0$,
 and  the  Bethe equations~\eqref{eq:BAmu}  again  decouple and can be studied
 by  using the  procedure outlined in  Section~\ref{SubSec:BATASEP}. 
  This  case was completely  solved by   B. Derrida and J. L. Lebowitz in
  1998 \cite{DLeb}. 
 These authors  calculated   $E(\mu)$ 
 by Bethe Ansatz to all orders  in $\mu$. More precisely,  they obtained 
 the following  representation
 of the function  $E(\mu)$  in terms  of an auxiliary  parameter $B$: 
\begin{eqnarray}
      E(\mu)  &=&  -N   \sum_{k=1}^\infty 
             \left(\begin{array}{c} kL-1\\kN \end{array} \right)
  \frac{B^k}{kL-1}    \, ,  \label{eq:EofY}\\
           \mu  &=&  - \sum_{k=1}^\infty   
  \left(\begin{array}{c} kL\\kN \end{array} \right)  \frac{B^k}{kL} 
        \label{eq:gamY} \, . 
\end{eqnarray}
These expressions allow to calculate the cumulants of $Y_t$, for example the mean-current $J$
 and the diffusion constant $D$:
\begin{eqnarray}
J =  \lim_{t \to \infty} \frac{ \langle Y_t \rangle}{t} &=&
 \frac{ {\rm d}  E(\mu)}{ {\rm d} \mu} 
 \Big|_{\mu =0} =  \frac{ N(L-N)}{L-1}\, ,    \\ 
 \,\,\, D = \lim_{t \to \infty} 
  \frac{ \langle Y_t^2 \rangle - \langle Y_t \rangle^2 }{t}  &=&
  \frac{ {\rm d}^2  E(\mu)}{ {\rm d} \mu^2} 
 \Big|_{\mu =0} =
 \frac{N^2\; (2L-3)!\; (N-1)!^2 \;(L-N)!^2 }
{ (L-1)!^2 \; (2N-1)! \; (2L-2N-1)! } \, .
 \end{eqnarray}
  When   $ L \to \infty$,
  with a fixed  density $\rho = L/N$ and $ |j - L\rho ( 1 -\rho) |  \ll L$,   the 
  large deviation function $G(j)$  can be written in the 
 following scaling form: 
\begin{equation}
 G(j) =     \sqrt{ \frac{\rho ( 1 -\rho)} {\pi N^3}}
 H \Big( \frac{j -  L\rho ( 1 -\rho)}{\rho ( 1 -\rho)}   \Big) 
 \label{eq:scalf}
  \end{equation}
with 
\begin{eqnarray}
   H(y) \simeq  - \frac{ 2 \sqrt{3}}{5 \sqrt{\pi}} y ^{5/2} \,\,\,\,
 &\hbox{ for }&  \,\,\,\,   y \to +\infty \, , \\
     H(y) \simeq -  \frac{ 4 \sqrt{\pi}}{3} |y| ^{3/2} \,\,\,\,
 &\hbox{ for }&  \,\,\,\,   y \to -\infty \, . 
\end{eqnarray}
 This   large deviation function is not a quadratic polynomial, even in the vicinity
 of the steady state.  Moreover, the shape of this
 function is  skew: it decays as the exponential of a power law
 with an exponent $5/2$ for  $y \to +\infty$ 
 and with an exponent $3/2$ for  $y \to -\infty$.

\subsubsection{The periodic ASEP: Functional Bethe Ansatz}

  In the general case $x\neq0$, the  Bethe Ansatz equations do not
  decouple  and a procedure for solving them was lacking.
  For example, it did not even seem  possible to extract from the
  Bethe equations~\eqref{eq:BAmu} a formula for the mean stationary current
  (which can be obtained very easily by other means from the fact  that the stationary measure
  is uniform).   This problem  was solved  in a few steps  and the
 complete  solution was given by S. Prolhac in 2010 \cite{Sylvain4}

 The  periodic ASEP  can be  analysed   by rewriting  the  Bethe Ansatz as a functional
  equation and  restating it as  a purely algebraic problem, as we shall now explain. 
 First, we perform the following    change of variables, 
\begin{equation}
  y_i =  \frac{ 1 - {\rm e}^{-\mu} z_i} { 1 - x {\rm e}^{-\mu} z_i} \, .
  \end{equation}
  In terms of the variables $y_i$ the Bethe equations read
\begin{equation}
     {\rm e}^{L\mu} \left( \frac{ 1-y_i}{1- x y_i} \right)^L =
      - \prod_{j =1}^N \frac{ y_i -  x y_j }{x y_i - y_j }
 \,\,\,\,   {\rm  for }\,\,\,\,  i =1 \ldots N\, .  
\label{eq:BAyi}
 \end{equation}
  Here again the  equations do not decouple as soon as $x \neq 0$. However,
 these equations are now built from first order monomials in the $y_i$'s
 and they are symmetrical in  these variables. This observation
 suggests to introduce an {\it  auxiliary variable} ${\bf T}$ that plays
 the same role with respect to all the  $y_i$'s and allows to define
 the auxiliary  equation: 
\begin{equation}
{\rm e}^{L\mu} \left( \frac{ 1-{\bf T} }{1- x{\bf T} } \right)^L =
      - \prod_{j =1}^N \frac{ {\bf T} -  x y_j }{x {\bf T}  - y_j }
 \,\,\,\,   {\rm  for }\,\,\,\,  i =1 \ldots N\, . 
\label{eq:BAT}
 \end{equation}
 This equation, in which  ${\bf T}$ is the unknown, and the  $y_i$'s 
 are parameters, can be rewritten as a polynomial equation: 
\begin{equation}
   P(T) = {\rm e}^{L\mu} (1 -   T)^L {\prod_{i =1}^N (x T - y_i)}
 + (1 - x T)^L  { \prod_{i =1}^N  (T -  x y_i)} = 0 \, .
 \end{equation}
  Because  the Bethe equations~\eqref{eq:BAyi} imply that  $P(y_i) = 0$ for
  $i =1 \ldots N\,$,   the polynomial $Q(T)$, defined as
 \begin{equation}
Q(T)=\prod\limits_{i=1}^{N}(T-y_{i}) \, , 
 \end{equation}
 must divide  the polynomial $P(T)$.  Now, if we examine  closely  the expression
 of  $P(T)$, we observe that the factors  that contain the $y_i$'s
 inside the  products over $i$  can be written  in terms of  $Q(T)$.
 Therefore, we conclude that    $Q(T)$  DIVIDES 
  $e^{L\mu}(1-T)^{L}Q(xT)+(1-xT)^{L}x^{N}Q(T/x).$
 Equivalently, there  exists a   polynomial   ${R(T)}$    such that
\begin{equation}
 Q(T)R(T)=e^{L\mu}(1-T)^{L}Q(xT)
  +x^{N}(1-xT)^{L}Q(T/x)  \,.
\label{eq:FBA}
 \end{equation}
 This functional equation is equivalent to the Bethe Ansatz equations
  (it is also known as Baxter's TQ equation). It can be used to determine
   the polynomial   ${Q(T)}$ of degree ${N}$  that  vanishes at the Bethe roots.
 In the present case, equation~\eqref{eq:FBA}
 can be  solved perturbatively  w.r.t. ${\mu}$  to any desired order.
  Knowing   ${Q(T)}$  perturbatively  an expansion of 
   ${E(\mu)}$ is derived, leading to the cumulants  of the current
 and to the  large deviation function. 
\hfill\break
 For example, this method allows to calculate the following cumulants
 of the total current:

    ${\bullet}$  {\it Mean Current $J$:}
    $ J=(1-x)\frac{N(L-N)}{L-1}\sim(1-x)L\rho(1-\rho)  \, \hbox{ for } \,\,
L\rightarrow\infty \, . $  \hfill\break

     ${\bullet}$  {\it Diffusion Constant $D$:} 
$${ D=(1-x)\frac{2L}{L-1}\sum_{k>0}{k^{2}\frac{C_{L}^{N+k}}{C_{L}^{N}}
\frac{C_{L}^{N-k}}{C_{L}^{N}}\left(\frac{1+x^{k}}{1-x^{k}}\right)} } \, . $$
In the limit of a large system size, ${L\rightarrow\infty}$, with  asymmetry
 parameter  ${x\rightarrow 1}$ and  with a  {\it fixed value}  of
 ${ \phi=\frac{(1-x)\sqrt{L\rho(1-\rho)}}{2}}$, the diffusion constant assumes
 a simple expression
$$ {D\sim  4\phi L\rho(1-\rho)\int_{0}^{\infty}du\frac{u^{2}}{\tanh{\phi u}}e^{-u^{2}} }.$$

  ${\bullet}$   {\it  Third cumulant:}  the  Skewness measures the non-Gaussian
 character of the fluctuations.  An exact combinatorial expression of the third
 moment,  valid for any values
 of $L$, $N$ and $x$,   was calculated by S. Prolhac in 2008. It is given by 
\begin{eqnarray}
\frac{E_{3}}{6L^{2}}
&=&\frac{1-x}{L-1}\sum_{i>0}\sum_{j>0}\frac{C_{L}^{N+i}C_{L}^{N-i}C_{L}^{N+j}C_{L}^{N-j}}
{(C_{L}^{N})^{4}}(i^{2}+j^{2})\frac{1+x^{i}}{1-x^{i}}\frac{1+x^{j}}{1-x^{j}} 
    \nonumber \\
&-&\frac{1-x}{L-1}\sum_{i>0}\sum_{j>0}
 \frac{C_{L}^{N+i}C_{L}^{N+j}C_{L}^{N-i-j}}
{(C_{L}^{N})^{3}}
\frac{i^{2}+ij+j^{2}}{2}\frac{1+x^{i}}{1-x^{i}}\frac{1+x^{j}}{1-x^{j}} 
 \nonumber  \\
&-&\frac{1-x}{L-1}\sum_{i>0}\sum_{j>0}\frac{C_{L}^{N-i}C_{L}^{N-j}C_{L}^{N+i+j}}{(C_{L}^{N})^{3}}
\frac{i^{2}+ij+j^{2}}{2}\frac{1+x^{i}}{1-x^{i}}\frac{1+x^{j}}{1-x^{j}}
  \nonumber  \\
&-& \frac{1-x}{L-1}\sum_{i>0}\frac{C_{L}^{N+i}C_{L}^{N-i}}
{(C_{L}^{N})^{2}} \frac{i^{2}}{2}\left(\frac{1+x^{i}}{1-x^{i}}\right)^{2} 
  \nonumber  \\
&+& (1-x)\frac{N(L-N)}{4(L-1)(2L-1)}\frac{C_{2L}^{2N}}{(C_{L}^{N})^{2}} \nonumber  \\
 &-&(1-x)\frac{N(L-N)}{6(L-1)(3L-1)}\frac{C_{3L}^{3N}}{(C_{L}^{N})^{3}}  \, . 
\nonumber \end{eqnarray}
     For ${L\rightarrow\infty}$,   ${x\rightarrow 1}$   and keeping 
 ${ \phi=\frac{(1-x)\sqrt{L\rho(1-\rho)}}{2}}$ fixed, this formula  becomes
\begin{eqnarray}
&& \frac{E_{3}}{ \phi(\rho(1-\rho))^{3/2}L^{5/2}} 
  \simeq  -\frac{4\pi}{3\sqrt{3}} + 
  \nonumber  \\ 
&& 12  \int_{0}^{\infty}dudv
\frac{(u^{2}+v^{2}) e^{-u^{2}-v^{2}}-(u^{2}+uv+v^{2})e^{-u^{2}-uv-v^{2}}}
 {\tanh{\phi u}\tanh{\phi v}}   \, .
\nonumber \end{eqnarray} 
  This shows that the  fluctuations display  a  non-Gaussian behaviour.
 We remark that for   $\phi \to \infty$ the  TASEP limit is recovered:
 $$  E_{3} \simeq 
 \left(  \frac{3}{2} - \frac{8}{3\sqrt{3}} \right)
  \pi (\rho(1-\rho))^{2}L^{3} \, . $$
 
\hfill\break 

\subsubsection{The  weakly  asymmetric limit}

\begin{figure}[ht]
\begin{center}
   \makebox{
     \rotatebox{0} {
\begin{tabular}{ccc}
    \begin{tabular}{c}
      \includegraphics[height=3.75cm,angle =0]{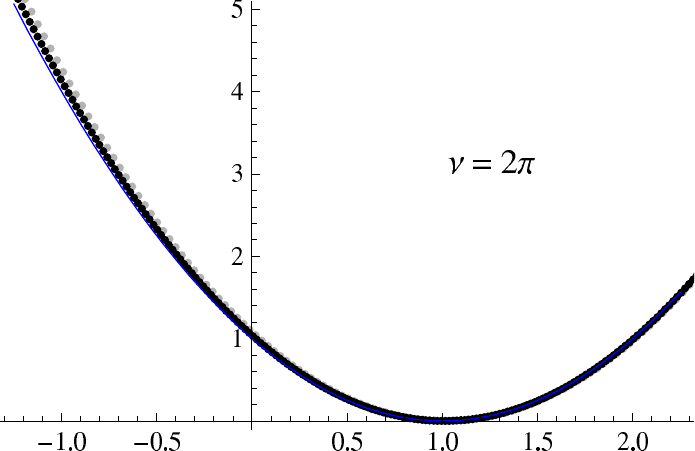} \\
       \includegraphics[height=3.75cm,angle =0]{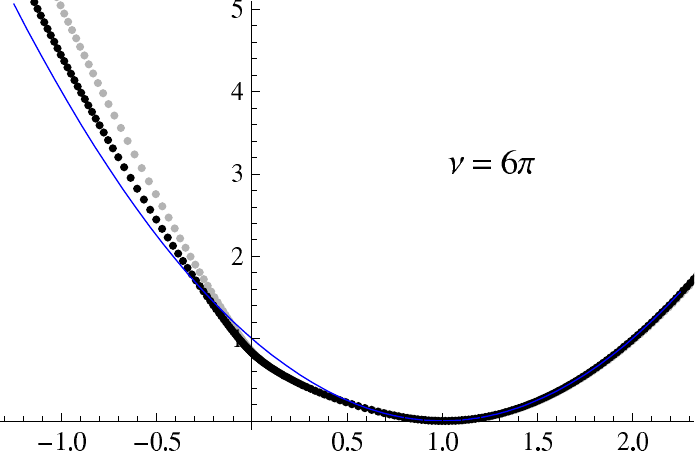} 
    \end{tabular} 
    & \quad  &
    \begin{tabular}{c}
   \includegraphics[height=3.75cm,angle =0]{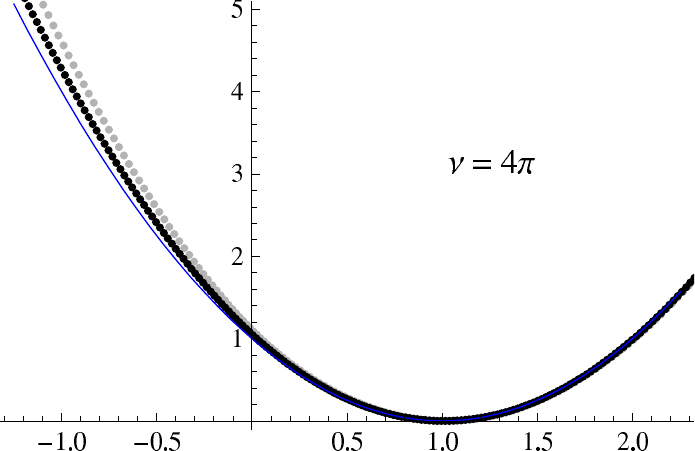} \\
         \includegraphics[height=3.75cm,angle =0]{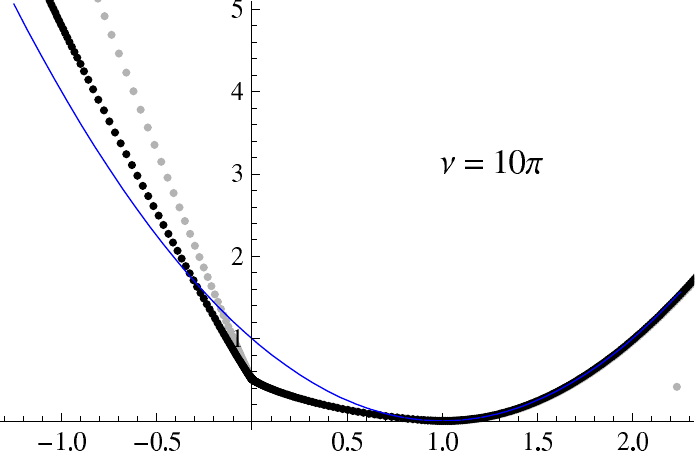} 
    \end{tabular} 
\end{tabular} 
 } } 
\end{center}
\caption{Behaviour of the large deviation function as a function
 of the current  $j/(\nu\rho(1-\rho))$  for different values of $\nu$.
  The grey dots correspond to  $L=50, N=25$ and the  black dots
 correspond to  $L=100, N=50$. They are obtained by solving numerically
 the functional Bethe Ansatz equation \eqref{eq:FBA}. The thin  blue curve
 represents the leading Gaussian behaviour~\eqref{LDFGaussian}.}
 \label{Fig:LDF}
 \end{figure}

 In this section, we make  some
 remarks specific to the  weakly  asymmetric case, for which the
 asymmetry parameter scales as  ${x=1-\frac{\nu}{L}}$ 
 in the  limit of  large system sizes  ${L\rightarrow\infty}$. 
 In this  case, we also need to rescale the fugacity parameter
 as ${\mu}/{L}$ and 
 the  following  asymptotic formula for   the cumulant generating function
 can be derived 
 \begin{eqnarray}
  \tilde{E}(\mu,\nu) \equiv
   E\left(\frac{\mu}{L},  1- \frac{\nu}{L} \right)
 &\simeq&  \frac{\rho(1-\rho) (\mu^{2} + \mu\nu)}{L}
     -\frac{\rho(1-\rho)\mu^{2}\nu}{2L^{2}}
  + \frac{1}{L^{2}} \phi [ \rho(1-\rho)(\mu^{2} + \mu\nu)] \, ,
 \label{eq:series1WASEP} \\
       \hbox{ with }   \,\,\,\,\,\,
   \phi(z) &=& \sum_{k=1}^{\infty} \frac{B_{2k-2}}{k!(k-1)!}z^k  \, ,
 \label{eq:series2WASEP}
 \end{eqnarray}
 and  where the ${B_{j}}$'s  are  Bernoulli Numbers. We observe
 that the leading order (in  ${1/L}$) is quadratic in $\mu$
 and describes Gaussian  fluctuations. It is only in the subleading
 correction (in $1/L^{2}$) that the  non-Gaussian character arises.
   We observe that the series
 that defines the function  $\phi(z)$ has a finite radius of convergence
 and that  $\phi(z)$ has a singularity for $z = -\pi^2$.  Thus,
 non-analyticities appear in $\tilde{E}(\mu,\nu)$ as soon as 
   $$ \nu \ge \nu_c = \frac{2\pi }{\sqrt{\rho(1-\rho)}} \, . $$
  By  Legendre transform,  non-analyticities  also occur 
 in the large deviation function $G(j)$.
 At half-filling, the singularity appears at $\nu_c = {4\pi }$ as can be 
 seen in   Figure~\ref{Fig:LDF}.  For  $\nu < \nu_c$ the leading
 behaviour of  $G(j)$ is quadratic (corresponding
 to Gaussian fluctuations)  and is given by
 \begin{equation}
 G(j) = \frac{( j -\nu \rho(1-\rho))^2}{ 4 L \rho(1-\rho) } \, .
\label{LDFGaussian}
 \end{equation}
 For  $\nu > \nu_c$, the series expansions~\eqref{eq:series1WASEP}
 and  ~\eqref{eq:series2WASEP}  break  down  and 
 the  large deviation function $G(j)$  becomes
 non-quadratic even at leading order. This phase transition
 was predicted by  T.  Bodineau and B. Derrida using
 macroscopic fluctuation theory (see  \cite{Bodineau} for a general discussion).
 One can observe in  Figure~\ref{Fig:LDF} that for $\nu \ge  \nu_c$,
 the   large deviation function  $G(j)$  becomes non-quadratic and develops
 a kink at a  special value of the total current $j$. 

\subsubsection{The general structure of the solution}

 A systematic expansion procedure that completely   solves the problem to all orders 
 and  yields exact expressions for all
 the cumulants of the current,   for an arbitrary value of the asymmetry parameter $x$,
 was carried out  by S. Prolhac in \cite{Sylvain4}.

 Using  the functional Bethe Ansatz,
 S. Prolhac  derived  a  parametric representation
 of the cumulant generating function $E(\mu)$ similar to the
 one given  for the TASEP in equations~\eqref{eq:EofY} and~\eqref{eq:gamY}, 
\begin{eqnarray}
            \mu  &=&  - \sum_{k \ge 1}  C_k \frac{B^k}{k}   \nonumber \\ 
              E  &=&  -(1 -x)\sum_{k \ge 1}   D_k \frac{B^k}{k}   
\end{eqnarray}
 where   $C_k$ and $D_k$ can be expressed as residues of a complex
 function $\phi_k(z)$: 
$$  C_k   =    \oint_{\mathcal C} \frac{dz}{2\, i\, \pi} \frac{\phi_k(z)}{z}  
  \,\,\,  \,\,\, \hbox{  and }  \,\,\,   \,\,\, 
   D_k   =  \oint_{\mathcal C} \frac{dz}{2\, i\, \pi}
 \frac{\phi_k(z)}{(z+1)^2} \, .$$
 It can be shown that  $C_k$ and $D_k$  are combinatorial factors
  that are  enumerate 
 to some tree structures.  To find these numbers, we need the functions 
 $\phi_k(z)$. These functions can be embodied into a generating function $W_B(z)$ 
 defined as 
 \begin{equation}
W_B(z) =  \sum_{k \ge 1}  \phi_k(z)  \frac{B^k}{k}
\end{equation}
 This  function  $W_B(z)$ encodes   the full  information about the 
 statistics of the current: if we know how to determine  it, the problem is solved.

 It can be proved that   $W_B(z)$ is the solution of the following  functional
 Bethe  equation:
\begin{equation}
\textcolor{red}{\fbox{ $ \ca{ W_B(z)=- \ln\Bigl(1-B F(z) e^{X[W_B](z)}\Bigr) }
 \,\, \quad  \hbox{ with }\quad \ca{ \,\, F(z)= \frac{(1 +z)^L}{z^N}\,\, }   $}}
\label{Eq:FonctionnelleW}
\end{equation}
 \begin{equation}
 \hbox{ and } \quad  X[W_B](z_1)=\oint_{\mathcal C}
\frac{d{z_2}}{\imath2\pi \, {z_2}} W_B({z_2}) K(z_1, z_2) 
\label{def:X}
\end{equation}
 where   the kernel    $K(z_1, z_2)$  is given  by 
 \begin{equation}
\textcolor{red}{\fbox{ $  \ca{
 K(z_1, {z_2}) =
 2\sum_{k=1}^{\infty}\frac{x^k}{1-x^k} \, \left\{ \left(\frac{z_1}{z_2}\right)^k
 + \left(\frac{z_2}{z_1}\right)^{k}\right\} } $}} 
\label{eq:NoyauK}
\end{equation}
 Equations~(\ref{Eq:FonctionnelleW}, \ref{def:X} and \ref{eq:NoyauK}) determine
 the unknown function  $W_B(z)$. They are  the core of the problem's solution.
 Using them, a  closed expansion of  $E(\mu)$  w.r.t. $\mu$ was
  derived. This  expansion, valid for any finite values of $L$, $N$
 and $x$, can be used to study the large system size limit with various scalings
 of the asymmetry. Various  regimes were found and the corresponding expressions for the cumulants
  were fully worked out: 
\begin{itemize}
 \item  For  $ 1 - x \ll \frac{1}{L}$,  the model falls into the Edward-Wilkinson universality class.
 \item   The  range $ 1 - x  \sim  \frac{\nu}{L}$, where $\nu$ is a finite number, defines
 the weakly asymmetric regime (to be discussed below). 
  \item The intermediate  regime,  
 corresponding  to  $\frac{1}{L} \ll   1 - x  \ll   \frac{1}{\sqrt{L}}$,  exhibits 
 a specific scaling behaviour that,  to our knowledge, can not be represented by a continuous
 stochastic equation.
  \item  For
  $ 1 - x \sim  \frac{\Phi}{\sqrt{\rho(1-\rho)L}}$ the system is in  the strongly asymmetric regime.
 \item  The range  $ 1 - x \gg  \frac{1}{\sqrt{L}}$ corresponds to the KPZ universality class,
 which contains  the TASEP. 
\end{itemize}

\hfill\break

\subsection{Open System with reservoirs}
\label{SubSec:openASEP}

 Finally, we can explain how  to calculate the large deviation of the current
 in the open ASEP in contact with two reservoirs \cite{Gorissen}. Now, the
  observable 
 $Y_t$  counts the total number of  particles  exchanged between
 the system and  the left  reservoir between times  $0$ and $t$. During a time interval $dt$, 
 $Y_t$ increases by 1 if a particle enters at site 1 (at rate $\alpha$),
 it decreases by if  a particle  exits  from 1 (at rate
 $\gamma$) and is unchanged if no  particle exchange with  the left  reservoir
 has occurred during  $dt$.

 We know that $\frac{\langle Y_t \rangle} {t}$ converges towards the 
average current  $J(x, \alpha,\beta, \gamma, \delta, L)$  when $t \to \infty$. We wish to determine
 the large deviation function associated with $Y_t$ or, equivalently, its 
cumulant generating function, that we recall is defined by 
 $$    \left\langle  {\rm e}^{\mu Y_t}   \right\rangle  \simeq  
   {\rm e}^{ E(\mu) t}   \quad    \hbox{ for } \quad  t \to  \infty $$

 The function $ E(\mu) $ can be calculated by the Matrix Ansatz method.
  The solution
 obtained has structure very similar to the one obtained in 
 Equations~(\ref{Eq:FonctionnelleW}, \ref{def:X} and \ref{eq:NoyauK})
 for the periodic case. For arbitrary values of ${x}$
 and   ${ (\alpha, \beta, \gamma,\delta)}$,
 and for any system size  $ L $  the parametric representation
 of ${E(\mu)}$ is given by
 \begin{eqnarray}
   \quad  \quad   \quad  \quad  \quad  \quad   \quad  \quad 
 \ca{  \mu }  &\ca{=}& \ca{ - \sum_{k=1}^\infty 
  C_k(x; \alpha,\beta, \gamma,\delta,L)
 \frac{B^k}{2 k} \,\, }   \nonumber \\
\ca{   E } &\ca{=}& \ca{ - \sum_{k=1}^\infty 
   D_k(x; \alpha,\beta, \gamma,\delta,L)
  \frac{B^k}{2 k}   \,\,}     \nonumber
 \end{eqnarray} 
 The  coefficients  $C_k$  and   $D_k$  are given   by contour integrals
 in the complex plane:
$$   { C_k   = 
   \oint_{\mathcal C} \frac{dz}{2\, i\, \pi} \frac{\phi_k(z)}{z}  } 
  \,\,\,  \,\,\, \hbox{  and }  \,\,\,   \,\,\, 
  {   D_k   =  \oint_{\mathcal C} \frac{dz}{2\, i\, \pi}
 \frac{\phi_k(z)}{(z+1)^2}   }  $$ 
 where the complex contour  ${{\mathcal C} }$
encircles   0,
 $x^k a_{+},x^k a_{-},x^k b_{+}$, $x^k b_{-}$   for $k \ge 0$ [$a_{\pm},b_{\pm}$ were defined
 in equation~(\ref{def:abpm})].
 Here again, there  exists an auxiliary  function
 ${  W_B(z) =  \sum_{k \ge 1}  \phi_k(z)  \frac{B^k}{k}} $
  that  contains  the full  information about the 
 statistics of the current. The functional equation for $ W_B(z)$ is given by
\begin{equation}
\textcolor{red}{\fbox{ $ \ca{ W_B(z)=-  2 \ln\Bigl(1-B F(z) e^{X[W_B](z)}\Bigr) }  $}}
\label{Eq:FonctionnelleWOPEN}
\end{equation}
 $X[W_B]$ being  the same operator defined  in  equations~(\ref{def:X})  and 
 (\ref{eq:NoyauK}). 
  But the  { function $\ca{F(z)}$} is   now  given by
  $$\textcolor{red}{\fbox{ $
\ca{F(z) = \frac{(1+z)^L(1+z^{-1})^L(z^2)_{\infty}(z^{-2})_{\infty}}{(a_{+}z)_{\infty}(a_{+}z^{-1})_{\infty}(a_{-}z)_{\infty}(a_{-}z^{-1})_{\infty}(b_{+}z)_{\infty}(b_{+}z^{-1})_{\infty}(b_{-}z)_{\infty}(b_{-}z^{-1})_{\infty}}    } 
\,\, $}}$$
where  we use the notation  $(A)_{\infty}=\prod_{k=0}^{\infty}(1-x^k A)$.
 This function,  related to Askey-Wilson polynomials, was used by T. Sasamoto in his study
 of the stationary state of the ASEP with open boundaries (see \cite{MartinRev2} for references).

\vskip 0.3cm

If we specialize these formulae to the special TASEP case $x = \gamma = \delta =0$
 with  non-zero rates equal to  unity $\alpha = \beta =1$, the expression becomes
 rather explicit. The   parametric representation of the
  cumulant generating function  $E(\mu)$ is given by 
\begin{eqnarray}
  \mu &=& - \sum_{k=1}^\infty
 \frac{(2k)!}{k!} \frac{[2k(L+1)]!}{[k(L+1)]! \, [k(L+2)]!}
 \frac{B^k}{2 k}   \,\, , \nonumber  \\
  E  &=& - \sum_{k=1}^\infty
 \frac{(2k)!}{k!} \frac{[2k(L+1)-2]!}{[k(L+1)-1]! \, [k(L+2)-1]!}
 \frac{B^k}{2 k}   \,\,  . \nonumber 
 \end{eqnarray} 
 By eliminating recursively $B$ in terms of $\mu$ in second equation and substituting
 in the first one, we can calculate the first few  cumulants  of the current:
\begin{itemize}
  \item  {Mean Value  :} $ J  = \frac{L+2}{2(2L+1)} $ 

  \item {Variance :} $ \Delta  = \frac{3}{2}
 \frac{(4L+1)! [ L! (L+2)!]^2 }{[(2L+1)!]^3 (2L+3)! }$ 

  \item {Skewness :}
  $ E_3 = 12 \,  \frac{ [ (L+1)!]^2 [(L+2)!]^4}{(2L+1) [(2L+2)!]^3 }\,
 \Big\{ 9 \frac{(L+1)!(L+2)! (4L+2)!(4L+4)!}{(2L+1)! [(2L+2)!]^2  [(2L+4)!]^2}
- 20 \frac{ (6L+4)!}{(3L+2)!(3L+6)!}  \Big\}  $ 

 For large systems:
 $ { E_3 \to  \frac{2187 - 1280 \sqrt{3}}{10368}
 \, \pi \sim -0.0090978...}$
\end{itemize}

\vskip 0.3cm
More generally, in the limit of large system sizes,
 the asymptotic behaviour of   $E(\mu)$ can be carried out, for the general ASEP,
 in the different phases.  In the Low Density (and High Density)  Phases, the
 large deviation function, obtained after a Legendre transform, takes a
 particularly simple form: 
 $$ \textcolor{red}{\fbox{ $ \,\, \ca{   \Phi(j) =(1 -q) \left\{ \rho_a - r  + 
  r ( 1 - r) \ln \left( \frac{1 - \rho_a}{\rho_a}
  \frac{r}{1-r} \right) \right\} }  \,\, $}}$$ 
 where the current $j$ is parametrized as  $\ca{  j = (1 -q) r ( 1 - r) .}$

 The exact expressions can also be compared with numerical results obtained
 by the Density Matrix Renormalization Group technique. 
\begin{figure}[ht]
 \begin{center}
\begin{tabular}{cc}
       \includegraphics[height=5cm,angle =0]{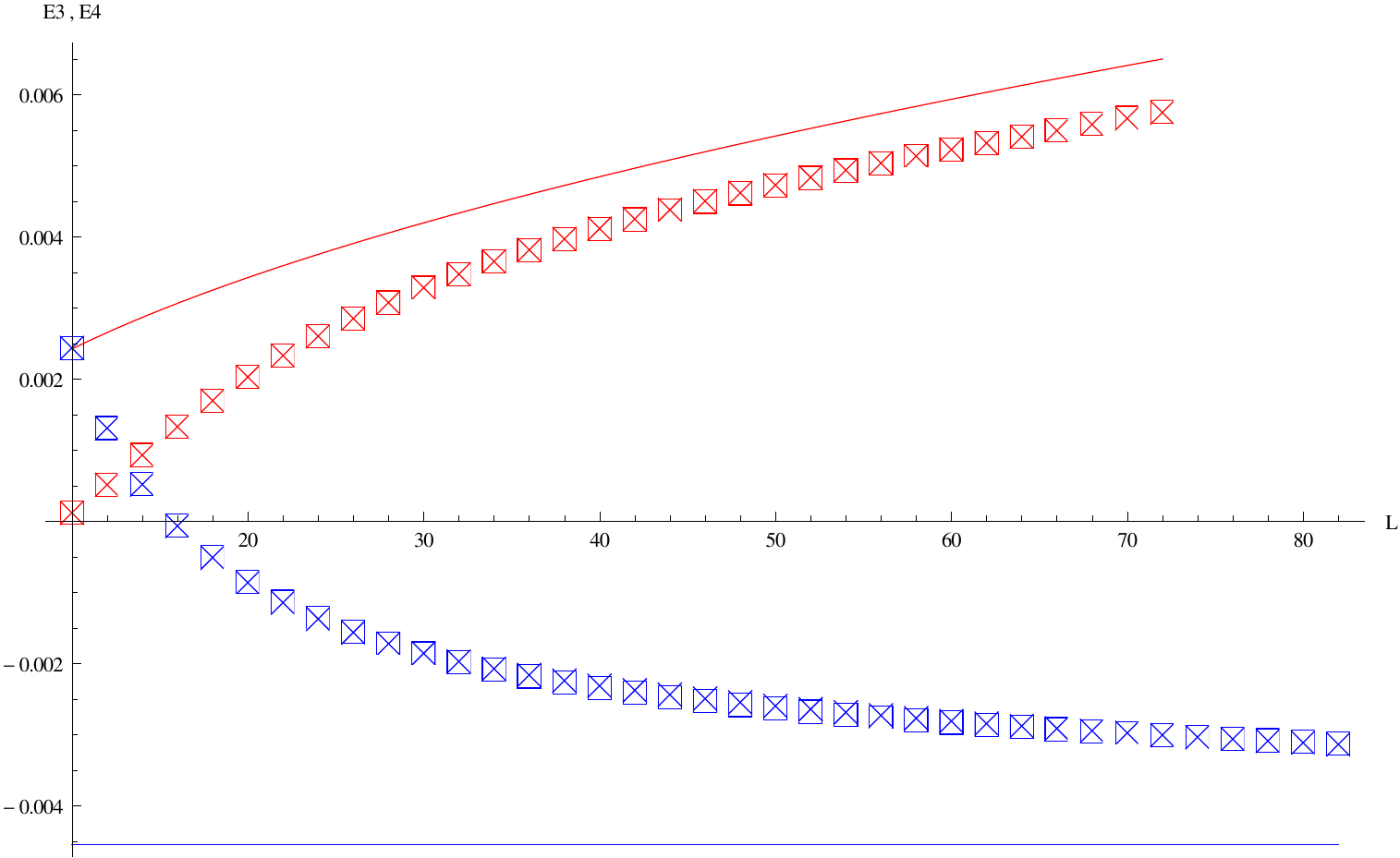}   \quad   &   
   \quad       \includegraphics[height=5cm,angle =0]{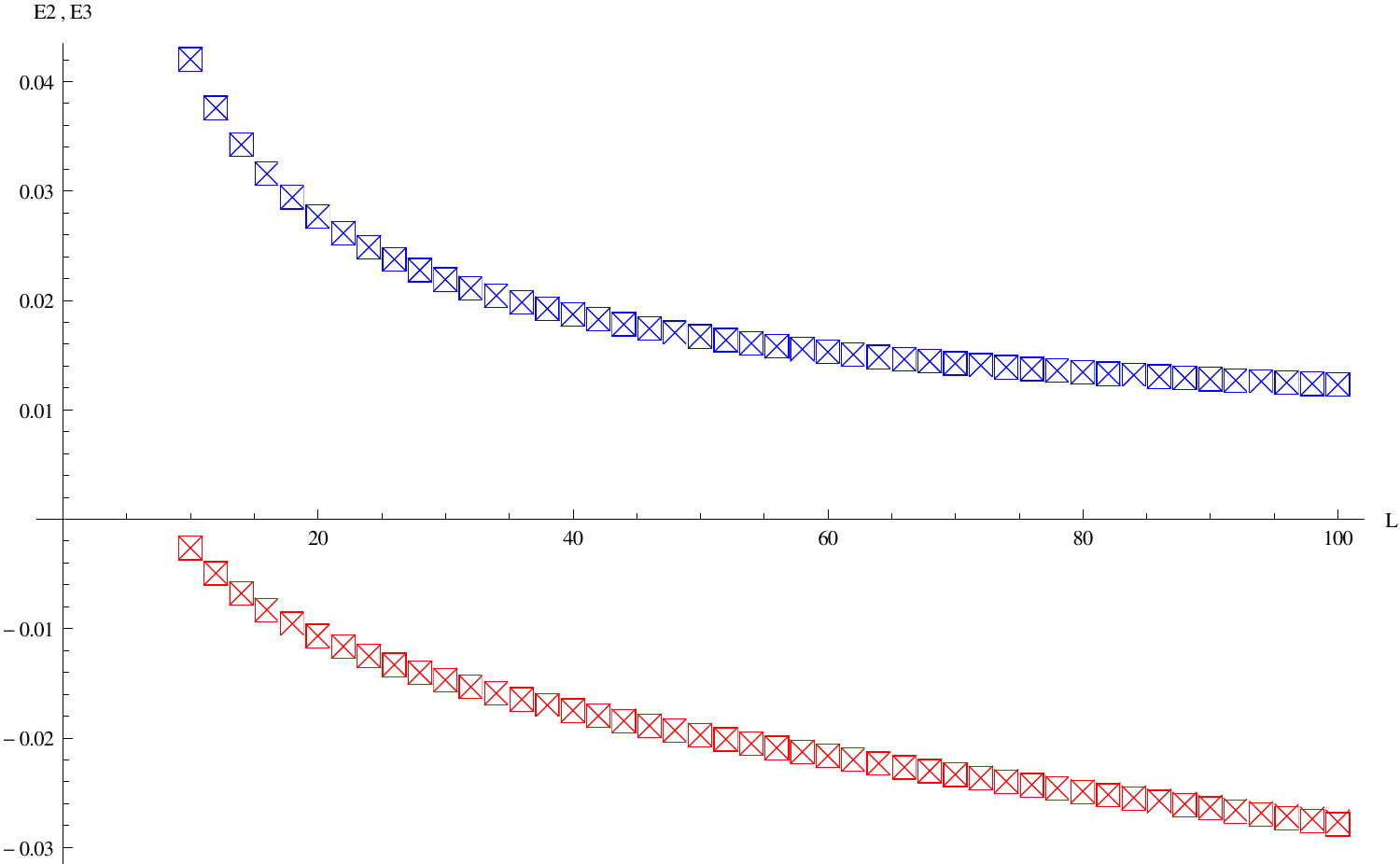}
\end{tabular} 
 \end{center}
 \caption{ {\it Left:}  Third  and Fourth  cumulants  plotted against
 the system size in the Maximal Current phase with 
$x=0.5$, $a_+=b_+=0.65$, $a_-=b_-=0.6$.  {\it Right:} 
   {Second}  and {Third} cumulants in  the  High Density phase
 with   $x=0.5$, $a_+=0.28$, $b_+=1.15$, $a_-=-0.48$ and $b_-=-0.27$.
 The crosses and the squares compare numerical with analytical results. The continuous curves
 correspond to asymptotic behaviours \cite{Gorissen}. }
\end{figure}

\section{Concluding remarks: Towards a fluctuating Hydrodynamics description}
\label{Sec:MFT}

 In these lectures, we have explained that non-equilibrium processes
 can be explored by studying  the large deviations of some physical observable
 such as the total current transported through a wire or the density profile.
 Indeed, large deviation functions  appear to be to right generalization
 of the thermodynamic potentials to  non-equilibrium systems.
  They exhibit  remarkable properties such as the Fluctuation Theorem
 that is valid far away from equilibrium. Large deviation functions 
 are likely to play a key-role in the future of  non-equilibrium
 statistical mechanics.

 The asymmetric exclusion process is a paradigm for  non-equilibrium
 behaviour in low dimensions. Just as the Ising model for phase transitions,
 the ASEP is the simplest non-trivial model that embodies the minimal
 ingredients to study mathematically statistical physics far from equilibrium.

 For the specific case of the ASEP, we have shown how  large
  deviation function can be calculated exactly by using techniques
 borrowed from the theory of integrable systems, such as the Bethe Ansatz
 or the Matrix Representation method. The results obtained are exact
 and mathematically appealing, but the calculations required are rather involved
 and can only be applied to very specific models. It  would be highly desirable
 to have at our disposal  a  more physical picture and a a set of tools
 more versatile that would allow us to study very general non-equilibrium
 systems.

  It seems that such a theory has been  emerging in the recent years.
This approach is  based upon a fluctuating hydrodynamic description.
 As a conclusion
 of these lectures, we shall briefly describe this theory  and its relation to 
 the problems studied here. 

 We consider  again a diffusive system in contact with two reservoirs
 and we adopt the  hydrodynamic description of Figure~\ref{fig:Courant2}
 consider  the case with no applied field $(\nu =0).$  On average,
 the  evolution of the profile and the current is  governed
 by Burgers equation~(\ref{eq:Burgers}).  But,  what if 
 we are interested in stochastic properties of 
the profile and the current? 
 Suppose we want to know  the probability to observe an 
{\it atypical} current  $j(x, t)$
 and the corresponding density profile $\rho(x, t)$  
 during $ 0 \le s \le L^2 \, T$? 
 For many systems, this  probability will follow  a large deviation
 behaviour:
$$ {\rm Pr} \{j(x, t),  \rho(x, t) \} \sim
 {\rm e}^{ - L \, {\mathcal I}(j,\rho)}   $$ 
 (This is highly non-trivial. We simply assume it here). The
{\it  functional}  ${\mathcal I}(j,\rho)$
 is the large deviation functional. 

For driven diffusive systems, G. Jona-Lasinio and coworkers have 
 developed a formalism to calculate  ${\mathcal I}(j,\rho)$,
 known as the Macroscopic Fluctuation Theory (MFT)
(see  \cite{Bodineau, DerrReview, Bertini} for  reviews). 
The goal is to replace the
 deterministic Burgers equation by a stochastic  equation  that describes
 correctly the fluctuations of the system in the diffusive scaling limit
 of large systems and long times. We shall formulate the MFT 
 using  once again the pipe  model language:
 Consider $Y_t$ the total number of particles transferred from the
 left reservoir to the right  reservoir during time $t$.
 Then one has
 \begin{itemize}
    \item  $\lim_{t \to \infty} \frac{ \langle Y_t \rangle}{t} = \ca{D(\rho)}
 \frac{\rho_1 - \rho_2}{L} 
 + \ca{ \sigma(\rho)}  \frac{\nu }{L} \quad$ for $\quad(\rho_1 - \rho_2)$ small
      \item $ \lim_{t \to \infty} \frac{ \langle Y_t^2\rangle}{t}  =
\displaystyle{ \frac{  \ca{ \sigma(\rho)} }{L} } \quad $ for $\rho_1 = \rho_2 =\rho$ 
 and $\nu = 0$.
\end{itemize}
 where the two   `phenomenological' coefficients encode
 the value of the average current and its  quadratic fluctuations.
For the symmetric exclusion process, they are given by 
$$ D(\rho) = 1  \quad   \hbox{ and   }   \quad
  \sigma(\rho ) =   2 \rho (1 -\rho )   \, .$$

 The stochastic  equation of motion is obtained as  
 $$ { \partial_t \rho =  -  \partial_x j } \quad   \hbox{ with  }  \quad  
 { j}  \cc{= -}  \ca{D(\rho)} { \nabla  \rho  +  \nu}  \ca{ \sigma(\rho)}
   {+ \sqrt{\sigma(\rho)} \xi(x,t) }$$
  where $\xi(x,t)$ is a Gaussian white noise with variance
 $$ {\langle \xi(x',t') \xi(x,t) \rangle = \ca{\frac{1}{L}} \delta(x -x')
 \delta(t -t') }$$
If we discard the noise term, we recover the 
Burgers equation~(\ref{eq:Burgers}). The important fact is that
 in the Macroscopic limit, fluctuations are accurately described by
 a {\it multiplicative} Gaussian white noise, the amplitude of this noise
 being proportional  to  the conductivity $ \sigma$. Besides, the fact that
 this noise is vanishing small (its amplitude has a $1/L$ factor) will
 allow us to use saddle point/WKB methods.

Indeed, the equation of  fluctuating hydrodynamics allows us to express
 the  large-deviation functional as a path-integral. Since 
the current and the density evolve $(\rho(x,t), j(x,t))$
 according to a stochastic dynamics, 
 the  weight of a trajectory  between 0 and $t$ can be  written as
\begin{eqnarray}
 && \hbox{Proba}\left(\rho(x,t),j(x,t)| \rho_0(x), j_0(x)\right)
=  \nonumber \\&&  \int\limits_{\substack{ \rho_0 \to \rho_t \\ j_0 \to \j_t}} 
{\mathcal D}\rho {\mathcal D}j
\prod\limits_{\substack{0 \le x \le 1 \\ 0 \le t' \le t}} \delta\left( 
 \frac{\partial \rho}{\partial t'}+ \frac{\partial j}{\partial x}  \right)
  {\exp\left(  -\frac{L}{2} \int_0^t  dt' \int_0^1 dx  
\frac{( j + D(\rho)  \frac{\partial \rho}{\partial x}  - \nu \sigma(\rho))^2 }{\sigma(\rho)}
 \right) } \,  \nonumber
\end{eqnarray} 
 In  the large ${L}$ limit, 
 the integral will be dominated
 by the optimal value of the exponent (saddle-point).
 Hence, the large deviation functional can be written as the solution
 of an optimal path problem: 
\begin{equation}
   \ca{  {\mathcal I}(j,\rho) = \min_{\rho,j} \Big\{  \int_0^T dt   \int_0^1 dx
    \frac{ \left(j - \nu  \sigma(\rho ) +   D(\rho) \nabla\rho  \right)^2}
      {2 \sigma(\rho )}   \Big\} }     \label{LDF:Optimal}
\end{equation}
 with the {constraint:}  $ { \partial_t \rho = -\nabla.j} \,. \, $  
 Then, knowing  ${\mathcal I}(j,\rho)$ one can deduce 
 (by the contraction principle  \cite{Touchette})
 the LDF of the current or the profile. For example, for the current:
 ${\Phi(j) = \min_\rho\{ {\mathcal I}(j,\rho) \} } \, .$

 This variational problem~(\ref{LDF:Optimal}) has a Hamiltonian structure and can be expressed
 by using a pair of conjugate variables $(p,q)$.
Mathematically, one has to solve  the corresponding Euler-Lagrange
equations. After some transformations,  a set of coupled non-linear 
 PDEs is obtained:
\begin{eqnarray}
\ca{\partial_t q } &\ca{=}&   \ca{\partial_x [D(q) \partial_x q] - \partial_x [\sigma(q)  \partial_x p]}
 \nonumber \\
 \ca{\partial_t p }&\ca{=}& \ca{ - D(q) \partial_{xx}p - \frac{1}{2}  \sigma'(q)  (\partial_x p)^2  }\nonumber
\end{eqnarray}
  where $q(x,t)$ is the density-field and $p(x,t)$ is a conjugate field.
  These equations have to be completed by suitable boundary conditions.
 We emphasize that this formalism can be applied to general diffusive systems:
 Physics is embodied   in the
  transport coefficients $D$ and $\sigma$  that
 carry  the  information of the microscopic 
 dynamics relevant at the macroscopic scale.

  In principle, the MFT provides us with a  general framework but
  the non-linear PDEs obtained are very difficult to  solve  in general.
  If we can solve them, we should be able to calculate large deviation functions
  directly at the macroscopic level, without having to cope with the intricate
  combinatorics at the microscopic scale.  For the moment being, very few 
  solutions of these equations exist. The exact results we have
  described in these lectures can be used as benchmarks for the 
   analysis of this new set of hydrodynamic equations,  a field of research
  that   has just opened.

\end{document}